\documentclass[jcp,aip,amsmath,amssymb,reprint,superscriptaddress]{revtex4-2}

\usepackage[dvipdfmx]{graphicx}
\usepackage{color}
\usepackage{epsf}
\usepackage{bm}
\usepackage{physics}
\usepackage{lipsum}
\usepackage[normalem]{ulem}
\usepackage{txfonts}
\usepackage{booktabs}
\usepackage{multirow}

\graphicspath{{./fig/}{../fig/}}

\begin{document}

%%%%%%%%%%%%%%%%%%%%%%%%%%%%%%%%%%%%%%%%%%%%%%%%%%%%%%%%%%%%%%%%%%%%%%%%%%%%%%%%%%%%%%%%%%%%%%%%%%%%%%%%%%%%%%%%%%%%%
\title{Atomistic analysis of nematic phase transition in
 4-cyano-4$^{\prime}$-$n$-alkyl biphenyl liquid crystals: 
Sampling for the first-order phase transition and the free-energy decomposition}

%%%%%%%%%%%%%%%%%%%%%%%%%%%%%%%%%%%%%%%%%%%%%%%%%%%%%%%%%%%%%%%%%%%%%%%%%%%%%%%%%%%%%%%%%%%%%%%%%%%%%%%%%%%%%%%%%%%%%
\author{Shunsuke Ogita}
\affiliation{Division of Chemical Engineering, Department of Materials Engineering Science, Graduate School of Engineering Science, Osaka University, Toyonaka, Osaka 560-8531, Japan}

\author{Yoshiki Ishii}
\affiliation{Department of Data Science, School of Frontier Engineering, Kitasato University, Sagamihara, Kanagawa 252-0373, Japan}

\author{Go Watanabe}
\affiliation{Department of Data Science, School of Frontier Engineering, Kitasato University, Sagamihara, Kanagawa 252-0373, Japan}

\author{Hitoshi Washizu}
\affiliation{Graduate School of Information Science, University of Hyogo, Hyogo 650-0047, Japan}

\author{Kang Kim}
\email{kk@cheng.es.osaka-u.ac.jp}
\affiliation{Division of Chemical Engineering, Department of Materials Engineering Science, Graduate School of Engineering Science, Osaka University, Toyonaka, Osaka 560-8531, Japan}

\author{Nobuyuki Matubayasi}
\email{nobuyuki@cheng.es.osaka-u.ac.jp}
\affiliation{Division of Chemical Engineering, Department of Materials Engineering Science, Graduate School of Engineering Science, Osaka University, Toyonaka, Osaka 560-8531, Japan}

%%%%%%%%%%%%%%%%%%%%%%%%%%%%%%%%%%%%%%%%%%%%%%%%%%%%%%%%%%%%%%%%%%%%%%%%%%%%%%%%%%%%%%%%%%%%%%%%%%%%%%%%%%%%%%%%%%%%%
\date{\today}

%%%%%%%%%%%%%%%%%%%%%%%%%%%%%%%%%%%%%%%%%%%%%%%%%%%%%%%%%%%%%%%%%%%%%%%%%%%%%%%%%%%%%%%%%%%%%%%%%%%%%%%%%%%%%%%%%%%%%
\begin{abstract}

Molecular dynamics (MD) simulations were conducted using the generalized
 replica exchange method (gREM) on the
 4-cyano-4$^{\prime}$-$n$-alkylbiphenyl ($n$CB) system with $n=5$, 6, 7, and 8, which exhibits a
 nematic-isotropic (NI) phase transition. 
Sampling near the phase transition temperature in systems undergoing
 first-order phase transitions, such as the NI phase transition, is
demanding due to the substantial energy gap between the two phases. 
To address this, gREM, specifically designed for first-order
 phase transitions, was utilized to enhance sampling near the NI phase
 transition temperature. 
Free-energy calculations based
 on the energy representation (ER) theory were employed to characterize the
 NI phase transition. 
ER evaluates the insertion free energy of $n$CB molecule for both nematic
 and isotropic phases, revealing a change in the temperature dependence 
 across the NI phase transition.
Further decomposition into
 intermolecular interaction energetic and entropic terms shows
 quantitatively the balance
 between these
 contributions at the NI phase
 transition temperature.
\end{abstract}
%%%%%%%%%%%%%%%%%%%%%%%%%%%%%%%%%%%%%%%%%%%%%%%%%%%%%%%%%%%%%%%%%%%%%%%%%%%%%%%%%%%%%%%%%%%%%%%%%%%%%%%%%%%%%%%%%%%%%
\maketitle
%%%%%%%%%%%%%%%%%%%%%%%%%%%%%%%%%%%%%%%%%%%%%%%%%%%%%%%%%%%%%%%%%%%%%%%%%%%%%%%%%%%%%%%%%%%%%%%%%%%%%%%%%%%%%%%%%%%%%

\section{Introduction}

Liquid crystals are composed of anisotropic molecules known as mesogens, exhibiting
various intermediate phases.
A typical example is the nematic phase, characterized by long-range
orientational order of the molecules without translational order of their
centers of mass. 
The smectic phase, on the other hand, exhibits both
orientational and translational order, with the molecules forming a layered
structure. 
Additional intermediate phases include the columnar and
cholesteric phases.~\cite{chandrasekhar1992Liquid, gennes1995Physics}

The most common mesogens are 4-cyano-$4^{\prime}$-$n$-alkylbiphenyl ($n$CB),
which undergoes a nematic-isotropic (NI) phase transition at room
temperature.
An orientation order parameter is introduced to characterize the nematic
phase in relation to the director, which is the orientation
axis of the entire system. 
The orientation order parameter decreases with increasing temperature,
typically from 0.6 to 0.4 as the NI phase transition 
approaches, and drops discontinuously to zero at the 
transition temperature. 

The discontinuous behavior is indicative of a first-order phase transition,
as described by the mean-field theory of the NI phase
transition.~\cite{stephen1974Physicsa, singh2000Phase, andrienko2018Introduction}
The microscopic models explaining this transition are those of 
Onsager and Maier--Saupe (MS).
The Onsager model is a mean-field model, where the NI transition
is derived by the competition between
excluded volume and rotational entropy effects between
rigid and cylindrical molecules.~\cite{onsager1949EFFECTS}
The MS model is another mean-field model that describes the NI phase
transition based on an orientation-dependent interaction
as the effective intermolecular potential.~\cite{maier1958Einfache}
The fundamental solutions of the Onsager and MS models are
equivalent, as derived by 
minimizing the free energy with respect to
the orientation order parameter at a given temperature.

To obtain a more precise molecular-level understanding beyond mean-field
models, it is essential to analyze the free energy of mesogens
inserted into both isotropic and nematic phases by 
varying the temperature across the NI phase
transition of $n$CB, particularly through molecular dynamics (MD)
simulations.
However, this free-energy calculation remains challenging and has not
been thoroughly elucidated, despite 
numerous MD simulations conducted on the NI phase transition of
liquid crystal systems.~\cite{zannoni2001Molecular,
cacelli2002Stability, berardi2004Can,
care2005Computer, capar2006Molecular, wilson2007Molecular,
cacelli2007Liquida, 
berardi2008Computer, cifelli2008Atomistic, tiberio2009Silico, zhang2011Atomistic,
palermo2013Atomistic, ju2016Prediction, pelaez2007Molecular, 
zannoni2018Idealised, sidky2018Silico, 
allen2019Molecular, sasaki2020Atomistic, shi2020Automated,
sheavly2020Molecular, zannoni2022Liquid, watanabe2023Missing, sarkar2024Calculation}

Here, the free energy of mesogens is defined as the insertion free energy
associated with transferring a
mesogen, considered as a solute molecule, from vacuum into a solution
system composed of identical mesogens; the molecules other than the tagged one is the solvent.
This is equivalent to the 
work required to `switch on' the solute-solvent intermolecular interaction. 
In principle, the free-energy calculation is performed via thermodynamic
integration in MD simulations, which 
requires an ensemble average over
specific arrangements of the translational and orientational degrees of
freedom between the solute and solvent. 
In the atomistic description of
mesogen molecules, the
degrees of freedom exceed six
dimensions when accounting for the molecule's internal degrees of
freedom. 
Thus, a free-energy analysis is a key to understanding an NI phase transition,
and in the analysis, it is desirable to address the roles of such
intermolecular interactions as electrostatic and van der Waals with keeping the atomic-level resolution.

%%%%%%%%% Fig. 1 %%%%%%%%%%%%%%%%%%%%%%%%%%%%%%%%%%%%%%%%%%%%%%%%%%%%%%%%%%%%%%%%%%%%%%%%%
\begin{figure*}[t]
\centering
\includegraphics[width=0.4\textwidth]{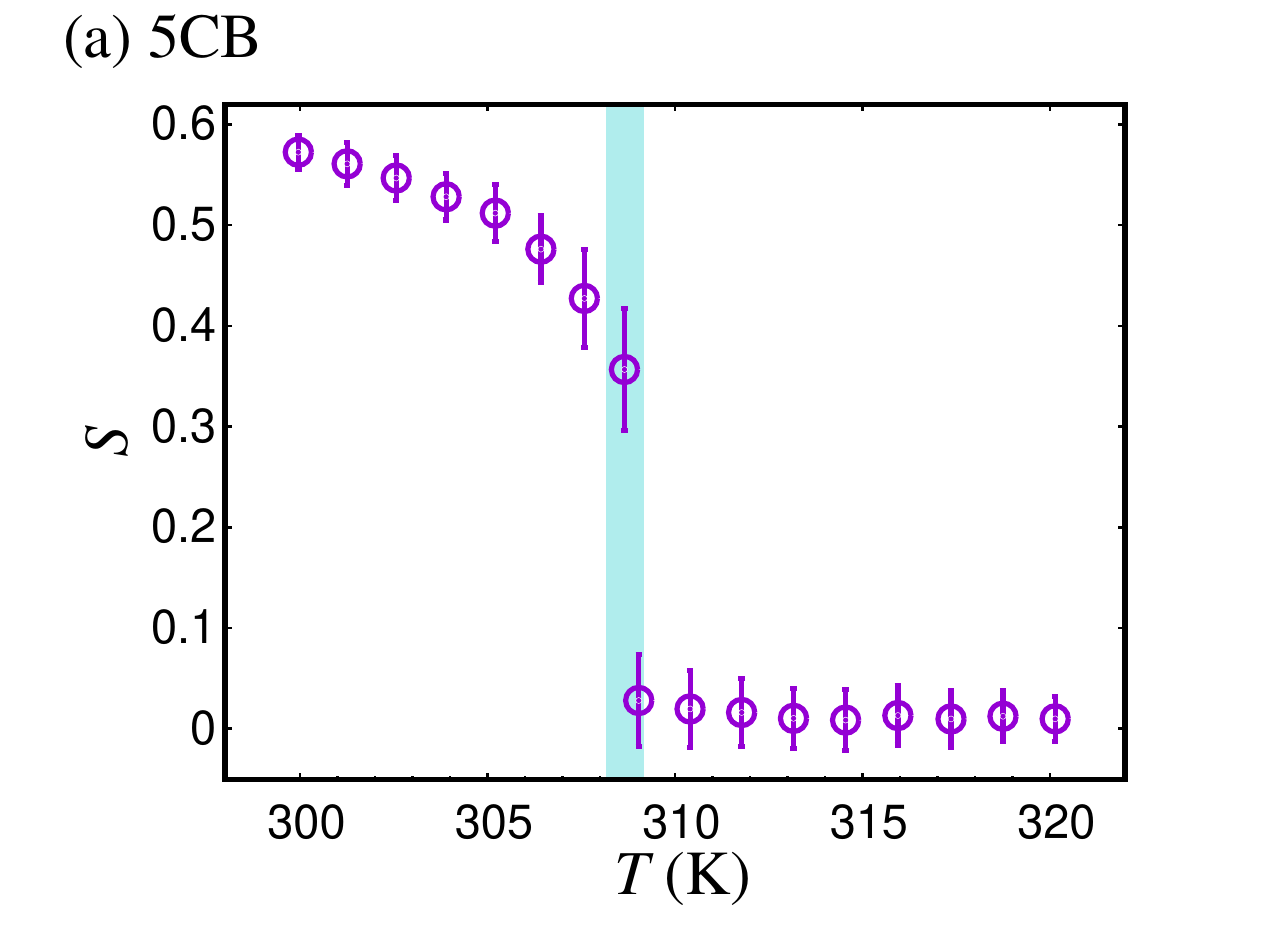}
\includegraphics[width=0.4\textwidth]{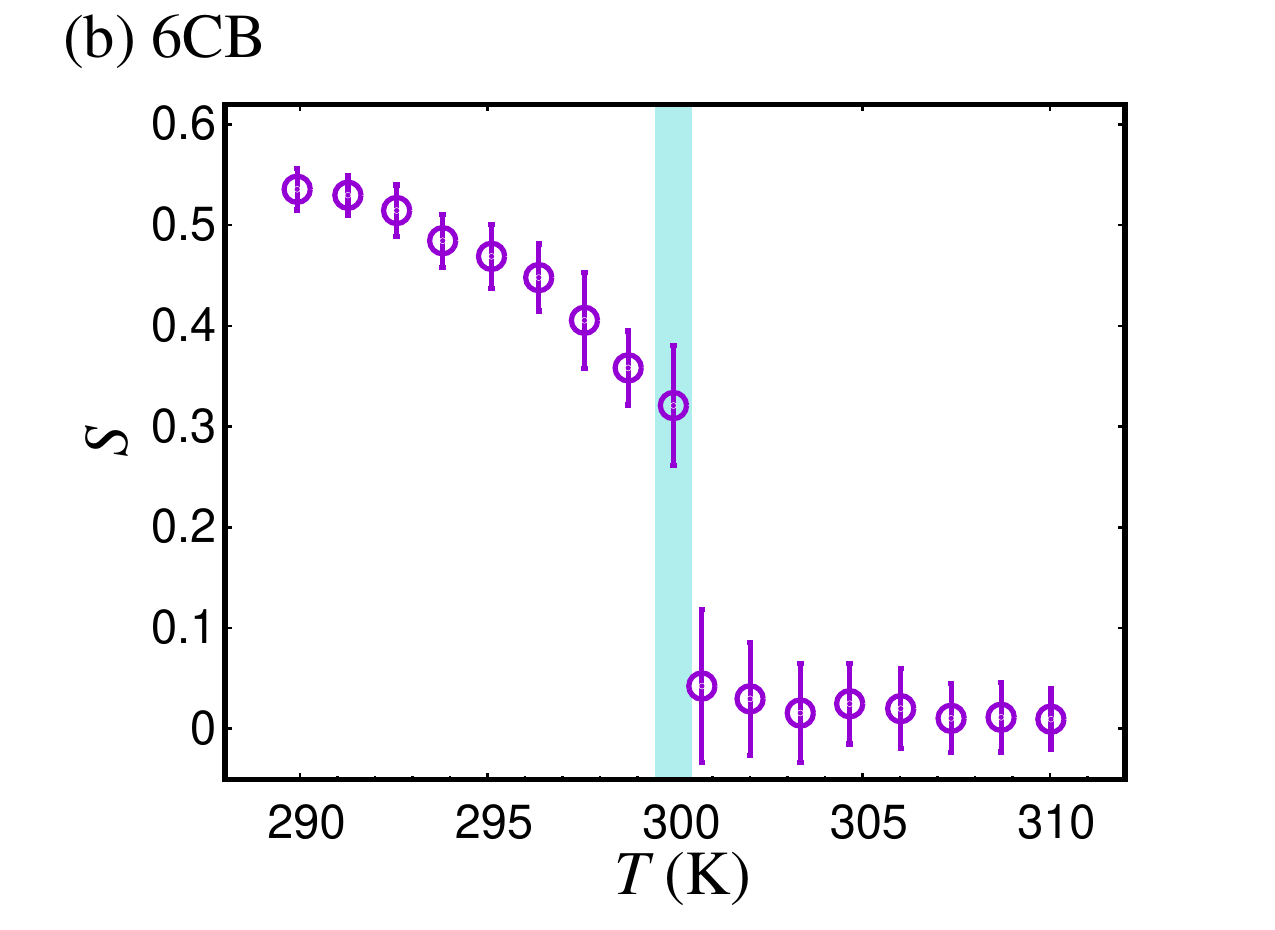}
\includegraphics[width=0.4\textwidth]{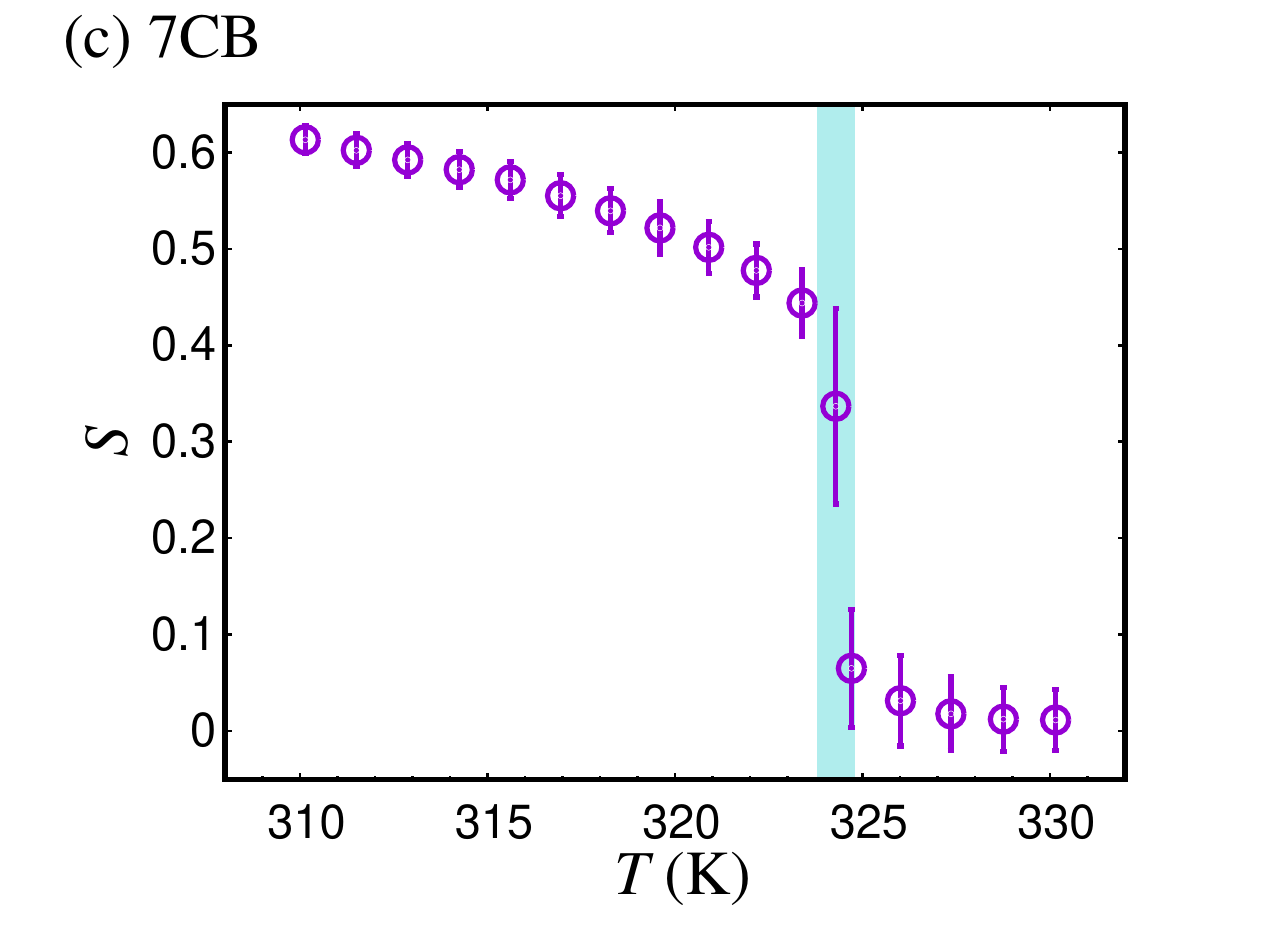}
\includegraphics[width=0.4\textwidth]{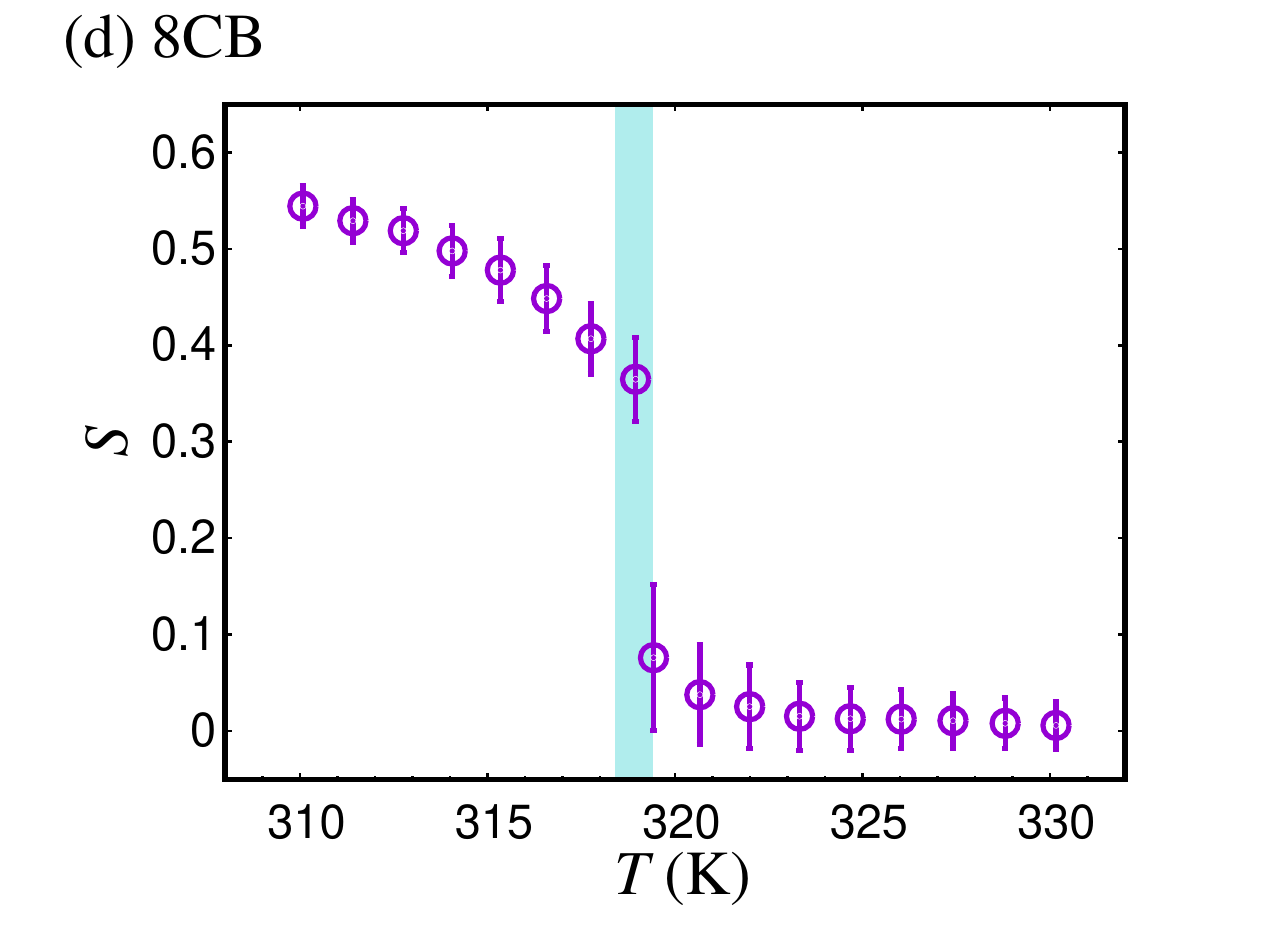}
\caption{Temperature dependence of orientational order parameter $S$ for
 5CB (a), 6CB (b), 7CB (c), and 8CB (d).
The molecular axis is defined using the CN bond, represented by the
 purple vector in Fig.~S1 of the supplementary material.
The error bar at each temperature corresponds to the standard deviation.
The vertical color bar indicates the NI transition temperature $T_\mathrm{NI}$.
}
\label{fig:P2}
\end{figure*}
%%%%%%%%%%%%%%%%%%%%%%%%%%%%%%%%%%%%%%%%%%%%%%%%%%%%%%%%%%%%%%%%%%%%%%%%%%%%%%%%%%%%%%%%%%

In this study, we performed MD simulations of $n$CB system with $n=5$,
6, 7, and 8 using the united-atom (UA) model.
We also 
applied the generalized replica-exchange method (gREM), 
designed to improve sampling of the states near the first-order phase
transition by utilizing a non-Boltzmann weight.~\cite{kim2010Generalized,
lu2012Exploring, lu2013Order, 
lu2014Investigating, 
malolepsza2015Isobaric, malolepsza2015Water, malolepsza2015Entropic, 
lu2016Freezing, 
stelter2017Enhanced, piskulich2022Machine}
Note that gREM simulations of 5CB liquid crystals~\cite{takemoto2022Simulating} 
and other applications of the replica-exchange method to coarse-grained
liquid crystalline systems were conducted.~\cite{berardi2009Softcore, 
kowaguchi2021Phase, kowaguchi2022Optimal,
kowaguchi2024Hysteresis}
The insertion free energy was subsequently evaluated using the energy
representation (ER) theory
for the states obtained through gREM, enabling the
assessment of the thermodynamic 
stability of mesogens in both isotropic and nematic phases across varying temperatures.
Within the ER framework, it is also possible to decompose the insertion
free energy into the energetic and entropic contributions. 
We conduct
free-energy decomposition and address the roles of these contributions.

%%%%%%%%% Fig. 2 %%%%%%%%%%%%%%%%%%%%%%%%%%%%%%%%%%%%%%%%%%%%%%%%%%%%%%%%%%%%%%%%%%%%%%%%%
\begin{figure}[t]
\centering
\includegraphics[width=0.4\textwidth]{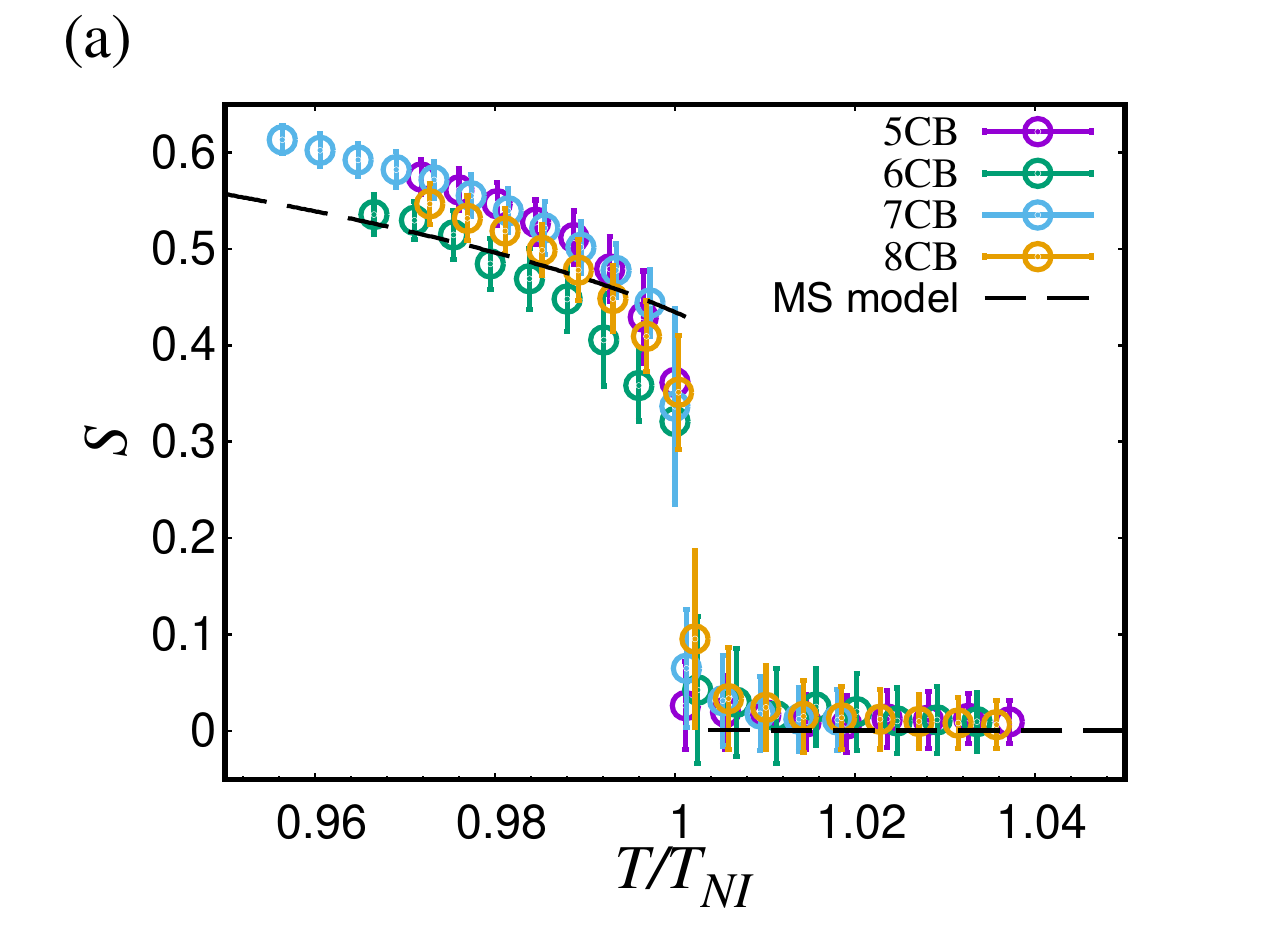}
\includegraphics[width=0.4\textwidth]{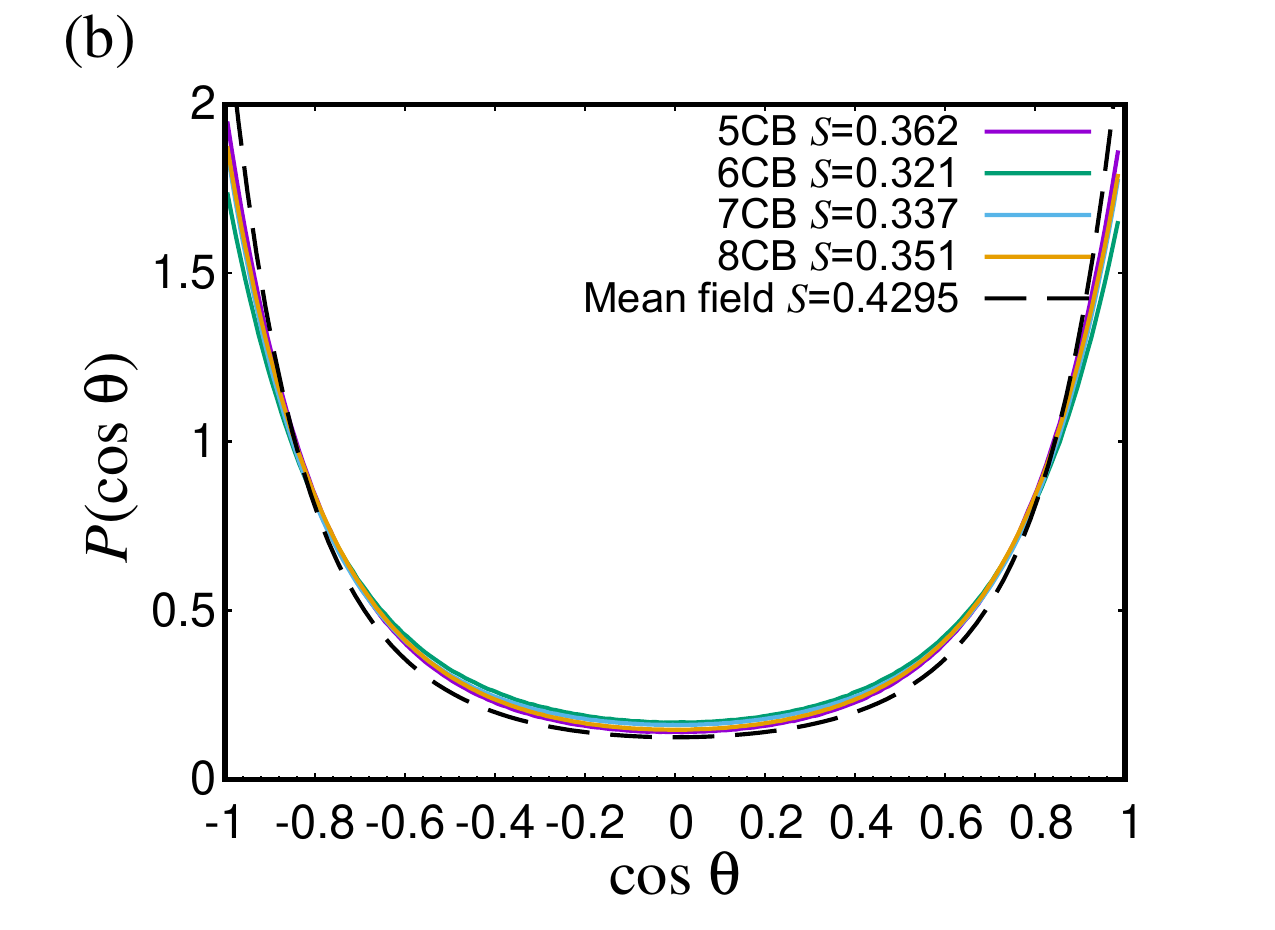}
\caption{(a) Orientational order parameter $S$ as a function of
 temperature scaled by $T_\mathrm{NI}$
from
 MD simulations for $n$CB ($n=5$, 6, 7, and 8).
For comparison, the result of the MS model is also plotted as black dashed curve.
(b) Orientation angle distribution function
 $P(\cos\theta)$ from
 MD simulations at $T_\mathrm{NI}$ for $n$CB ($n=5$, 6, 7, and 8).
The
 corresponding values of the orientation order parameter $S$ are presented.
For comparison, the MS model described by Eq.~\eqref{eq:pcosine} with $S
 = 0.4295$ is also plotted as black dashed curve.
}
\label{fig:pcosine}
\end{figure}
%%%%%%%%%%%%%%%%%%%%%%%%%%%%%%%%%%%%%%%%%%%%%%%%%%%%%%%%%%%%%%%%%%%%%%%%%%%%%%%%%%%%%%%%%%

\section{Model and methods}

The molecular model employed was the UA model of the AMBER force field,
developed by Tiberio \textit{et al.} to reproduce the experimental behavior near
the NI phase transition temperature.~\cite{tiberio2009Silico}
The molecular structures of $n$CB ($n=5$, 6, 7, and 8) are
illustrated in Fig.~S1 of the supplementary material.
The system consists of $N=4000$ molecules; it was seen in
Ref.~\onlinecite{takemoto2022Simulating} that the system size with $N = 4000$
is enough for treatments of the NI transition. 
The simulation box is a
cubic cell with periodic boundary conditions. 
All MD simulations in this study were performed using 
the Large-scale Atomic/Molecular Massively Parallel Simulator
(LAMMPS)~\cite{plimpton1995Fast}.
The time step was set to 2 fs. 
The total simulation times of gREM were 360 ns for 5CB and
6CB and 450 ns for 7CB and 8CB, respectively, starting
from an initial configuration with a random structure.
Temperature was maintained using Nos\'{e}--Hoover thermostat, while 
pressure was controlled by isotropic Nos\'{e}--Hoover barostat with a pressure of 1
atm.
For the isobaric version of gREM, 
the enthalpy $H(=U+pV)$ dependent non-Boltzmann weight $W_\alpha(H)$
$(\alpha=1, \cdots, M)$ is utilized.~\cite{malolepsza2015Isobaric,
malolepsza2015Water, malolepsza2015Entropic}
Here, $U$, $p$, and $V$ represent the potential energy, pressure, and
volume of the system, respectively.
In addition, $\alpha$ denotes the replica index and $M$ is the number of replicas.
The weight $w_\alpha$ is connected with the effective potential through
$w_\alpha = -k_\mathrm{B}\ln W_\alpha$, and is determined by inverse
mapping of the effective temperature:
\begin{align}
T_\alpha (H)=[{\partial w_\alpha}/{\partial H}]^{-1}.
\label{eq:eff_temp}
\end{align}
The statistical temperature is defined as $T_\mathcal{S}(H)=[\partial
\mathcal{S}/\partial H]^{-1}$, where $\mathcal{S}$ is the entropy.
At replica $\alpha$, the statistical temperature $T_\mathcal{S}$ is
evaluated by the most probable value of the enthalpy $H_\alpha^*$, 
with $T_\alpha(H_\alpha^*) = T_\mathcal{S}(H_\alpha^*)$.
The simplest expression for $T_\alpha$ is the linear effective
temperature, given by
\begin{align}
T_\alpha (H)=\lambda_\alpha+\gamma (H-H_0),
\label{eq:linear_eff_temp}
\end{align}
where 
$\lambda_\alpha$ and $\gamma$ are control parameters representing the
intercept and slope of the straight line, respectively, at an arbitrary
chosen enthalpy $H_0$.
Note that $\gamma$ is chosen to be negative so that $T_\alpha(H)$
intersects $T_\mathcal{S}(H)$ at only one point.
In practice, $\gamma$ is selected as
\begin{align}
  \gamma=\frac{T_M-T_1}{\tilde{H}_1-\tilde{H}_M},
  \label{eq:slope}
\end{align}
where $\hat{H}_1$ and $\hat{H}_M$ represent the average 
enthalpies from MD simulations at the predefined minimum and maximum temperatures, $T_1$ 
and $T_M$, respectively.
Furthermore, $H_0$ is set to $\hat{H}_1$, and the conditions
$\lambda_1=T_1$ and $\lambda_M=T_M-\gamma(\hat{H}_M-\hat{H}_1)$ leads to 
\begin{align}
  \lambda_\alpha=\lambda_1+(\alpha-1)\Delta \lambda,
  \label{eq:lambda}
\end{align}
where $\Delta \lambda=(\lambda_M-\lambda_1)/(M-1)$.
The exchange between neighbouring replicas, $\alpha$ (with enthalpy $H$)
and $\alpha'$ (with enthalpy $H'$), is determined by the
Metropolis method:
\begin{align}
    A_{\alpha, \alpha'}=\min[1,\exp(\Delta_{\alpha, \alpha'})],
    \label{eq:acceptance_ratio}
\end{align}
where
$\Delta_{\alpha, \alpha'}=w_{\alpha'}(H')-w_{\alpha'}(H)+w_{\alpha}(H)-w_\alpha(H')$.
In this study, the number of replicas was set to $M = 17$.
Other parameters, $T_1$, $T_M$, $\hat{H}_1$, $\hat{H}_M$, $\gamma$ are
summarized in Table~S1 of the supplementary material.
The LAMMPS input files used in this study and 1 ns trajectory
files are openly available at the repository of
Zenodo (see DATA AVAILABILITY statement).

Figure S2 of the supplementary material shows the
distribution of the 
 enthalpy normalized by the number of molecules, $H/N$, at each replica, 
for each replica for $n$CB ($n=5$, 6, 7, and 8).
These results were calculated from the final 36 ns (5CB and
6CB) and 108 ns (7CB and 8CB) trajectories in the gREM
calculation.
The findings demonstrate sufficient overlap of enthalpy
distributions between replicas. 
However, 
the distributions are broad in certain regions of enthalpy 
between specific replicas, which can be attributed to
enthalpy gaps caused by the NI phase transition. 
Despite this, we confirmed that the replica exchange acceptance ratio
remains maintained, as shown in Fig.~S3 of the supplementary material.
To enhance the replica exchange acceptance ratio, particularly for
larger system sizes, 
it is necessary to increase the number of replicas $M$, significantly
raising computational demands.
Figure S4 of the supplementary material displays the statistical
temperature $T_\mathcal{S}$ of each temperature as a function of $H/N$,
with the linear effective
temperature, 
as defined in Eq.~\eqref{eq:linear_eff_temp}, represented 
by straight lines.
Henceforth, the statistical temperature $T_\mathcal{S}$ is referred to
simply as $T$.

The ER theory provides the stable and efficient 
formalism for the free-energy calculation in MD simulations.~\cite{matubayasi2000Theory, matubayasi2002Theory,
matubayasi2003Theory, sakuraba2014Ermod, matubayasi2019EnergyRepresentation}
In the ER theory, a multidimensional coordinate that accounts for the
degrees of freedom, including
the positions and orientations of solvent and solute molecules, 
is projected to the intermolecular interaction energy
coordinate, $\varepsilon$.
Based on the Kirkwood's charging equation, 
the expression for the insertion free energy $\Delta \mu$ in the ER theory is given by
\begin{equation}
\Delta \mu = \int_0^1 d\lambda \int d \varepsilon \frac{\partial
 u_\lambda (\varepsilon)}{\partial \lambda} \rho_\lambda(\varepsilon),
\label{eq:Kirkwood}
\end{equation}
where 
$u_\lambda(\varepsilon)$ represents the potential energy, which
continuously varies from
0 to $u$ according to the coupling parameter
$\lambda$ $(0\le \lambda \le 1)$ such that
$u_0=0$ and $u_1=u$, where $u$ is the potential function of interest between the mesogen molecules.
Additionally, $\rho_\lambda(\varepsilon)$ denotes the 
energy distribution function of solvent-solute interaction energy at a
given coupling parameter $\lambda$.
By integrating by parts, Eq.~\eqref{eq:Kirkwood} reduces to 
\begin{align}
\Delta \mu &=\int_{}^{} d \varepsilon \varepsilon \rho \qty(\varepsilon)
                -\int_{0}^{1} d\lambda \int d \varepsilon u_\lambda (\varepsilon) \pdv{\rho_\lambda \qty(\varepsilon)}{\lambda},\label{eq:delta_mu}\\
&=\langle u \rangle+\int\mathcal{F}[\rho (\varepsilon), \rho_0 (\varepsilon)]d\varepsilon
\label{eq:ER}
\end{align}
where the first term represents
the average sum of solute-solvent interactions in the target system with $\lambda =
1$, and the second term accounts for the free-energy penalty 
due to the insertion of the solute molecule in Eq.~\eqref{eq:delta_mu}.
Furthermore, in the ER theory, the second term 
is approximated by a density functional form $\mathcal{F}$ using
$\rho_0(\varepsilon)$ with $\lambda=0$,
representing 
the reference solvent system, and $\rho(\varepsilon)$
with $\lambda=1$, representing the target system, as expressed in Eq.~\eqref{eq:ER}.
In other words, ER theory approximates the insertion free energy
from the energy distribution functions of the initial ($\lambda=0$) and
final ($\lambda=1$) states, thereby reducing the computational cost by avoiding 
MD simulations of intermediate states.
It has also been reported that the error due to the use of density
functional approximation is no greater than the inherent error in the
force field.~\cite{karino2013Interactioncomponent}
Specifically, 
the ER method was used to calculate
the solvation free energy of amino acid analogs 
in a pure-water solvent, showing 
a typical discrepancy of less than 1 kcal/mol
compared to the numerically exact values obtained using the Bennett
acceptance ratio (BAR) method.~\cite{karino2013Interactioncomponent}
In Eq.~\eqref{eq:ER}, the first term is the average interaction energy of the solute
with the solvent in the system of interest. The second term corresponds
to the free-energy penalty due to the reorganization of the solvent
structures with the insertion of the solute. When the solvent degrees of
freedom are more restricted by the solute, the second term is more
positive.
In the following, the first and second terms of Eq.~\eqref{eq:ER} are
called energetic and entropic terms, respectively.

%%%%%%%%% Fig. 3 %%%%%%%%%%%%%%%%%%%%%%%%%%%%%%%%%%%%%%%%%%%%%%%%%%%%%%%%%%%%%%%%%%%%%%%%%
\begin{figure*}[t]
\centering
\includegraphics[width=0.4\textwidth]{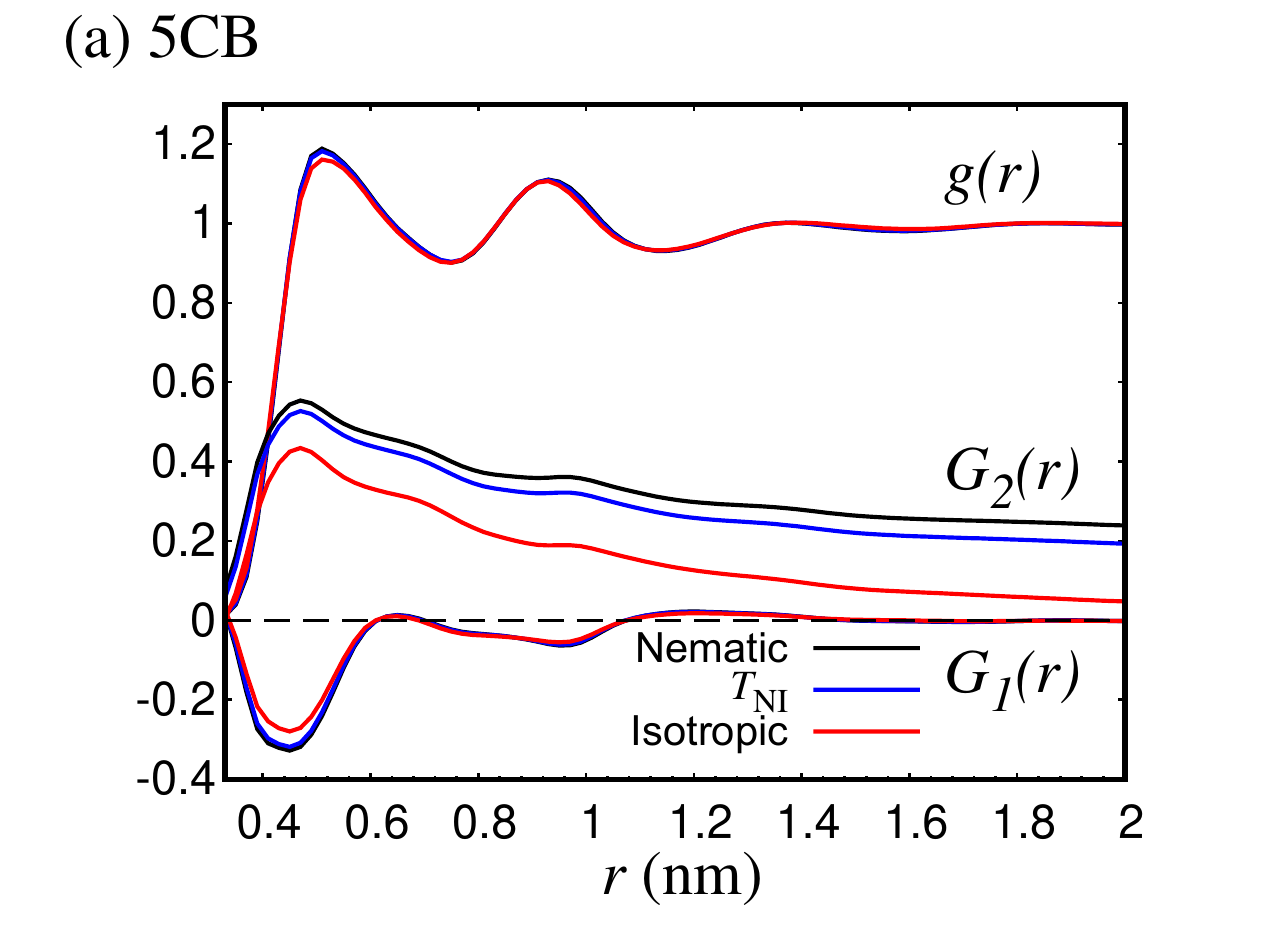}
\includegraphics[width=0.4\textwidth]{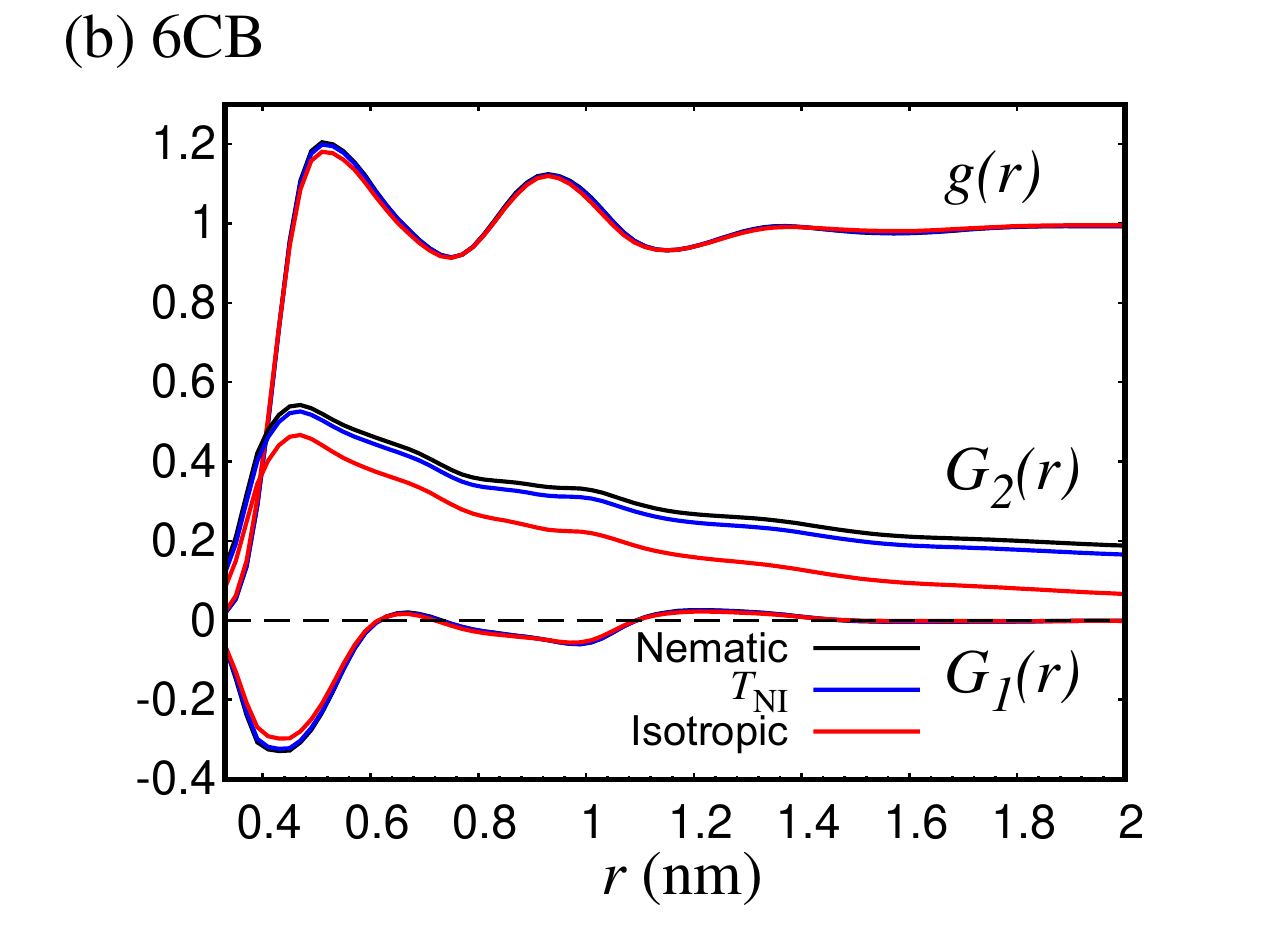}
\includegraphics[width=0.4\textwidth]{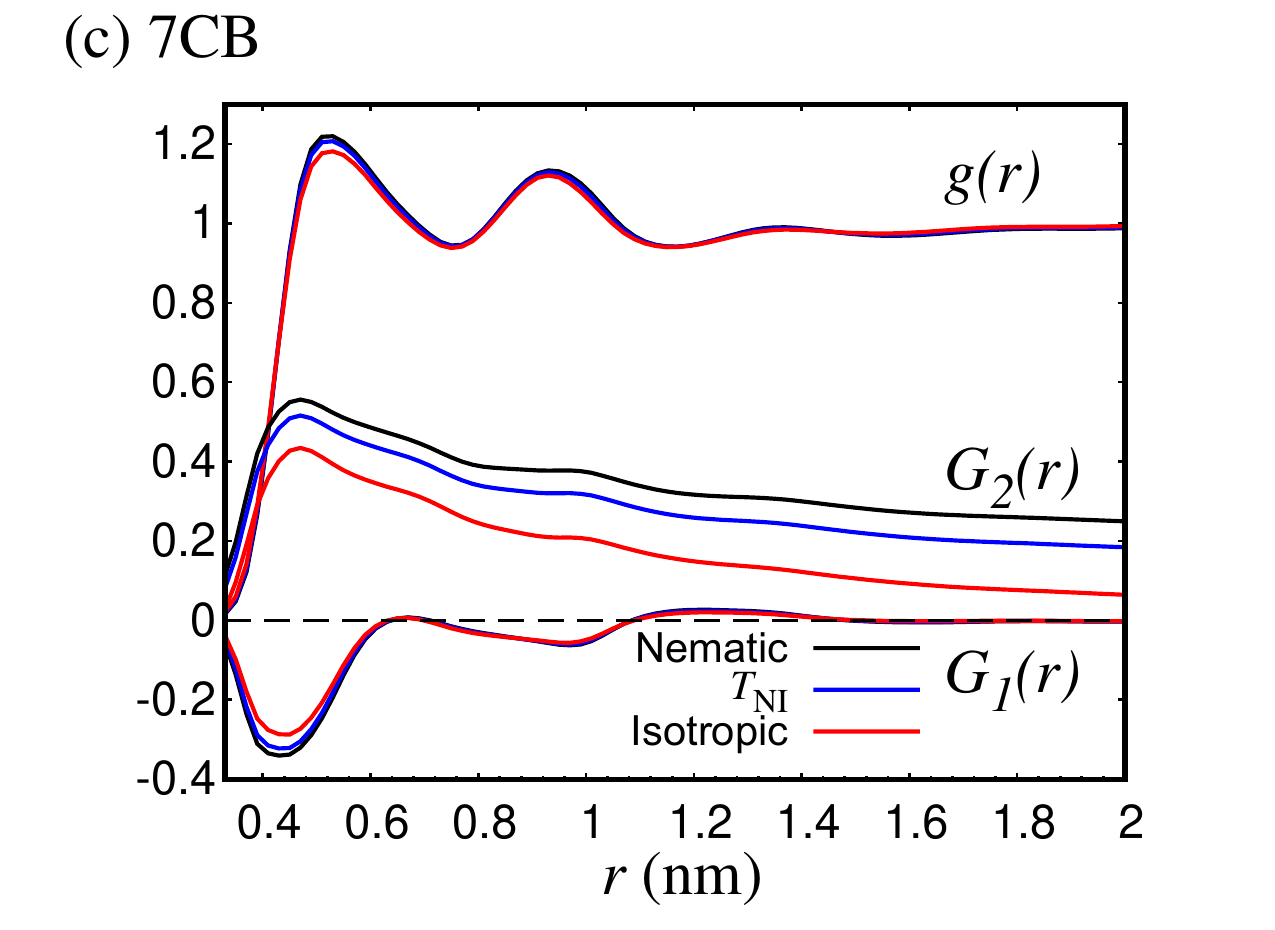}
\includegraphics[width=0.4\textwidth]{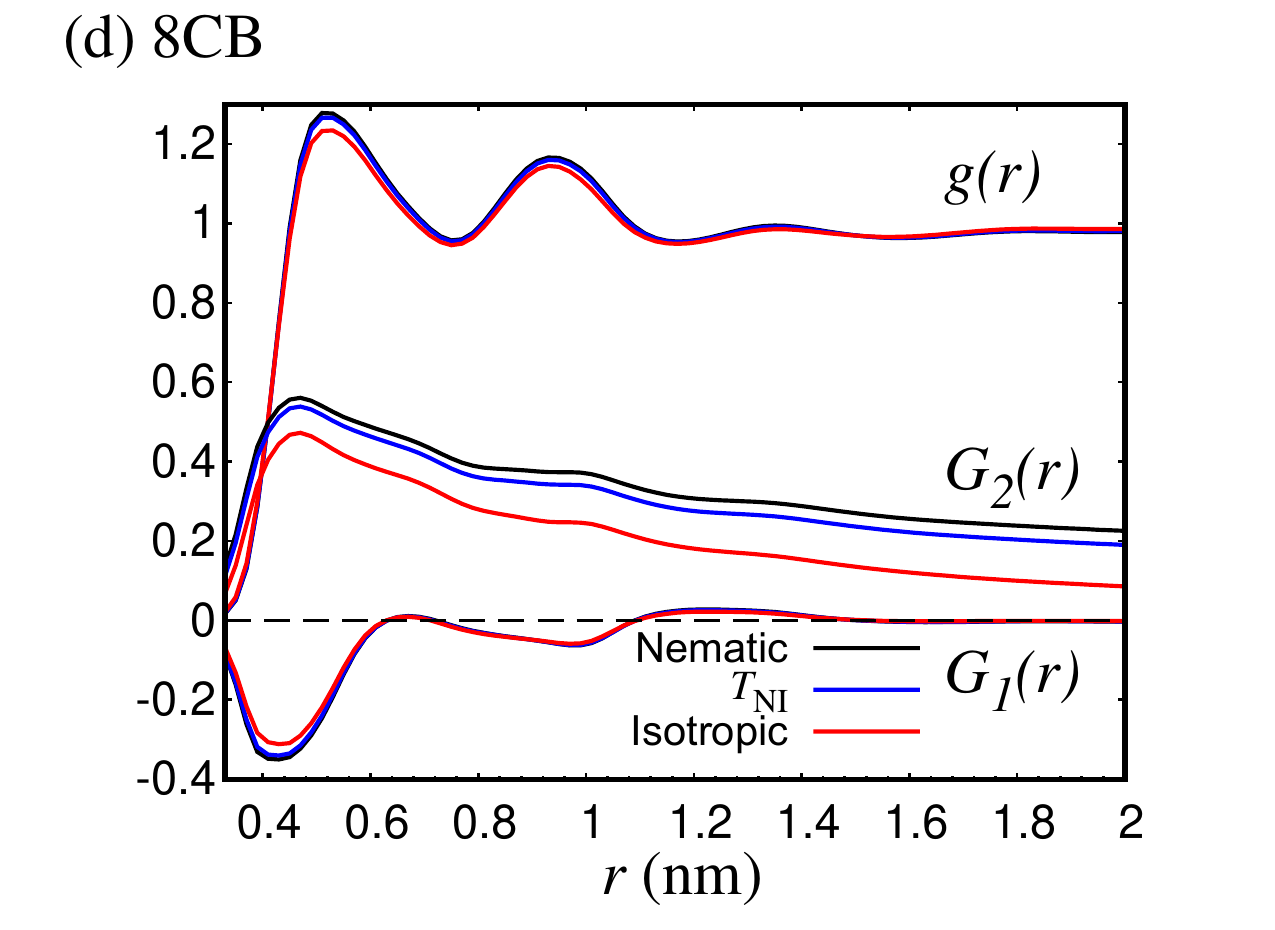}
\caption{Radial distribution function $g(r)$ and orientational
 correlation functions $G_1(r)$ and $G_2(r)$ for
 5CB (a), 6CB (b), 7CB (c), and 8CB (d).
The results are plotted for replicas at $T_\mathrm{NI}$, as well as for
 replicas above and below $T_\mathrm{NI}$ (isotropic and nematic phases).
}
\label{fig:G}
\end{figure*}
%%%%%%%%%%%%%%%%%%%%%%%%%%%%%%%%%%%%%%%%%%%%%%%%%%%%%%%%%%%%%%%%%%%%%%%%%%%%%%%%%%%%%%%%%%

\section{Results and discussion}

\subsection{Orientational order}

To assess the NI phase transition for $n$CB ($n=5$, 6, 7, and 8), we
calculated the temperature dependence of the orientational order
parameter, $S$, which is defined as 
\begin{align}
S=\frac{1}{N} \sum_{i=1}^{N} 
P_2(\cos\theta_i) 
%= \sum_{i=1}^{N} 
%\left(\frac{3}{2} \cos^2 \theta_i -
% \frac{1}{2}\right), 
\label{eq:2nd_Legendre}
\end{align}
where $\theta_i$ represents the orientational angle between the $i$-th molecular axis $\bm{u}_i$
and the director $\bm{n}$.
Furthermore, $P_2(x)=(3x^2-1)/2$ is 
the second-order Legendre polynomial.

In MD simulations, the direction of the director $\bm{n}$ cannot be
predetermined.
Alternatively, the second-rank order parameter 
tensor,
\begin{align}
\bm{Q}= \frac{1}{N}\sum_{i=1}^{N} \left(\frac{3}{2}\bm{u}_i \otimes \bm{u}_i - \frac{1}{2}\bm{I}\right),
\label{eq:order_tensor}
\end{align}
is employed to analyze the orientational order of biaxial
nematic liquid crystals.
Here, $\bm{I}$ denotes the identity matrix.
Due to the traceless property of $\bm{Q}$, three eigenvalues,
$\lambda_{-}$, $\lambda_{0}$, and $\lambda_{+}$ ($\lambda_{-}<
\lambda_{0} < \lambda_{+}$), satisfies the condition $\lambda_- + \lambda_0 + \lambda_+=0$.
The orientational order
parameter $S$ can be evaluated by 
the maximum eigenvalue $\lambda_{+}$ as $S = \lambda_+$, and 
the corresponding eigenvector represents the director
$\bm{n}$.
However, this definition results in positive values for $S$ even in the
isotropic phase due to the traceless property of $Q$. 
In this study, following the approach of Eppenga and Frenkel, 
$S$ was calculated by $-2\lambda_0$, considering $\lambda_0 \approx
\lambda_- = -\lambda_+/2$ in the nematic phase.~\cite{eppenga1984Monte}
This definition ensures that 
$S$ fluctuates around zero in the isotropic phase, indicating the absence
of orientational order.

The temperature dependence of the ensemble average of $S$ is
plotted in Fig.~\ref{fig:P2}.
In this study, the molecular axis was defined as the CN bond,
represented by the purple 
vector in Fig.~S1 of the supplementary material.
Here, $S$ was calculated from the final 36 ns (5CB and
6CB) and 108 ns (7CB and 8CB) trajectories in the gREM
simulation, consistent with the calculation of the enthalpy
distribution shown in Fig.~S2 of the supplementary material.
In all liquid crystal systems, $S$ exhibits a value near 0 at
high temperatures, indicating the isotropic phase. 
As the temperature decreases,
a discontinuous behavior is observed at a certain temperature, and $S$
reaches a value around 0.4, signaling a phase transition to the nematic
phase. 
The highest temperature at which $S$ reaches
approximately 
0.4 was defined as the NI phase transition temperature, $T_\mathrm{NI}$.
The $S$ value gradually increases as the temperature
decreases further, 
indicating a strengthening of the nematic phase's orientational order.
Although 8CB of the same UA model is known to undergo a smectic phase
transition upon further
cooling,~\cite{tiberio2009Silico, palermo2013Atomistic} this study
specifically targets the NI phase transition and does not
address the smectic-nematic phase transition.
Simulated $T_\mathrm{NI}$ values for $n$CB ($n=5$, 6, 7, and 8) are
308.7 K (replica 8), 300.0 K (replica 9), 324.3 K (replica 12), and
318.9 K (replica 8), respectively.
The results demonstrate that $n$CB exhibits odd-even effects as the number of carbons in
the alkyl chain increases, consistent with findings from other MD simulations.~\cite{berardi2004Can,
capar2006Molecular, cacelli2007Liquida, 
cifelli2008Atomistic, tiberio2009Silico} 
We also calculated the orientational order parameter $S$ using
an alternative definition of the molecular axis, defined by the line
connecting N and C of the phenyl ring, represented by the red vector in
Fig.~S1 of the supplementary material.
This definition was examined in a previous study.~\cite{sarkar2024Calculation}
The temperature dependence of $S$, shown in Fig.~S5 of the 
supplementary material, demonstrates that 
$S$ remains unaffected by the choice of the
molecular axis.

%%%%%%%%% Fig. 4 %%%%%%%%%%%%%%%%%%%%%%%%%%%%%%%%%%%%%%%%%%%%%%%%%%%%%%%%%%%%%%%%%%%%%%%%%
\begin{figure*}[t]
\centering
\includegraphics[width=0.4\textwidth]{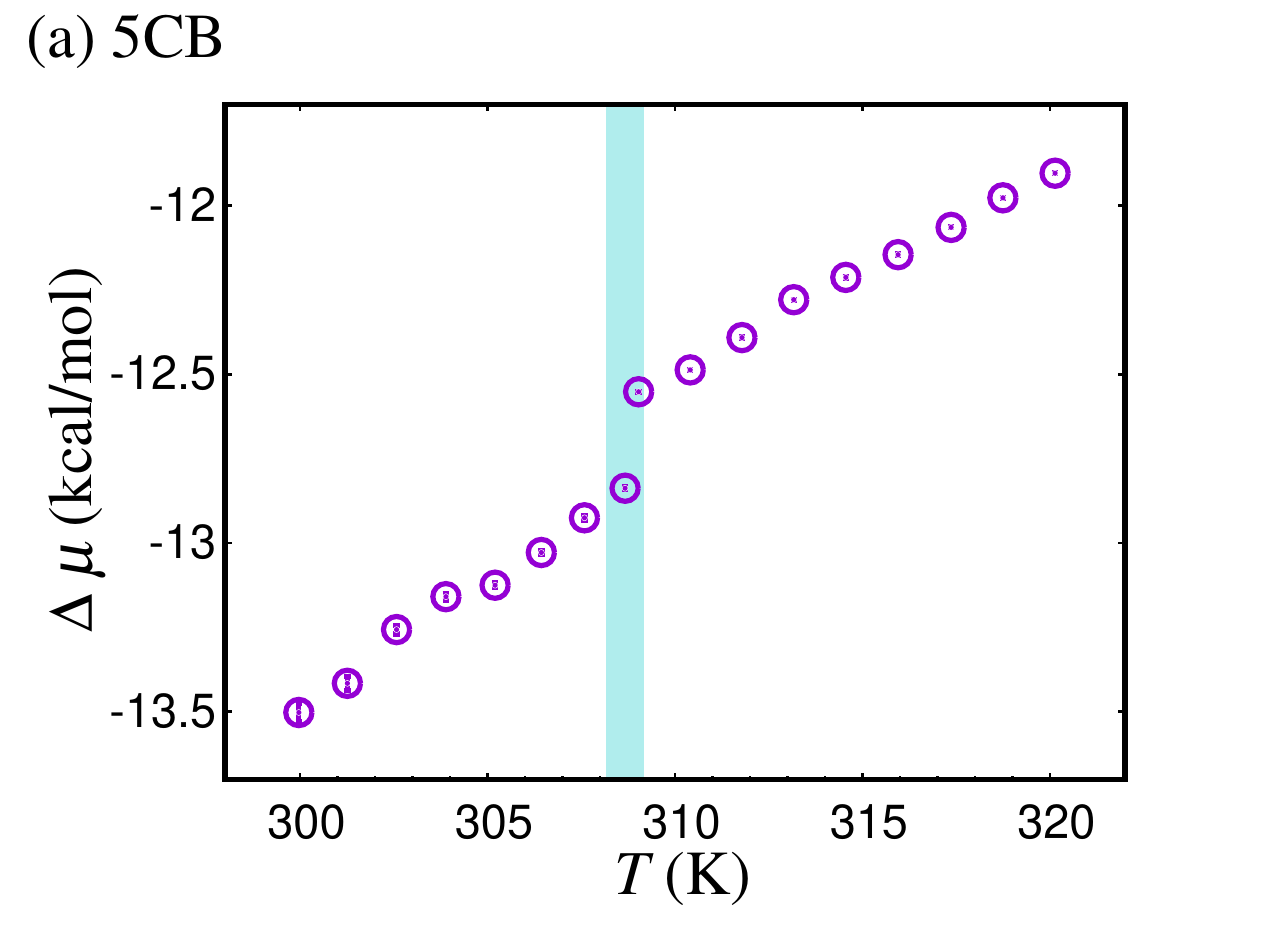}
\includegraphics[width=0.4\textwidth]{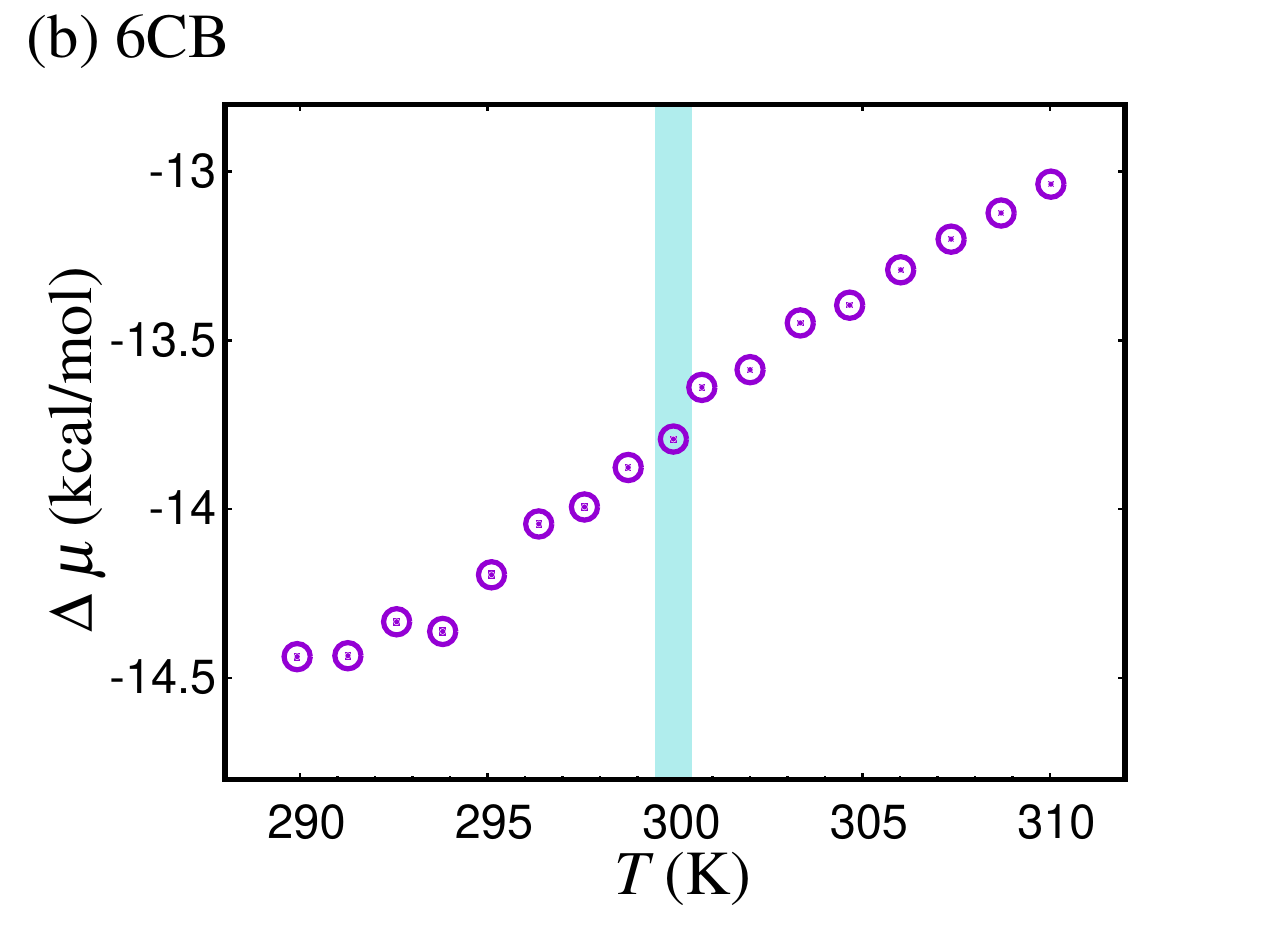}
\includegraphics[width=0.4\textwidth]{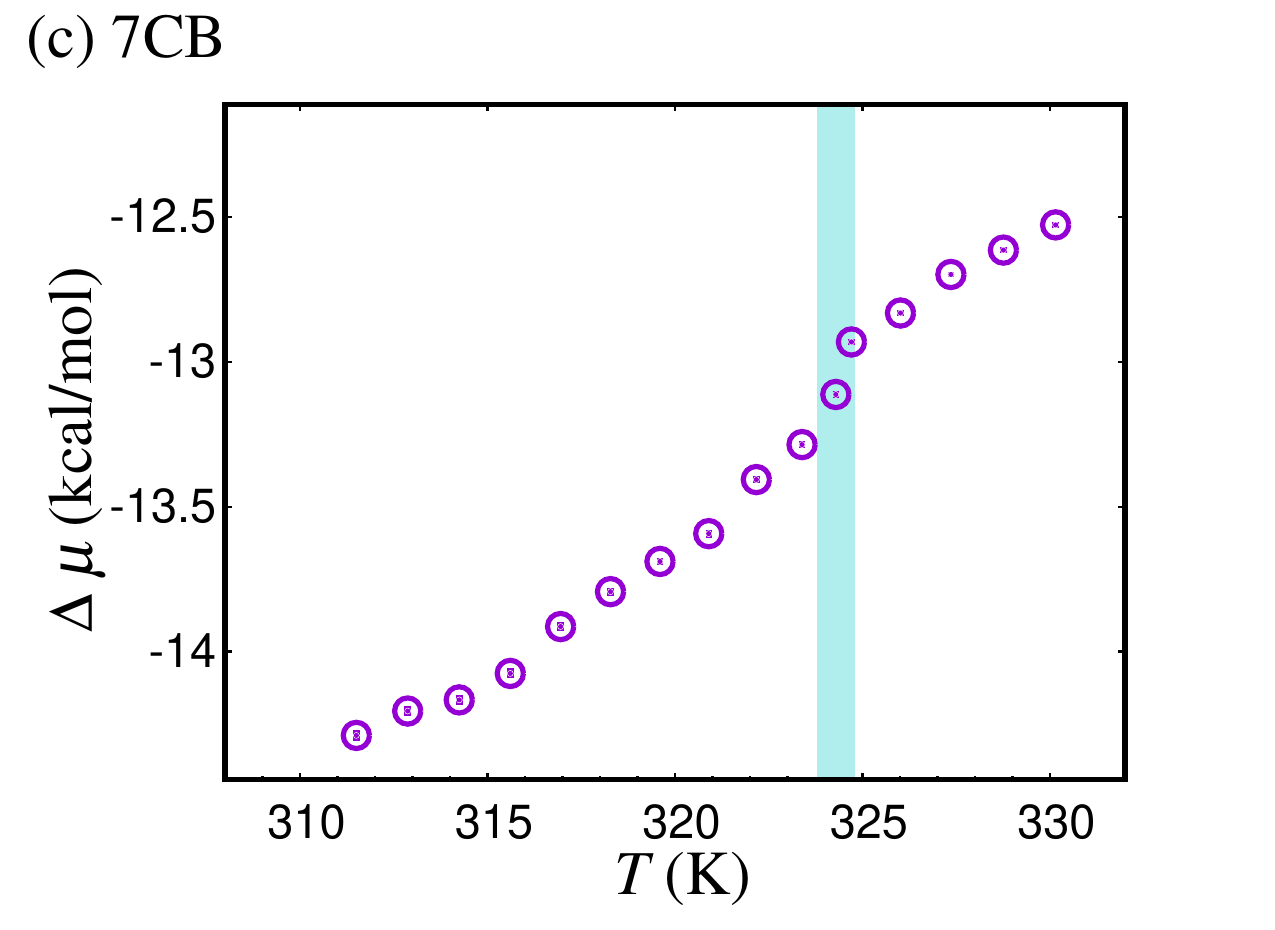}
\includegraphics[width=0.4\textwidth]{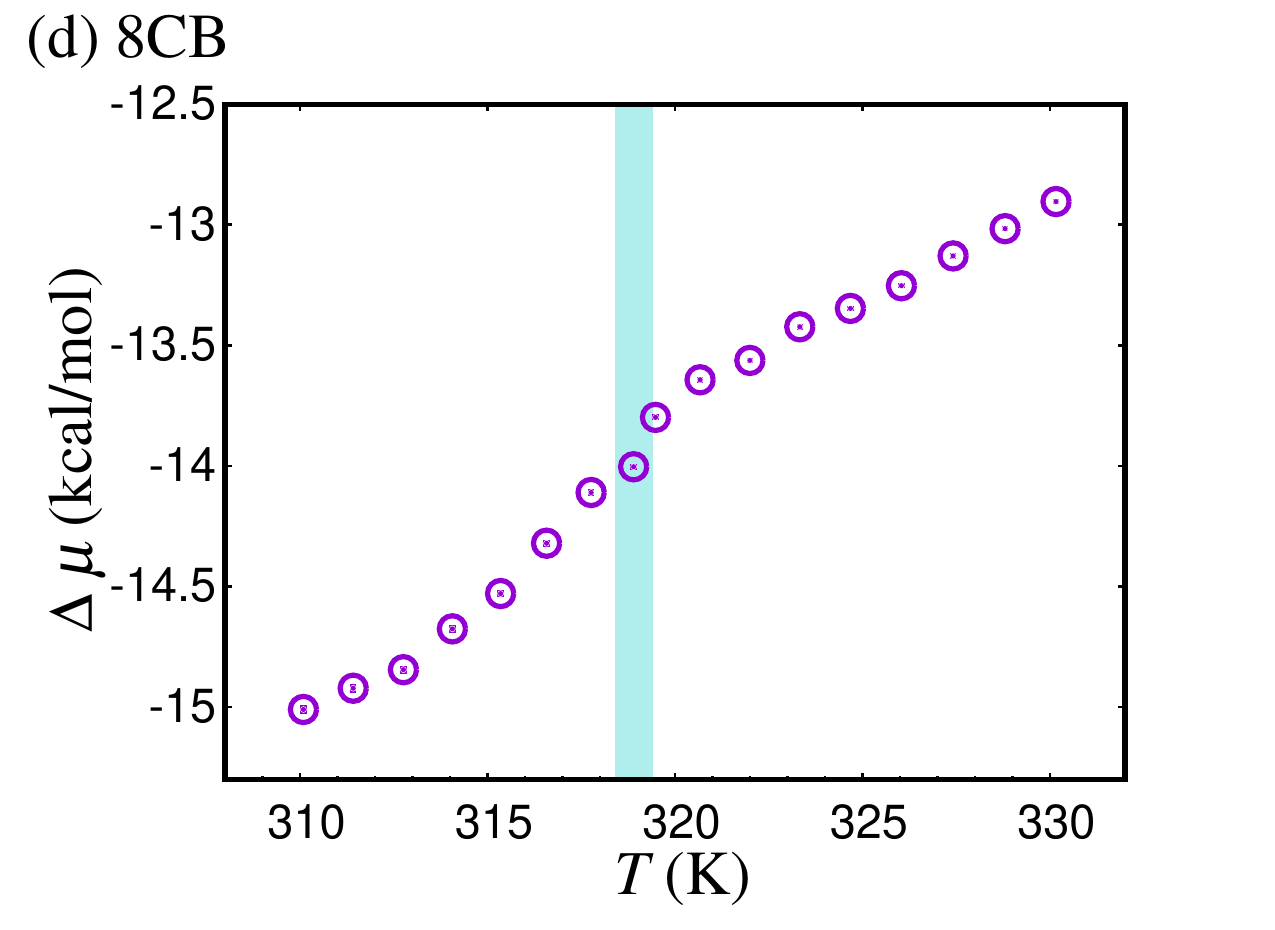}
\caption{Temperature dependence of the insertion free energy $\Delta
 \mu$ for 5CB (a), 6CB (b), 7CB (c), and 8CB (d).
The error bars at each temperature represent the standard deviation.
The vertical color bar indicates the NI transition temperature $T_\mathrm{NI}$.
}
\label{fig:fe}
\end{figure*}
%%%%%%%%%%%%%%%%%%%%%%%%%%%%%%%%%%%%%%%%%%%%%%%%%%%%%%%%%%%%%%%%%%%%%%%%%%%%%%%%%%%%%%%%%%

%%%%%%%%% Fig. 5 %%%%%%%%%%%%%%%%%%%%%%%%%%%%%%%%%%%%%%%%%%%%%%%%%%%%%%%%%%%%%%%%%%%%%%%%%
\begin{figure*}[t]
\centering
\includegraphics[width=0.4\textwidth]{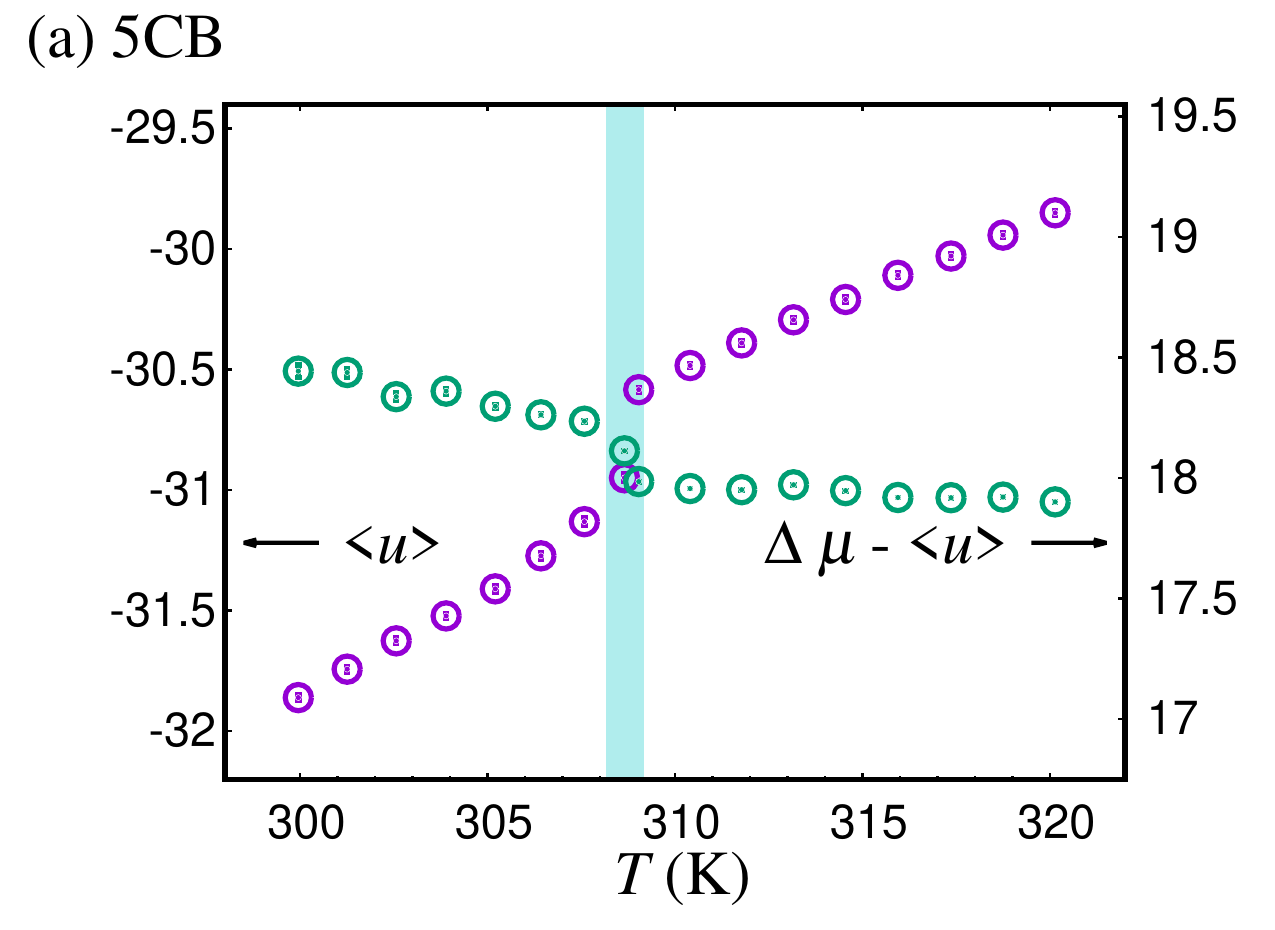}
\includegraphics[width=0.4\textwidth]{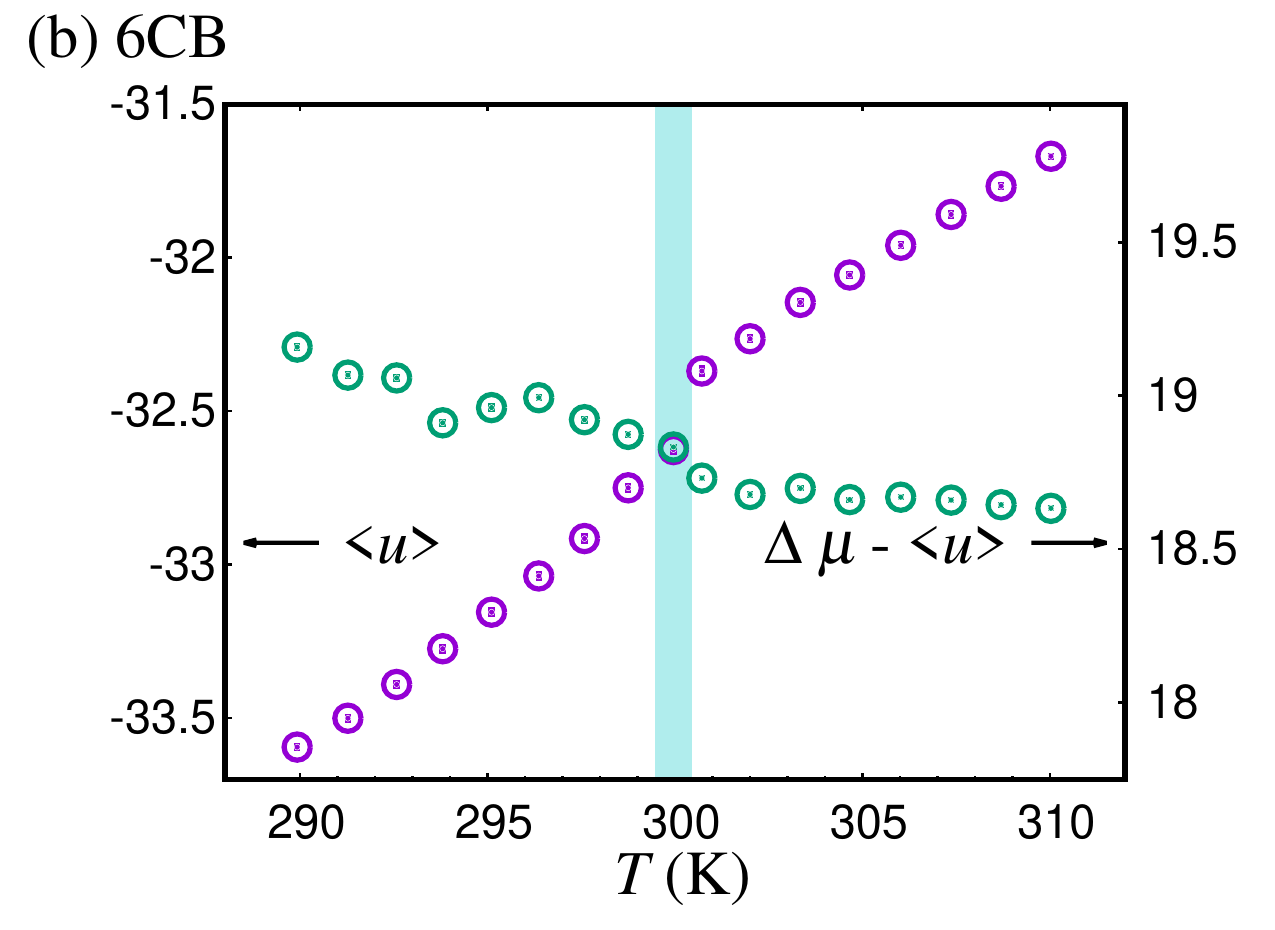}
\includegraphics[width=0.4\textwidth]{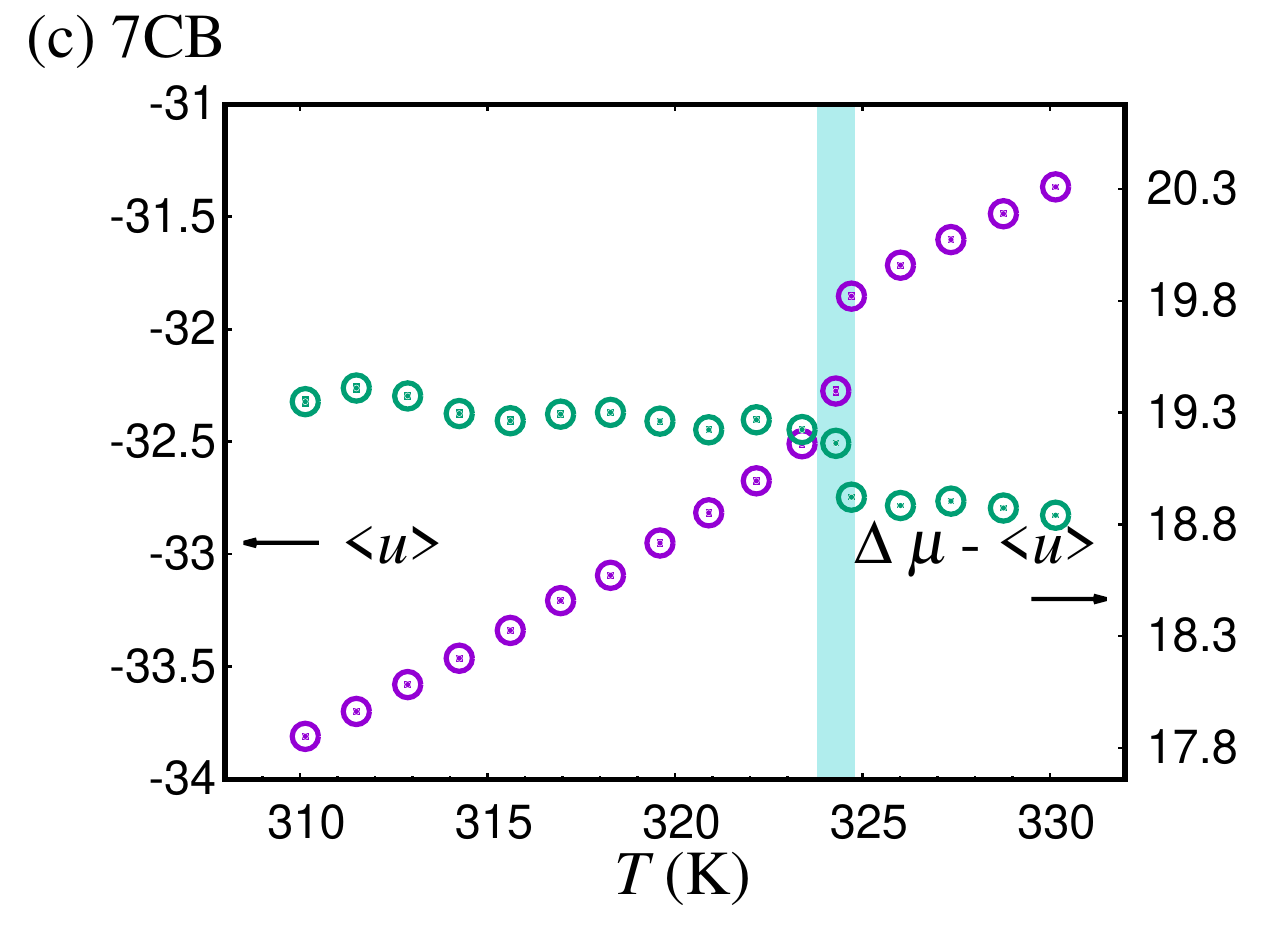}
\includegraphics[width=0.4\textwidth]{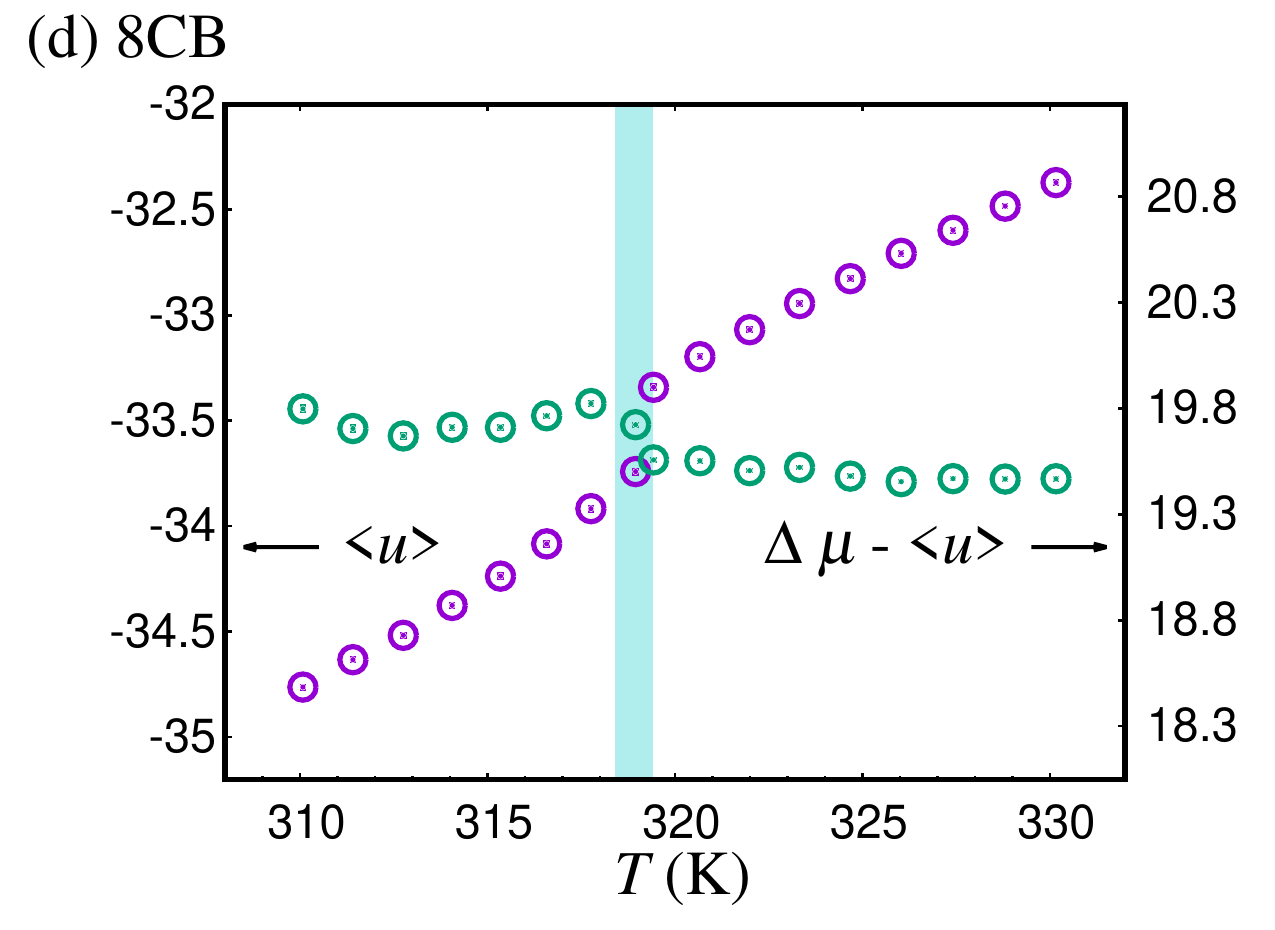}
\caption{Temperature dependence of the decomposition of $\Delta \mu $ into the average sum of the intermolecular
 interaction energy, $\langle u\rangle$ (purple) and the entropic contribution
 $\Delta \mu - \langle u\rangle$ (green) for for 5CB (a), 6CB (b), 7CB (c), and
 8CB (d).
Note that $\langle u\rangle$ and $\Delta \mu - \langle u\rangle$ are
 plotted on 
the left and right $y$-axes, respectively, in units of kcal/mol.
The error bars at each temperature represent the standard deviation.
The vertical color bar indicates the NI transition temperature $T_\mathrm{NI}$.
}
\label{fig:fe_decomposition}
\end{figure*}
%%%%%%%%%%%%%%%%%%%%%%%%%%%%%%%%%%%%%%%%%%%%%%%%%%%%%%%%%%%%%%%%%%%%%%%%%%%%%%%%%%%%%%%%%%

\subsection{Comparison with Maier--Saupe model}

It is interesting to compare the MD results with the Onsager and
MS models.
In these models, the distribution function of the cosine of the
orientation angle $\cos\theta$ is assumed to be the form of 
\begin{align}
P(\cos\theta) =\frac{1}{Z} \exp[\Gamma_0 S P_2(\cos\theta)],
\label{eq:pcosine}
\end{align}
where $\Gamma_0$ is the parameter characterizing the degree of the
nematic ordering.
The normalization factor $Z$ is given by
\begin{align}
Z = 2\pi \int_{-1}^{1} \exp[\Gamma_0 S P_2(\cos\theta)] d(\cos\theta),
\end{align}
which ensures that the distribution function $P(\cos\theta)$
integrates to 1.

In the Onsager model, $\Gamma_0$ is represented by the length $L$ and
diameter $D$ of the rigid and cylindrical molecule, as well as the volume
fraction $\phi$.
This model is considered to be a repulsive model that characterizes the
NI phase transition due to the excluded volume effect.
In contrast, the MS model incorporates an intermolecular potential between
molecules with 
the potential $V(\cos\theta, S)$ at temperature 
$T$ given by 
\begin{align}
V(\cos\theta, S) = - k_\mathrm{B}T\Gamma_0 S P_2(\cos\theta),
\end{align}
where $\Gamma_0$ characterizes
the strength of intermolecular interactions.
Equation~\eqref{eq:pcosine} represents the Boltzmann distribution of
$V(\cos\theta, S)$, thereby classifying the MS model as an attractive model.
In this context, the MS model considers
the combined effects of volume fraction and temperature on 
the NI phase transition, making it more suitable for describing thermotropic
liquid crystals compared to the Onsager model.

Note that these two models are common in that the solution is
obtained by minimizing the free energy $F$ with respect to $S$.
In addition, both the Onsager and MS models have the distribution
function of Eq.~\eqref{eq:pcosine} with a parameter $\Gamma_0$. 
Their difference lies in
the origin of $\Gamma_0$ and its connection to the molecular properties.
In the following, 
the free energy of the MS model is employed
to investigate the temperature dependence of the NI phase transition.
The expression of $F$ is given by 
\begin{align}
\frac{F}{k_\mathrm{B}T} = \frac{1}{2} \Gamma_0 S^2 + \ln\left(\frac{Z}{4\pi}\right).
\label{eq:MS}
\end{align}
The first term can be regarded as the ensemble average of the intermolecular potential,
$\langle V(\cos\theta, S)/k_\mathrm{B}T\rangle / 2 = \Gamma_0S^2/2$, where the factor $1/2$
compensates for the double-counting of interactions.
The second term corresponds to 
the Kullback--Leibler divergence, which
quantifies the difference between the distribution $P(\cos\theta)$ and
the uniform distribution characterizing the isotropic phase, $1/4\pi$.
By using $\Gamma_0$ as the control parameter, the
orientational order parameter $S$ can be
evaluated numerically in the self-consistent manner based on
Eq.~\eqref{eq:MS}.

At the NI phase transition, the parameter
$\Gamma_0 \approx 4.54$ yields an orientation order parameter
value of $S \approx 0.4295$, which coexists with the trivial solution of $S=0$.
Using the corresponding NI phase transition temperature $T_\mathrm{NI}$, 
$S$ as a function of $T/T_\mathrm{NI}$ for the MS model is plotted in
Fig.~\ref{fig:pcosine}(a).
A comparison with MD simulations for $n$CB ($n=5$, 6, 7, and 8) is also presented, 
showing overall consistency between MD simulations and MS model.
Furthermore, 
the orientation angle distributions, $P(\cos\theta)$, at $T_\mathrm{NI}$
for the MS model
is illustrated in Fig.~\ref{fig:pcosine}(b).
The result presents the agreement with MD simulation results 
for $n$CB, particularly highlighting
a significant proportion of molecules aligned parallel
to the director.
Slight deviations between MD simulations and the MS model are observed in
Fig.~\ref{fig:pcosine}, 
which are attributed to
the value of $S$ at $T_\mathrm{NI}$ being less than 
the mean-field value of $S= 0.4295$.

\subsection{Orientatinal correlations}

While our focus was on the orientational order parameter, which quantifies the nematic
ordering of the entire system, we turn our attention to the orientation
correlation functions, $G_1(r)$ and $G_2(r)$, to gain further insight,
particularly regarding short-range ordering.~\cite{pelaez2007Molecular,
tiberio2009Silico, palermo2013Atomistic}
The $G_1(r)$ and $G_2(r)$ are defined by 
\begin{align}
  G_1(r)&= \frac{\langle \delta(r-r_{ij}) (\bm{u}_i
 \cdot \bm{u}_j) \rangle}{\langle \delta(r-r_{ij}) \rangle}, 
  \label{eq:G_1}\\
  G_2(r)&=\frac{\left\langle \delta(r-r_{ij})
[(3/2)(\bm{u}_i \cdot \bm{u}_j)^2 - (1/2)]
 \right\rangle}{\langle \delta(r-r_{ij}) \rangle}, 
  \label{eq:G_2}
\end{align}
where
$\bm{u}_i$ is the unit dipole moment in the $i$-th molecule, aligned with
the CN bond, as used in the calculation of the orientation order
parameter $S$.
The distance between the charge centers of the $i$-th and $j$-th
molecules is denoted as $r_{ij}$
The functions $G_1(r)$ and $G_2(r)$ characterize the local polar order and local
orientational order between molecules, respectively.

$G_1(r)$, $G_2(r)$, and the radial distribution function
$g(r)$ of the replica at $T_\mathrm{NI}$ and its neighboring replicas
(isotropic and nematic phases) are
presented in Fig.~\ref{fig:G}. 
At the distance where $g(r)$ shows a peak, $G_1(r)$ indicates a
polar-ordered correlation even within the isotropic phase. 
The presence of oscillation
at shorter ranges suggests that 
neighboring molecules tend to adopt anti-parallel configurations.
However, at longer distances, this correlation diminishes in both
nematic and isotropic phases, 
asymptotically approaching zero. 
Similarly, $G_2(r)$ reveals orientation ordering at short distances
within 
the isotropic
phase, akin to $G_1(r)$. 
At longer distances, $G_2(r)$ converges to finite values in the nematic
phase, while in the isotropic phase, the values approach zero,
indicating the disappearance of long-range 
orientation order.

\subsection{Energy representation free-energy analysis}

We analyze the thermodynamic stability of the mesogen using the
gREM sampling near the NI phase transition temperature. 
Specifically, we calculated the temperature dependence of the insertion
free energy, $\Delta \mu$, in the $n$CB system based on the ER thory (see Eq.~\eqref{eq:ER}).
The results are shown in Fig.~\ref{fig:fe}. 
Note that $\rho(\varepsilon)$ and $\rho_0(\varepsilon)$ were evaluated through
MD simulations of systems with $N=4000$ and $N=3999$ molecules, respectively.
The results of $\rho(\varepsilon)$ and $\rho_0(\varepsilon)$
of the replica at $T_\mathrm{NI}$ and its neighboring replicas
(isotropic and nematic phases) are
presented in 
Fig.~S6 and Fig.~S7 of the supplementary material, respectively.

For all LC systems, the insertion free energy $\Delta \mu$ decreases with decreasing
temperature, exhibiting a noticeable change of $\Delta \mu(T)$ around $T_\mathrm{NI}$.
This suggests the thermodynamic stability of the mesogen shifts
from the isotropic to the nematic phase, which is driven by the
competition between energetic and entropic contributions.
As expressed in Eq.~\eqref{eq:ER}, 
the insertion free energy $\Delta \mu$ consists of the average
sum of
the intermolecular interaction energies (the first term) and the
entropic contribution (the second term).
Specifically, the second term is regarded as the 
free-energy penalty for the reorganization of the solvent structure due to insertion of the solute,
evaluated through 
the contribution from all configurations of
position and orientation of the solvent molecules via the functional
form $\mathcal{F}$.
Figure~\ref{fig:fe_decomposition} illustrates the temperature dependence
of $\langle u\rangle$ and $\Delta \mu - \langle u\rangle = \int\mathcal{F}[\rho (\varepsilon), \rho_0
(\varepsilon)]d\varepsilon$.

The average sum of the intermolecular interaction energy, $\langle u\rangle$, decreases with
decreasing temperatures and exhibits a discontinuous change at $T_\mathrm{NI}$.
This indicates that the nematic phase is more stable than
the isotropic phase in terms of interaction energy.
$\langle u\rangle$ can be further decomposed into Lennard-Jones (LJ) and
electrostatic contributions, denoted as $\langle u^\mathrm{LJ}\rangle$
and $\langle u^\mathrm{ele}\rangle$, respectively.
The results are illustrated in Fig.~S8 of the supplementary material.
It is observed that the stabilization of intermolecular interactions is
dominated by the contribution of LJ interactions, as indicated by the
large changes in their values compared to those of electrostatic terms.
This can be attributed to the fact that the $n$CB mesogen is electrically
neutral overall, though it does possess a dipole moment at the CN bond.

Conversely, the second term, $\Delta \mu - \langle u\rangle =
\int\mathcal{F}[\rho (\varepsilon), \rho_0
(\varepsilon)]d\varepsilon$, increases as the temperature decreases, and
shows a discontinuous change at $T_\mathrm{NI}$.
This indicates that the nematic phase is entropically less stable than
the isotropic phase due to the orientational ordering.
At $T_\mathrm{NI}$, 
the intermolecular energy surpasses the entropy, 
leading to the
thermodynamic stability of the nematic phase.
In general, the total chemical potential of a target molecule, $\mu$,
consists of ideal term and excess terms, 
expressed by 
%the sum of $\Delta \mu$, $k_\mathrm{B} T$ times the
%logarithm of the density, expressed as 
$\mu = k_\mathrm{B}T \ln (\rho/\rho_0) + \Delta \mu$, where $\rho$ 
is the density and the subscript 0 represents the standard state.
Note that the first term is independent of whether is system
is nematic or isotropic phase, while
the latter corresponds to the insertion free energy, as
demonstrated in Fig.~\ref{fig:fe}.
The total chemical potential $\mu$ remains continuous across the NI phase transition.
However, since the NI phase
transition is first-order and the density is discontinuous at
$T_\mathrm{NI}$, $\Delta \mu$ is also discontinuous.

Finally, we discuss the odd-even effect
based on our free-energy calculation of Fig.~\ref{fig:fe_decomposition}.
For odd-numbered alkyl chains, the terminal alkyl group aligns parallel
to the molecular axis, while for even-numbered alkyl chains, the angle
to the molecular axis tends to be larger.
This results in reduced molecular anisotropy in the even-numbered
compounds, leading to a lower NI phase transition temperature. 
We analyzed the energy gap and entropy gap 
across the NI transition temperature, denoted as
$\Delta \langle u \rangle$ and $\Delta (\Delta \mu-\langle u\rangle)$.
As illustrated in Fig.~S9 of the supplementary material, the results
indicate that odd-numbered compounds exhibit more stability of the
nematic ordering with a
significantly large energy loss, accompanied by a large entropy loss.
This suggests that the nematic phase in odd-numbered compounds is
stabilized by a strong molecular anisotropy.
Conversely, the even-numbered compounds show smaller energy and entropy
gaps, reflecting their reduced anisotropy and lower transition
temperature to the nematic phase.

\section{Conclusions}

In this study, we performed MD simulation of the
NI phase transition of the UA model $n$CB ($n=5$, 6, 7, and 8), combined with
with gREM.
The gREM introducing the linear effective temperature
was useful for simulating the NI phase transition, because the
thermodynamically unstable states,  which originates from the
first-order phase transition, can be efficiently sampled.
The temperature dependence of the orientation order parameter $S$ was
evaluated.

The orientation distribution function calculated from the simulation
trajectories is found to be in good agreement with the mean-field MS
model.
Furthermore, the local orientational order was characterized by the
orientation correlation functions, $G_1(r)$ and $G_2(r)$.
In both the nematic and isotropic phases, 
short-range order was observed, 
in which the molecules were arranged in antiparallel
configurations.

The NI phase transition was described thermodynamically by the free-energy
analysis based on the ER theory. 
The temperature dependence of the insertion free energy $\Delta \mu$
reveals the change in thermodynamic stability associated with the NI
phase transition. 
Notably, the temperature dependence of $\Delta \mu$
changes at $T_\mathrm{NI}$, indicating a
shift in the temperature dependence of the free energy at this point.
Further decomposition into intermolecular interaction energies and
entropic terms suggests the competition between these
factors, which drives the NI phase transition.
These findings are consistent with the MS model, which describe the NI phase
transition in terms of the competition between the intermolecular attractive force and entropy.
It is important to emphasize
that our free-energy analysis based on the ER theory exceeds
the mean-field treatment by providing a molecular-level description
that incorporates LJ, electrostatic, and entropic contributions.

\section*{SUPPLEMENTARY MATERIAL}

The supplementary material includes parameters of gREM simulations
(Table~S1); 
molecular structures of $n$CB (Fig.~S1); 
probability distribution of enthalpy of replicas in gREM
simulations (Fig.~S2); 
acceptance ratios of replica exchanges (Fig.~S3); 
statistical temperature as a function of enthalpy
in gREM simulations (Fig.~S4); 
orientational order parameter $S$ using another molecular axis (Fig.~S5);
density profiles of intermolecular
interaction energy for ER calculations, $\rho(\varepsilon)$ and
$\rho_0(\varepsilon)$ (Fig.~S6 and Fig.~S7); and temperature
dependence of LJ and electrostatic energies (Fig.~S8);
dependence of 
energy gap $\Delta \langle u\rangle$ and entropy gap $\Delta (\Delta
\mu- \langle u\rangle)$ on the number of carbons in
the alkyl chain (Fig.~S9).

\begin{acknowledgments}
This work was supported by 
JSPS KAKENHI Grant-in-Aid 
Grant Nos.~\mbox{JP24H01719}, \mbox{JP22H04542}, \mbox{JP22K03550},
 \mbox{JP23K27313}, \mbox{JP23H02622},
 \mbox{JP24K00792}, \mbox{JP19H05718},
 \mbox{JP22K18953},
 \mbox{JP24K17576}, and \mbox{JP24H01727}.
Y.~I. was also supported by JST ACT-X Grant No.~\mbox{JPMJAX23D3}.
We are grateful to 
the Fugaku Supercomputing Project (Nos.~JPMXP1020230325 and JPMXP1020230327) and 
the Data-Driven Material Research Project (No.~\mbox{JPMXP1122714694})
from the
Ministry of Education, Culture, Sports, Science, and Technology and to
 Maruho Collaborative Project for Theoretical Pharmaceutics.
The numerical calculations were performed at Research Center for
Computational Science, Okazaki Research Facilities, National Institutes
 of Natural Sciences (Project: \mbox{24-IMS-C051}).
\end{acknowledgments}

\section*{AUTHOR DECLARATIONS}

\section*{Conflict of Interest}
The authors have no conflicts to disclose.

\section*{Data availability statement}

The LAMMPS input files and initial configurations and 1 ns trajectories are
available in Zenodo at https://doi.org/10.5281/zenodo.13690896.
Further data that support the findings of this study are available from the
corresponding author upon reasonable request.

%\bibliography{lc}

\begin{thebibliography}{53}%
\makeatletter
\providecommand \@ifxundefined [1]{%
 \@ifx{#1\undefined}
}%
\providecommand \@ifnum [1]{%
 \ifnum #1\expandafter \@firstoftwo
 \else \expandafter \@secondoftwo
 \fi
}%
\providecommand \@ifx [1]{%
 \ifx #1\expandafter \@firstoftwo
 \else \expandafter \@secondoftwo
 \fi
}%
\providecommand \natexlab [1]{#1}%
\providecommand \enquote  [1]{``#1''}%
\providecommand \bibnamefont  [1]{#1}%
\providecommand \bibfnamefont [1]{#1}%
\providecommand \citenamefont [1]{#1}%
\providecommand \href@noop [0]{\@secondoftwo}%
\providecommand \href [0]{\begingroup \@sanitize@url \@href}%
\providecommand \@href[1]{\@@startlink{#1}\@@href}%
\providecommand \@@href[1]{\endgroup#1\@@endlink}%
\providecommand \@sanitize@url [0]{\catcode `\\12\catcode `\$12\catcode
  `\&12\catcode `\#12\catcode `\^12\catcode `\_12\catcode `\%12\relax}%
\providecommand \@@startlink[1]{}%
\providecommand \@@endlink[0]{}%
\providecommand \url  [0]{\begingroup\@sanitize@url \@url }%
\providecommand \@url [1]{\endgroup\@href {#1}{\urlprefix }}%
\providecommand \urlprefix  [0]{URL }%
\providecommand \Eprint [0]{\href }%
\providecommand \doibase [0]{https://doi.org/}%
\providecommand \selectlanguage [0]{\@gobble}%
\providecommand \bibinfo  [0]{\@secondoftwo}%
\providecommand \bibfield  [0]{\@secondoftwo}%
\providecommand \translation [1]{[#1]}%
\providecommand \BibitemOpen [0]{}%
\providecommand \bibitemStop [0]{}%
\providecommand \bibitemNoStop [0]{.\EOS\space}%
\providecommand \EOS [0]{\spacefactor3000\relax}%
\providecommand \BibitemShut  [1]{\csname bibitem#1\endcsname}%
\let\auto@bib@innerbib\@empty
%</preamble>
\bibitem [{\citenamefont {Chandrasekhar}(1992)}]{chandrasekhar1992Liquid}%
  \BibitemOpen
  \bibfield  {author} {\bibinfo {author} {\bibfnamefont {S.}~\bibnamefont
  {Chandrasekhar}},\ }\href {https://doi.org/10.1017/CBO9780511622496} {\emph
  {\bibinfo {title} {Liquid {{Crystals}}}}},\ \bibinfo {edition} {2nd}\ ed.\
  (\bibinfo  {publisher} {Cambridge University Press},\ \bibinfo {address}
  {Cambridge},\ \bibinfo {year} {1992})\BibitemShut {NoStop}%
\bibitem [{\citenamefont {de~Gennes}\ and\ \citenamefont
  {Prost}(1995)}]{gennes1995Physics}%
  \BibitemOpen
  \bibfield  {author} {\bibinfo {author} {\bibfnamefont {P.-G.}\ \bibnamefont
  {de~Gennes}}\ and\ \bibinfo {author} {\bibfnamefont {J.}~\bibnamefont
  {Prost}},\ }\href@noop {} {\emph {\bibinfo {title} {The Physics of Liquid
  Crystals}}},\ \bibinfo {edition} {2nd}\ ed.\ (\bibinfo  {publisher} {Oxford
  University Press},\ \bibinfo {address} {Oxford},\ \bibinfo {year}
  {1995})\BibitemShut {NoStop}%
\bibitem [{\citenamefont {Stephen}\ and\ \citenamefont
  {Straley}(1974)}]{stephen1974Physicsa}%
  \BibitemOpen
  \bibfield  {author} {\bibinfo {author} {\bibfnamefont {M.~J.}\ \bibnamefont
  {Stephen}}\ and\ \bibinfo {author} {\bibfnamefont {J.~P.}\ \bibnamefont
  {Straley}},\ }\bibfield  {title} {\enquote {\bibinfo {title} {Physics of
  liquid crystals},}\ }\href {https://doi.org/10.1103/RevModPhys.46.617}
  {\bibfield  {journal} {\bibinfo  {journal} {Rev. Mod. Phys.}\ }\textbf
  {\bibinfo {volume} {46}},\ \bibinfo {pages} {617--704} (\bibinfo {year}
  {1974})}\BibitemShut {NoStop}%
\bibitem [{\citenamefont {Singh}(2000)}]{singh2000Phase}%
  \BibitemOpen
  \bibfield  {author} {\bibinfo {author} {\bibfnamefont {S.}~\bibnamefont
  {Singh}},\ }\bibfield  {title} {\enquote {\bibinfo {title} {Phase transitions
  in liquid crystals},}\ }\href {https://doi.org/10.1016/S0370-1573(99)00049-6}
  {\bibfield  {journal} {\bibinfo  {journal} {Phys. Rep.}\ }\textbf {\bibinfo
  {volume} {324}},\ \bibinfo {pages} {107--269} (\bibinfo {year}
  {2000})}\BibitemShut {NoStop}%
\bibitem [{\citenamefont {Andrienko}(2018)}]{andrienko2018Introduction}%
  \BibitemOpen
  \bibfield  {author} {\bibinfo {author} {\bibfnamefont {D.}~\bibnamefont
  {Andrienko}},\ }\bibfield  {title} {\enquote {\bibinfo {title} {Introduction
  to liquid crystals},}\ }\href {https://doi.org/10.1016/j.molliq.2018.01.175}
  {\bibfield  {journal} {\bibinfo  {journal} {J. Mol. Liq.}\ }\textbf {\bibinfo
  {volume} {267}},\ \bibinfo {pages} {520--541} (\bibinfo {year}
  {2018})}\BibitemShut {NoStop}%
\bibitem [{\citenamefont {Onsager}(1949)}]{onsager1949EFFECTS}%
  \BibitemOpen
  \bibfield  {author} {\bibinfo {author} {\bibfnamefont {L.}~\bibnamefont
  {Onsager}},\ }\bibfield  {title} {\enquote {\bibinfo {title} {{{THE EFFECTS
  OF SHAPE ON THE INTERACTION OF COLLOIDAL PARTICLES}}},}\ }\href
  {https://doi.org/10.1111/j.1749-6632.1949.tb27296.x} {\bibfield  {journal}
  {\bibinfo  {journal} {Ann. NY Acad. Sci.}\ }\textbf {\bibinfo {volume}
  {51}},\ \bibinfo {pages} {627--659} (\bibinfo {year} {1949})}\BibitemShut
  {NoStop}%
\bibitem [{\citenamefont {Maier}\ and\ \citenamefont
  {Saupe}(1958)}]{maier1958Einfache}%
  \BibitemOpen
  \bibfield  {author} {\bibinfo {author} {\bibfnamefont {W.}~\bibnamefont
  {Maier}}\ and\ \bibinfo {author} {\bibfnamefont {A.}~\bibnamefont {Saupe}},\
  }\bibfield  {title} {\enquote {\bibinfo {title} {Eine einfache molekulare
  {{Theorie}} des nematischen kristallinfl{\"u}ssigen {{Zustandes}}},}\ }\href
  {https://doi.org/10.1515/zna-1958-0716} {\bibfield  {journal} {\bibinfo
  {journal} {Z. Naturforsch. A}\ }\textbf {\bibinfo {volume} {13}},\ \bibinfo
  {pages} {564--566} (\bibinfo {year} {1958})}\BibitemShut {NoStop}%
\bibitem [{\citenamefont {Zannoni}(2001)}]{zannoni2001Molecular}%
  \BibitemOpen
  \bibfield  {author} {\bibinfo {author} {\bibfnamefont {C.}~\bibnamefont
  {Zannoni}},\ }\bibfield  {title} {\enquote {\bibinfo {title} {Molecular
  design and computer simulations of novel mesophases},}\ }\href
  {https://doi.org/10.1039/b103923g} {\bibfield  {journal} {\bibinfo  {journal}
  {J. Mater. Chem.}\ }\textbf {\bibinfo {volume} {11}},\ \bibinfo {pages}
  {2637--2646} (\bibinfo {year} {2001})}\BibitemShut {NoStop}%
\bibitem [{\citenamefont {Cacelli}\ \emph {et~al.}(2002)\citenamefont
  {Cacelli}, \citenamefont {Campanile}, \citenamefont {Prampolini},\ and\
  \citenamefont {Tani}}]{cacelli2002Stability}%
  \BibitemOpen
  \bibfield  {author} {\bibinfo {author} {\bibfnamefont {I.}~\bibnamefont
  {Cacelli}}, \bibinfo {author} {\bibfnamefont {S.}~\bibnamefont {Campanile}},
  \bibinfo {author} {\bibfnamefont {G.}~\bibnamefont {Prampolini}},\ and\
  \bibinfo {author} {\bibfnamefont {A.}~\bibnamefont {Tani}},\ }\bibfield
  {title} {\enquote {\bibinfo {title} {Stability of the nematic phase of 4-
  {\emph{n}} -pentyl-4{$\prime$}-cyanobiphenyl studied by computer simulation
  using a hybrid model},}\ }\href {https://doi.org/10.1063/1.1482702}
  {\bibfield  {journal} {\bibinfo  {journal} {J. Chem. Phys.}\ }\textbf
  {\bibinfo {volume} {117}},\ \bibinfo {pages} {448--453} (\bibinfo {year}
  {2002})}\BibitemShut {NoStop}%
\bibitem [{\citenamefont {Berardi}, \citenamefont {Muccioli},\ and\
  \citenamefont {Zannoni}(2004)}]{berardi2004Can}%
  \BibitemOpen
  \bibfield  {author} {\bibinfo {author} {\bibfnamefont {R.}~\bibnamefont
  {Berardi}}, \bibinfo {author} {\bibfnamefont {L.}~\bibnamefont {Muccioli}},\
  and\ \bibinfo {author} {\bibfnamefont {C.}~\bibnamefont {Zannoni}},\
  }\bibfield  {title} {\enquote {\bibinfo {title} {Can {{Nematic Transitions Be
  Predicted By Atomistic Simulations}}? {{A Computational Study}} of {{The
  Odd}}--{{Even Effect}}},}\ }\href {https://doi.org/10.1002/cphc.200300908}
  {\bibfield  {journal} {\bibinfo  {journal} {ChemPhysChem}\ }\textbf {\bibinfo
  {volume} {5}},\ \bibinfo {pages} {104--111} (\bibinfo {year}
  {2004})}\BibitemShut {NoStop}%
\bibitem [{\citenamefont {Care}\ and\ \citenamefont
  {Cleaver}(2005)}]{care2005Computer}%
  \BibitemOpen
  \bibfield  {author} {\bibinfo {author} {\bibfnamefont {C.~M.}\ \bibnamefont
  {Care}}\ and\ \bibinfo {author} {\bibfnamefont {D.~J.}\ \bibnamefont
  {Cleaver}},\ }\bibfield  {title} {\enquote {\bibinfo {title} {Computer
  simulation of liquid crystals},}\ }\href
  {https://doi.org/10.1088/0034-4885/68/11/R04} {\bibfield  {journal} {\bibinfo
   {journal} {Rep. Prog. Phys.}\ }\textbf {\bibinfo {volume} {68}},\ \bibinfo
  {pages} {2665--2700} (\bibinfo {year} {2005})}\BibitemShut {NoStop}%
\bibitem [{\citenamefont {Capar}\ and\ \citenamefont
  {Cebe}(2006)}]{capar2006Molecular}%
  \BibitemOpen
  \bibfield  {author} {\bibinfo {author} {\bibfnamefont {M.~I.}\ \bibnamefont
  {Capar}}\ and\ \bibinfo {author} {\bibfnamefont {E.}~\bibnamefont {Cebe}},\
  }\bibfield  {title} {\enquote {\bibinfo {title} {Molecular dynamic study of
  the odd-even effect insome 4-{$n$}-alkyl-4{$^\prime$}-cyanobiphenyls},}\
  }\href {https://doi.org/10.1103/PhysRevE.73.061711} {\bibfield  {journal}
  {\bibinfo  {journal} {Phys. Rev. E}\ }\textbf {\bibinfo {volume} {73}},\
  \bibinfo {pages} {061711} (\bibinfo {year} {2006})}\BibitemShut {NoStop}%
\bibitem [{\citenamefont {Wilson}(2007)}]{wilson2007Molecular}%
  \BibitemOpen
  \bibfield  {author} {\bibinfo {author} {\bibfnamefont {M.~R.}\ \bibnamefont
  {Wilson}},\ }\bibfield  {title} {\enquote {\bibinfo {title} {Molecular
  simulation of liquid crystals: Progress towards a better understanding of
  bulk structure and the prediction of material properties},}\ }\href
  {https://doi.org/10.1039/b612799c} {\bibfield  {journal} {\bibinfo  {journal}
  {Chem. Soc. Rev.}\ }\textbf {\bibinfo {volume} {36}},\ \bibinfo {pages}
  {1881} (\bibinfo {year} {2007})}\BibitemShut {NoStop}%
\bibitem [{\citenamefont {Cacelli}\ \emph {et~al.}(2007)\citenamefont
  {Cacelli}, \citenamefont {De~Gaetani}, \citenamefont {Prampolini},\ and\
  \citenamefont {Tani}}]{cacelli2007Liquida}%
  \BibitemOpen
  \bibfield  {author} {\bibinfo {author} {\bibfnamefont {I.}~\bibnamefont
  {Cacelli}}, \bibinfo {author} {\bibfnamefont {L.}~\bibnamefont {De~Gaetani}},
  \bibinfo {author} {\bibfnamefont {G.}~\bibnamefont {Prampolini}},\ and\
  \bibinfo {author} {\bibfnamefont {A.}~\bibnamefont {Tani}},\ }\bibfield
  {title} {\enquote {\bibinfo {title} {Liquid {{Crystal Properties}} of the
  n-{{Alkyl-cyanobiphenyl Series}} from {{Atomistic Simulations}} with {{Ab
  Initio Derived Force Fields}}},}\ }\href {https://doi.org/10.1021/jp065806l}
  {\bibfield  {journal} {\bibinfo  {journal} {J. Phys. Chem. B}\ }\textbf
  {\bibinfo {volume} {111}},\ \bibinfo {pages} {2130--2137} (\bibinfo {year}
  {2007})}\BibitemShut {NoStop}%
\bibitem [{\citenamefont {Berardi}\ \emph {et~al.}(2008)\citenamefont
  {Berardi}, \citenamefont {Muccioli}, \citenamefont {Orlandi}, \citenamefont
  {Ricci},\ and\ \citenamefont {Zannoni}}]{berardi2008Computer}%
  \BibitemOpen
  \bibfield  {author} {\bibinfo {author} {\bibfnamefont {R.}~\bibnamefont
  {Berardi}}, \bibinfo {author} {\bibfnamefont {L.}~\bibnamefont {Muccioli}},
  \bibinfo {author} {\bibfnamefont {S.}~\bibnamefont {Orlandi}}, \bibinfo
  {author} {\bibfnamefont {M.}~\bibnamefont {Ricci}},\ and\ \bibinfo {author}
  {\bibfnamefont {C.}~\bibnamefont {Zannoni}},\ }\bibfield  {title} {\enquote
  {\bibinfo {title} {Computer simulations of biaxial nematics},}\ }\href
  {https://doi.org/10.1088/0953-8984/20/46/463101} {\bibfield  {journal}
  {\bibinfo  {journal} {J. Phys.: Condens. Matter}\ }\textbf {\bibinfo {volume}
  {20}},\ \bibinfo {pages} {463101} (\bibinfo {year} {2008})}\BibitemShut
  {NoStop}%
\bibitem [{\citenamefont {Cifelli}\ \emph {et~al.}(2008)\citenamefont
  {Cifelli}, \citenamefont {De~Gaetani}, \citenamefont {Prampolini},\ and\
  \citenamefont {Tani}}]{cifelli2008Atomistic}%
  \BibitemOpen
  \bibfield  {author} {\bibinfo {author} {\bibfnamefont {M.}~\bibnamefont
  {Cifelli}}, \bibinfo {author} {\bibfnamefont {L.}~\bibnamefont {De~Gaetani}},
  \bibinfo {author} {\bibfnamefont {G.}~\bibnamefont {Prampolini}},\ and\
  \bibinfo {author} {\bibfnamefont {A.}~\bibnamefont {Tani}},\ }\bibfield
  {title} {\enquote {\bibinfo {title} {Atomistic {{Computer Simulation}} and
  {{Experimental Study}} on the {{Dynamics}} of the n-{{Cyanobiphenyls
  Mesogenic Series}}},}\ }\href {https://doi.org/10.1021/jp802935q} {\bibfield
  {journal} {\bibinfo  {journal} {J. Phys. Chem. B}\ }\textbf {\bibinfo
  {volume} {112}},\ \bibinfo {pages} {9777--9786} (\bibinfo {year}
  {2008})}\BibitemShut {NoStop}%
\bibitem [{\citenamefont {Tiberio}\ \emph {et~al.}(2009)\citenamefont
  {Tiberio}, \citenamefont {Muccioli}, \citenamefont {Berardi},\ and\
  \citenamefont {Zannoni}}]{tiberio2009Silico}%
  \BibitemOpen
  \bibfield  {author} {\bibinfo {author} {\bibfnamefont {G.}~\bibnamefont
  {Tiberio}}, \bibinfo {author} {\bibfnamefont {L.}~\bibnamefont {Muccioli}},
  \bibinfo {author} {\bibfnamefont {R.}~\bibnamefont {Berardi}},\ and\ \bibinfo
  {author} {\bibfnamefont {C.}~\bibnamefont {Zannoni}},\ }\bibfield  {title}
  {\enquote {\bibinfo {title} {Towards {\emph{in }}{{{\emph{Silico}}}} {{Liquid
  Crystals}}. {{Realistic Transition Temperatures}} and {{Physical Properties}}
  for {\emph{n}} -{{Cyanobiphenyls}} via {{Molecular Dynamics Simulations}}},}\
  }\href {https://doi.org/10.1002/cphc.200800231} {\bibfield  {journal}
  {\bibinfo  {journal} {ChemPhysChem}\ }\textbf {\bibinfo {volume} {10}},\
  \bibinfo {pages} {125--136} (\bibinfo {year} {2009})}\BibitemShut {NoStop}%
\bibitem [{\citenamefont {Zhang}, \citenamefont {Su},\ and\ \citenamefont
  {Guo}(2011)}]{zhang2011Atomistic}%
  \BibitemOpen
  \bibfield  {author} {\bibinfo {author} {\bibfnamefont {J.}~\bibnamefont
  {Zhang}}, \bibinfo {author} {\bibfnamefont {J.}~\bibnamefont {Su}},\ and\
  \bibinfo {author} {\bibfnamefont {H.}~\bibnamefont {Guo}},\ }\bibfield
  {title} {\enquote {\bibinfo {title} {An {{Atomistic Simulation}} for
  4-{{Cyano-4}}{$\prime$}-pentylbiphenyl and {{Its Homologue}} with a
  {{Reoptimized Force Field}}},}\ }\href {https://doi.org/10.1021/jp111408n}
  {\bibfield  {journal} {\bibinfo  {journal} {J. Phys. Chem. B}\ }\textbf
  {\bibinfo {volume} {115}},\ \bibinfo {pages} {2214--2227} (\bibinfo {year}
  {2011})}\BibitemShut {NoStop}%
\bibitem [{\citenamefont {Palermo}\ \emph {et~al.}(2013)\citenamefont
  {Palermo}, \citenamefont {Pizzirusso}, \citenamefont {Muccioli},\ and\
  \citenamefont {Zannoni}}]{palermo2013Atomistic}%
  \BibitemOpen
  \bibfield  {author} {\bibinfo {author} {\bibfnamefont {M.~F.}\ \bibnamefont
  {Palermo}}, \bibinfo {author} {\bibfnamefont {A.}~\bibnamefont {Pizzirusso}},
  \bibinfo {author} {\bibfnamefont {L.}~\bibnamefont {Muccioli}},\ and\
  \bibinfo {author} {\bibfnamefont {C.}~\bibnamefont {Zannoni}},\ }\bibfield
  {title} {\enquote {\bibinfo {title} {An atomistic description of the nematic
  and smectic phases of 4-n-octyl-4{$\prime$} cyanobiphenyl ({{8CB}})},}\
  }\href {https://doi.org/10.1063/1.4804270} {\bibfield  {journal} {\bibinfo
  {journal} {J. Chem. Phys.}\ }\textbf {\bibinfo {volume} {138}},\ \bibinfo
  {pages} {204901} (\bibinfo {year} {2013})}\BibitemShut {NoStop}%
\bibitem [{\citenamefont {Ju}\ \emph {et~al.}(2016)\citenamefont {Ju},
  \citenamefont {Huang}, \citenamefont {Lin}, \citenamefont {Chen},\ and\
  \citenamefont {Shen}}]{ju2016Prediction}%
  \BibitemOpen
  \bibfield  {author} {\bibinfo {author} {\bibfnamefont {S.-P.}\ \bibnamefont
  {Ju}}, \bibinfo {author} {\bibfnamefont {S.-C.}\ \bibnamefont {Huang}},
  \bibinfo {author} {\bibfnamefont {K.-H.}\ \bibnamefont {Lin}}, \bibinfo
  {author} {\bibfnamefont {H.-Y.}\ \bibnamefont {Chen}},\ and\ \bibinfo
  {author} {\bibfnamefont {T.-K.}\ \bibnamefont {Shen}},\ }\bibfield  {title}
  {\enquote {\bibinfo {title} {Prediction of {{Optical}} and {{Dielectric
  Properties}} of 4-{{Cyano-4-pentylbiphenyl Liquid Crystals}} by {{Molecular
  Dynamics Simulation}}, {{Coarse-Grained Dynamics Simulation}}, and {{Density
  Functional Theory Calculation}}},}\ }\href
  {https://doi.org/10.1021/acs.jpcc.5b12222} {\bibfield  {journal} {\bibinfo
  {journal} {J. Phys. Chem. C}\ }\textbf {\bibinfo {volume} {120}},\ \bibinfo
  {pages} {14277--14288} (\bibinfo {year} {2016})}\BibitemShut {NoStop}%
\bibitem [{\citenamefont {Pel{\'a}ez}\ and\ \citenamefont
  {Wilson}(2007)}]{pelaez2007Molecular}%
  \BibitemOpen
  \bibfield  {author} {\bibinfo {author} {\bibfnamefont {J.}~\bibnamefont
  {Pel{\'a}ez}}\ and\ \bibinfo {author} {\bibfnamefont {M.}~\bibnamefont
  {Wilson}},\ }\bibfield  {title} {\enquote {\bibinfo {title} {Molecular
  orientational and dipolar correlation in the liquid crystal mixture {{E7}}: A
  molecular dynamics simulation study at a fully atomistic level},}\ }\href
  {https://doi.org/10.1039/B614422E} {\bibfield  {journal} {\bibinfo  {journal}
  {Phys. Chem. Chem. Phys.}\ }\textbf {\bibinfo {volume} {9}},\ \bibinfo
  {pages} {2968--2975} (\bibinfo {year} {2007})}\BibitemShut {NoStop}%
\bibitem [{\citenamefont {Zannoni}(2018)}]{zannoni2018Idealised}%
  \BibitemOpen
  \bibfield  {author} {\bibinfo {author} {\bibfnamefont {C.}~\bibnamefont
  {Zannoni}},\ }\bibfield  {title} {\enquote {\bibinfo {title} {From idealised
  to predictive models of liquid crystals},}\ }\href
  {https://doi.org/10.1080/02678292.2018.1512170} {\bibfield  {journal}
  {\bibinfo  {journal} {Liquid Crystals}\ }\textbf {\bibinfo {volume} {45}},\
  \bibinfo {pages} {1880--1893} (\bibinfo {year} {2018})}\BibitemShut {NoStop}%
\bibitem [{\citenamefont {Sidky}, \citenamefont {{de Pablo}},\ and\
  \citenamefont {Whitmer}(2018)}]{sidky2018Silico}%
  \BibitemOpen
  \bibfield  {author} {\bibinfo {author} {\bibfnamefont {H.}~\bibnamefont
  {Sidky}}, \bibinfo {author} {\bibfnamefont {J.~J.}\ \bibnamefont {{de
  Pablo}}},\ and\ \bibinfo {author} {\bibfnamefont {J.~K.}\ \bibnamefont
  {Whitmer}},\ }\bibfield  {title} {\enquote {\bibinfo {title} {{\emph{In
  }}{{{\emph{Silico}}}} {{Measurement}} of {{Elastic Moduli}} of {{Nematic
  Liquid Crystals}}},}\ }\href {https://doi.org/10.1103/PhysRevLett.120.107801}
  {\bibfield  {journal} {\bibinfo  {journal} {Phys. Rev. Lett.}\ }\textbf
  {\bibinfo {volume} {120}},\ \bibinfo {pages} {107801} (\bibinfo {year}
  {2018})}\BibitemShut {NoStop}%
\bibitem [{\citenamefont {Allen}(2019)}]{allen2019Molecular}%
  \BibitemOpen
  \bibfield  {author} {\bibinfo {author} {\bibfnamefont {M.~P.}\ \bibnamefont
  {Allen}},\ }\bibfield  {title} {\enquote {\bibinfo {title} {Molecular
  simulation of liquid crystals},}\ }\href
  {https://doi.org/10.1080/00268976.2019.1612957} {\bibfield  {journal}
  {\bibinfo  {journal} {Mol. Phys.}\ }\textbf {\bibinfo {volume} {117}},\
  \bibinfo {pages} {2391--2417} (\bibinfo {year} {2019})}\BibitemShut {NoStop}%
\bibitem [{\citenamefont {Sasaki}\ \emph {et~al.}(2020)\citenamefont {Sasaki},
  \citenamefont {Takahashi}, \citenamefont {Hayashi},\ and\ \citenamefont
  {Kawauchi}}]{sasaki2020Atomistic}%
  \BibitemOpen
  \bibfield  {author} {\bibinfo {author} {\bibfnamefont {R.}~\bibnamefont
  {Sasaki}}, \bibinfo {author} {\bibfnamefont {Y.}~\bibnamefont {Takahashi}},
  \bibinfo {author} {\bibfnamefont {Y.}~\bibnamefont {Hayashi}},\ and\ \bibinfo
  {author} {\bibfnamefont {S.}~\bibnamefont {Kawauchi}},\ }\bibfield  {title}
  {\enquote {\bibinfo {title} {Atomistic {{Mechanism}} of {{Anisotropic Heat
  Conduction}} in the {{Liquid Crystal}}
  4-{{Heptyl-4}}{$\prime$}-cyanobiphenyl: {{All-Atom Molecular Dynamics}}},}\
  }\href {https://doi.org/10.1021/acs.jpcb.9b08158} {\bibfield  {journal}
  {\bibinfo  {journal} {J. Phys. Chem. B}\ }\textbf {\bibinfo {volume} {124}},\
  \bibinfo {pages} {881--889} (\bibinfo {year} {2020})}\BibitemShut {NoStop}%
\bibitem [{\citenamefont {Shi}, \citenamefont {Sidky},\ and\ \citenamefont
  {Whitmer}(2020)}]{shi2020Automated}%
  \BibitemOpen
  \bibfield  {author} {\bibinfo {author} {\bibfnamefont {J.}~\bibnamefont
  {Shi}}, \bibinfo {author} {\bibfnamefont {H.}~\bibnamefont {Sidky}},\ and\
  \bibinfo {author} {\bibfnamefont {J.~K.}\ \bibnamefont {Whitmer}},\
  }\bibfield  {title} {\enquote {\bibinfo {title} {Automated determination of
  {\emph{n}} -cyanobiphenyl and {\emph{n}} -cyanobiphenyl binary mixtures
  elastic constants in the nematic phase from molecular simulation},}\ }\href
  {https://doi.org/10.1039/C9ME00065H} {\bibfield  {journal} {\bibinfo
  {journal} {Mol. Syst. Des. Eng.}\ }\textbf {\bibinfo {volume} {5}},\ \bibinfo
  {pages} {1131--1136} (\bibinfo {year} {2020})}\BibitemShut {NoStop}%
\bibitem [{\citenamefont {Sheavly}\ \emph {et~al.}(2020)\citenamefont
  {Sheavly}, \citenamefont {Gold}, \citenamefont {Mavrikakis},\ and\
  \citenamefont {Van~Lehn}}]{sheavly2020Molecular}%
  \BibitemOpen
  \bibfield  {author} {\bibinfo {author} {\bibfnamefont {J.~K.}\ \bibnamefont
  {Sheavly}}, \bibinfo {author} {\bibfnamefont {J.~I.}\ \bibnamefont {Gold}},
  \bibinfo {author} {\bibfnamefont {M.}~\bibnamefont {Mavrikakis}},\ and\
  \bibinfo {author} {\bibfnamefont {R.~C.}\ \bibnamefont {Van~Lehn}},\
  }\bibfield  {title} {\enquote {\bibinfo {title} {Molecular simulations of
  analyte partitioning and diffusion in liquid crystal sensors},}\ }\href
  {https://doi.org/10.1039/C9ME00126C} {\bibfield  {journal} {\bibinfo
  {journal} {Mol. Syst. Des. Eng.}\ }\textbf {\bibinfo {volume} {5}},\ \bibinfo
  {pages} {304--316} (\bibinfo {year} {2020})}\BibitemShut {NoStop}%
\bibitem [{\citenamefont {Zannoni}(2022)}]{zannoni2022Liquid}%
  \BibitemOpen
  \bibfield  {author} {\bibinfo {author} {\bibfnamefont {C.}~\bibnamefont
  {Zannoni}},\ }\href {https://doi.org/10.1017/9781108539630} {\emph {\bibinfo
  {title} {Liquid {{Crystals}} and {{Their Computer Simulations}}}}}\ (\bibinfo
   {publisher} {Cambridge University Press},\ \bibinfo {address} {Cambridge},\
  \bibinfo {year} {2022})\BibitemShut {NoStop}%
\bibitem [{\citenamefont {Watanabe}, \citenamefont {Yamazaki},\ and\
  \citenamefont {Yoshida}(2023)}]{watanabe2023Missing}%
  \BibitemOpen
  \bibfield  {author} {\bibinfo {author} {\bibfnamefont {G.}~\bibnamefont
  {Watanabe}}, \bibinfo {author} {\bibfnamefont {A.}~\bibnamefont {Yamazaki}},\
  and\ \bibinfo {author} {\bibfnamefont {J.}~\bibnamefont {Yoshida}},\
  }\bibfield  {title} {\enquote {\bibinfo {title} {The {{Missing Relationship}}
  between the {{Miscibility}} of {{Chiral Dopants}} and the {{Microscopic
  Dynamics}} of {{Solvent Liquid Crystals}}: {{A Molecular Dynamics Study}}},}\
  }\href {https://doi.org/10.3390/sym15051092} {\bibfield  {journal} {\bibinfo
  {journal} {Symmetry}\ }\textbf {\bibinfo {volume} {15}},\ \bibinfo {pages}
  {1092} (\bibinfo {year} {2023})}\BibitemShut {NoStop}%
\bibitem [{\citenamefont {Sarkar}\ \emph {et~al.}(2024)\citenamefont {Sarkar},
  \citenamefont {May}, \citenamefont {Jost},\ and\ \citenamefont
  {{M{\"u}ller-Plathe}}}]{sarkar2024Calculation}%
  \BibitemOpen
  \bibfield  {author} {\bibinfo {author} {\bibfnamefont {S.}~\bibnamefont
  {Sarkar}}, \bibinfo {author} {\bibfnamefont {F.}~\bibnamefont {May}},
  \bibinfo {author} {\bibfnamefont {M.}~\bibnamefont {Jost}},\ and\ \bibinfo
  {author} {\bibfnamefont {F.}~\bibnamefont {{M{\"u}ller-Plathe}}},\ }\bibfield
   {title} {\enquote {\bibinfo {title} {Calculation of {{Dielectric Spectra}}
  by {{Molecular Dynamics}}: 4-{{Cyano-4}}{$\prime$}-hexylbiphenyl in its
  {{Nematic Phase}}},}\ }\href {https://doi.org/10.1021/acs.jpcb.4c05154}
  {\bibfield  {journal} {\bibinfo  {journal} {J. Phys. Chem. B}\ ,\ \bibinfo
  {pages} {acs.jpcb.4c05154}} (\bibinfo {year} {2024})}\BibitemShut {NoStop}%
\bibitem [{\citenamefont {Kim}, \citenamefont {Keyes},\ and\ \citenamefont
  {Straub}(2010)}]{kim2010Generalized}%
  \BibitemOpen
  \bibfield  {author} {\bibinfo {author} {\bibfnamefont {J.}~\bibnamefont
  {Kim}}, \bibinfo {author} {\bibfnamefont {T.}~\bibnamefont {Keyes}},\ and\
  \bibinfo {author} {\bibfnamefont {J.~E.}\ \bibnamefont {Straub}},\ }\bibfield
   {title} {\enquote {\bibinfo {title} {Generalized {{Replica Exchange
  Method}}},}\ }\href {https://doi.org/10.1063/1.3432176} {\bibfield  {journal}
  {\bibinfo  {journal} {J. Chem. Phys.}\ }\textbf {\bibinfo {volume} {132}},\
  \bibinfo {pages} {224107} (\bibinfo {year} {2010})}\BibitemShut {NoStop}%
\bibitem [{\citenamefont {Lu}, \citenamefont {Kim},\ and\ \citenamefont
  {Straub}(2012)}]{lu2012Exploring}%
  \BibitemOpen
  \bibfield  {author} {\bibinfo {author} {\bibfnamefont {Q.}~\bibnamefont
  {Lu}}, \bibinfo {author} {\bibfnamefont {J.}~\bibnamefont {Kim}},\ and\
  \bibinfo {author} {\bibfnamefont {J.~E.}\ \bibnamefont {Straub}},\ }\bibfield
   {title} {\enquote {\bibinfo {title} {Exploring the {{Solid}}--{{Liquid Phase
  Change}} of an {{Adapted Dzugutov Model Using Generalized Replica Exchange
  Method}}},}\ }\href {https://doi.org/10.1021/jp300406c} {\bibfield  {journal}
  {\bibinfo  {journal} {J. Phys. Chem. B}\ }\textbf {\bibinfo {volume} {116}},\
  \bibinfo {pages} {8654--8661} (\bibinfo {year} {2012})}\BibitemShut {NoStop}%
\bibitem [{\citenamefont {Lu}, \citenamefont {Kim},\ and\ \citenamefont
  {Straub}(2013)}]{lu2013Order}%
  \BibitemOpen
  \bibfield  {author} {\bibinfo {author} {\bibfnamefont {Q.}~\bibnamefont
  {Lu}}, \bibinfo {author} {\bibfnamefont {J.}~\bibnamefont {Kim}},\ and\
  \bibinfo {author} {\bibfnamefont {J.~E.}\ \bibnamefont {Straub}},\ }\bibfield
   {title} {\enquote {\bibinfo {title} {Order parameter free enhanced sampling
  of the vapor-liquid transition using the generalized replica exchange
  method},}\ }\href {https://doi.org/10.1063/1.4794786} {\bibfield  {journal}
  {\bibinfo  {journal} {J. Chem. Phys.}\ }\textbf {\bibinfo {volume} {138}},\
  \bibinfo {pages} {104119} (\bibinfo {year} {2013})}\BibitemShut {NoStop}%
\bibitem [{\citenamefont {Lu}\ \emph {et~al.}(2014)\citenamefont {Lu},
  \citenamefont {Kim}, \citenamefont {Farrell}, \citenamefont {Wales},\ and\
  \citenamefont {Straub}}]{lu2014Investigating}%
  \BibitemOpen
  \bibfield  {author} {\bibinfo {author} {\bibfnamefont {Q.}~\bibnamefont
  {Lu}}, \bibinfo {author} {\bibfnamefont {J.}~\bibnamefont {Kim}}, \bibinfo
  {author} {\bibfnamefont {J.~D.}\ \bibnamefont {Farrell}}, \bibinfo {author}
  {\bibfnamefont {D.~J.}\ \bibnamefont {Wales}},\ and\ \bibinfo {author}
  {\bibfnamefont {J.~E.}\ \bibnamefont {Straub}},\ }\bibfield  {title}
  {\enquote {\bibinfo {title} {Investigating the solid-liquid phase transition
  of water nanofilms using the generalized replica exchange method},}\ }\href
  {https://doi.org/10.1063/1.4896513} {\bibfield  {journal} {\bibinfo
  {journal} {J. Chem. Phys.}\ }\textbf {\bibinfo {volume} {141}},\ \bibinfo
  {pages} {18C525} (\bibinfo {year} {2014})}\BibitemShut {NoStop}%
\bibitem [{\citenamefont {Ma{\l}olepsza}, \citenamefont {Secor},\ and\
  \citenamefont {Keyes}(2015)}]{malolepsza2015Isobaric}%
  \BibitemOpen
  \bibfield  {author} {\bibinfo {author} {\bibfnamefont {E.}~\bibnamefont
  {Ma{\l}olepsza}}, \bibinfo {author} {\bibfnamefont {M.}~\bibnamefont
  {Secor}},\ and\ \bibinfo {author} {\bibfnamefont {T.}~\bibnamefont {Keyes}},\
  }\bibfield  {title} {\enquote {\bibinfo {title} {Isobaric {{Molecular
  Dynamics Version}} of the {{Generalized Replica Exchange Method}} ({{gREM}}):
  {{Liquid}}--{{Vapor Equilibrium}}},}\ }\href
  {https://doi.org/10.1021/acs.jpcb.5b07614} {\bibfield  {journal} {\bibinfo
  {journal} {J. Phys. Chem. B}\ }\textbf {\bibinfo {volume} {119}},\ \bibinfo
  {pages} {13379--13384} (\bibinfo {year} {2015})}\BibitemShut {NoStop}%
\bibitem [{\citenamefont {Ma{\l}olepsza}\ and\ \citenamefont
  {Keyes}(2015)}]{malolepsza2015Water}%
  \BibitemOpen
  \bibfield  {author} {\bibinfo {author} {\bibfnamefont {E.}~\bibnamefont
  {Ma{\l}olepsza}}\ and\ \bibinfo {author} {\bibfnamefont {T.}~\bibnamefont
  {Keyes}},\ }\bibfield  {title} {\enquote {\bibinfo {title} {Water
  {{Freezing}} and {{Ice Melting}}},}\ }\href
  {https://doi.org/10.1021/acs.jctc.5b00637} {\bibfield  {journal} {\bibinfo
  {journal} {J. Chem. Theory Comput.}\ }\textbf {\bibinfo {volume} {11}},\
  \bibinfo {pages} {5613--5623} (\bibinfo {year} {2015})}\BibitemShut {NoStop}%
\bibitem [{\citenamefont {Ma{\l}olepsza}, \citenamefont {Kim},\ and\
  \citenamefont {Keyes}(2015)}]{malolepsza2015Entropic}%
  \BibitemOpen
  \bibfield  {author} {\bibinfo {author} {\bibfnamefont {E.}~\bibnamefont
  {Ma{\l}olepsza}}, \bibinfo {author} {\bibfnamefont {J.}~\bibnamefont {Kim}},\
  and\ \bibinfo {author} {\bibfnamefont {T.}~\bibnamefont {Keyes}},\ }\bibfield
   {title} {\enquote {\bibinfo {title} {Entropic {{Description}} of {{Gas
  Hydrate Ice-Liquid Equilibrium}} via {{Enhanced Sampling}} of {{Coexisting
  Phases}}},}\ }\href {https://doi.org/10.1103/PhysRevLett.114.170601}
  {\bibfield  {journal} {\bibinfo  {journal} {Phys. Rev. Lett.}\ }\textbf
  {\bibinfo {volume} {114}},\ \bibinfo {pages} {170601} (\bibinfo {year}
  {2015})}\BibitemShut {NoStop}%
\bibitem [{\citenamefont {Lu}\ and\ \citenamefont
  {Straub}(2016)}]{lu2016Freezing}%
  \BibitemOpen
  \bibfield  {author} {\bibinfo {author} {\bibfnamefont {Q.}~\bibnamefont
  {Lu}}\ and\ \bibinfo {author} {\bibfnamefont {J.~E.}\ \bibnamefont
  {Straub}},\ }\bibfield  {title} {\enquote {\bibinfo {title} {Freezing
  {{Transitions}} of {{Nanoconfined Coarse-Grained Water Show Subtle
  Dependence}} on {{Confining Environment}}},}\ }\href
  {https://doi.org/10.1021/acs.jpcb.5b10481} {\bibfield  {journal} {\bibinfo
  {journal} {J. Phys. Chem. B}\ }\textbf {\bibinfo {volume} {120}},\ \bibinfo
  {pages} {2517--2525} (\bibinfo {year} {2016})}\BibitemShut {NoStop}%
\bibitem [{\citenamefont {Stelter}\ and\ \citenamefont
  {Keyes}(2017)}]{stelter2017Enhanced}%
  \BibitemOpen
  \bibfield  {author} {\bibinfo {author} {\bibfnamefont {D.}~\bibnamefont
  {Stelter}}\ and\ \bibinfo {author} {\bibfnamefont {T.}~\bibnamefont
  {Keyes}},\ }\bibfield  {title} {\enquote {\bibinfo {title} {Enhanced
  {{Sampling}} of {{Phase Transitions}} in {{Coarse-Grained Lipid
  Bilayers}}},}\ }\href {https://doi.org/10.1021/acs.jpcb.6b11711} {\bibfield
  {journal} {\bibinfo  {journal} {J. Phys. Chem. B}\ }\textbf {\bibinfo
  {volume} {121}},\ \bibinfo {pages} {5770--5780} (\bibinfo {year}
  {2017})}\BibitemShut {NoStop}%
\bibitem [{\citenamefont {Piskulich}\ and\ \citenamefont
  {Cui}(2022)}]{piskulich2022Machine}%
  \BibitemOpen
  \bibfield  {author} {\bibinfo {author} {\bibfnamefont {Z.~A.}\ \bibnamefont
  {Piskulich}}\ and\ \bibinfo {author} {\bibfnamefont {Q.}~\bibnamefont
  {Cui}},\ }\bibfield  {title} {\enquote {\bibinfo {title} {Machine
  {{Learning-Assisted Phase Transition Temperatures}} from {{Generalized
  Replica Exchange Simulations}} of {{Dry Martini Lipid Bilayers}}},}\ }\href
  {https://doi.org/10.1021/acs.jpclett.2c01654} {\bibfield  {journal} {\bibinfo
   {journal} {J. Phys. Chem. Lett.}\ ,\ \bibinfo {pages} {6481--6486}}
  (\bibinfo {year} {2022})}\BibitemShut {NoStop}%
\bibitem [{\citenamefont {Takemoto}\ \emph {et~al.}(2022)\citenamefont
  {Takemoto}, \citenamefont {Ishii}, \citenamefont {Washizu}, \citenamefont
  {Kim},\ and\ \citenamefont {Matubayasi}}]{takemoto2022Simulating}%
  \BibitemOpen
  \bibfield  {author} {\bibinfo {author} {\bibfnamefont {K.}~\bibnamefont
  {Takemoto}}, \bibinfo {author} {\bibfnamefont {Y.}~\bibnamefont {Ishii}},
  \bibinfo {author} {\bibfnamefont {H.}~\bibnamefont {Washizu}}, \bibinfo
  {author} {\bibfnamefont {K.}~\bibnamefont {Kim}},\ and\ \bibinfo {author}
  {\bibfnamefont {N.}~\bibnamefont {Matubayasi}},\ }\bibfield  {title}
  {\enquote {\bibinfo {title} {Simulating the nematic-isotropic phase
  transition of liquid crystal model via generalized replica-exchange
  method},}\ }\href {https://doi.org/10.1063/5.0073105} {\bibfield  {journal}
  {\bibinfo  {journal} {J. Chem. Phys.}\ }\textbf {\bibinfo {volume} {156}},\
  \bibinfo {pages} {014901} (\bibinfo {year} {2022})}\BibitemShut {NoStop}%
\bibitem [{\citenamefont {Berardi}\ \emph {et~al.}(2009)\citenamefont
  {Berardi}, \citenamefont {Zannoni}, \citenamefont {Lintuvuori},\ and\
  \citenamefont {Wilson}}]{berardi2009Softcore}%
  \BibitemOpen
  \bibfield  {author} {\bibinfo {author} {\bibfnamefont {R.}~\bibnamefont
  {Berardi}}, \bibinfo {author} {\bibfnamefont {C.}~\bibnamefont {Zannoni}},
  \bibinfo {author} {\bibfnamefont {J.~S.}\ \bibnamefont {Lintuvuori}},\ and\
  \bibinfo {author} {\bibfnamefont {M.~R.}\ \bibnamefont {Wilson}},\ }\bibfield
   {title} {\enquote {\bibinfo {title} {A soft-core {{Gay}}--{{Berne}} model
  for the simulation of liquid crystals by {{Hamiltonian}} replica exchange},}\
  }\href {https://doi.org/10.1063/1.3254019} {\bibfield  {journal} {\bibinfo
  {journal} {J. Chem. Phys.}\ }\textbf {\bibinfo {volume} {131}},\ \bibinfo
  {pages} {174107} (\bibinfo {year} {2009})}\BibitemShut {NoStop}%
\bibitem [{\citenamefont {Kowaguchi}, \citenamefont {Brumby},\ and\
  \citenamefont {Yasuoka}(2021)}]{kowaguchi2021Phase}%
  \BibitemOpen
  \bibfield  {author} {\bibinfo {author} {\bibfnamefont {A.}~\bibnamefont
  {Kowaguchi}}, \bibinfo {author} {\bibfnamefont {P.~E.}\ \bibnamefont
  {Brumby}},\ and\ \bibinfo {author} {\bibfnamefont {K.}~\bibnamefont
  {Yasuoka}},\ }\bibfield  {title} {\enquote {\bibinfo {title} {Phase
  {{Transitions}} and {{Hysteresis}} for a {{Simple Model Liquid Crystal}} by
  {{Replica-Exchange Monte Carlo Simulations}}},}\ }\href
  {https://doi.org/10.3390/molecules26051421} {\bibfield  {journal} {\bibinfo
  {journal} {Molecules}\ }\textbf {\bibinfo {volume} {26}},\ \bibinfo {pages}
  {1421} (\bibinfo {year} {2021})}\BibitemShut {NoStop}%
\bibitem [{\citenamefont {Kowaguchi}\ \emph {et~al.}(2022)\citenamefont
  {Kowaguchi}, \citenamefont {Endo}, \citenamefont {Brumby}, \citenamefont
  {Nomura},\ and\ \citenamefont {Yasuoka}}]{kowaguchi2022Optimal}%
  \BibitemOpen
  \bibfield  {author} {\bibinfo {author} {\bibfnamefont {A.}~\bibnamefont
  {Kowaguchi}}, \bibinfo {author} {\bibfnamefont {K.}~\bibnamefont {Endo}},
  \bibinfo {author} {\bibfnamefont {P.~E.}\ \bibnamefont {Brumby}}, \bibinfo
  {author} {\bibfnamefont {K.}~\bibnamefont {Nomura}},\ and\ \bibinfo {author}
  {\bibfnamefont {K.}~\bibnamefont {Yasuoka}},\ }\bibfield  {title} {\enquote
  {\bibinfo {title} {Optimal {{Replica-Exchange Molecular Simulations}} in
  {{Combination}} with {{Evolution Strategies}}},}\ }\href
  {https://doi.org/10.1021/acs.jcim.2c00608} {\bibfield  {journal} {\bibinfo
  {journal} {J. Chem. Inf. Model.}\ ,\ \bibinfo {pages} {acs.jcim.2c00608}}
  (\bibinfo {year} {2022})}\BibitemShut {NoStop}%
\bibitem [{\citenamefont {Kowaguchi}, \citenamefont {Brumby},\ and\
  \citenamefont {Yasuoka}(2024)}]{kowaguchi2024Hysteresis}%
  \BibitemOpen
  \bibfield  {author} {\bibinfo {author} {\bibfnamefont {A.}~\bibnamefont
  {Kowaguchi}}, \bibinfo {author} {\bibfnamefont {P.~E.}\ \bibnamefont
  {Brumby}},\ and\ \bibinfo {author} {\bibfnamefont {K.}~\bibnamefont
  {Yasuoka}},\ }\bibfield  {title} {\enquote {\bibinfo {title} {Hysteresis
  {{Elimination}} for an {{Anisotropic Liquid-Crystal Model}} via {{Molecule
  Design}} and {{Replica-Exchange Optimization}}},}\ }\href
  {https://doi.org/10.1021/acs.jcim.4c00078} {\bibfield  {journal} {\bibinfo
  {journal} {J. Chem. Inf. Model.}\ }\textbf {\bibinfo {volume} {64}},\
  \bibinfo {pages} {4673--4686} (\bibinfo {year} {2024})}\BibitemShut {NoStop}%
\bibitem [{\citenamefont {Plimpton}(1995)}]{plimpton1995Fast}%
  \BibitemOpen
  \bibfield  {author} {\bibinfo {author} {\bibfnamefont {S.}~\bibnamefont
  {Plimpton}},\ }\bibfield  {title} {\enquote {\bibinfo {title} {Fast
  {{Parallel Algorithms}} for {{Short-Range Molecular Dynamics}}},}\ }\href
  {https://doi.org/10.1006/jcph.1995.1039} {\bibfield  {journal} {\bibinfo
  {journal} {J. Comput. Phys.}\ }\textbf {\bibinfo {volume} {117}},\ \bibinfo
  {pages} {1--19} (\bibinfo {year} {1995})}\BibitemShut {NoStop}%
\bibitem [{\citenamefont {Matubayasi}\ and\ \citenamefont
  {Nakahara}(2000)}]{matubayasi2000Theory}%
  \BibitemOpen
  \bibfield  {author} {\bibinfo {author} {\bibfnamefont {N.}~\bibnamefont
  {Matubayasi}}\ and\ \bibinfo {author} {\bibfnamefont {M.}~\bibnamefont
  {Nakahara}},\ }\bibfield  {title} {\enquote {\bibinfo {title} {Theory of
  solutions in the energetic representation. {{I}}. {{Formulation}}},}\ }\href
  {https://doi.org/10.1063/1.1309013} {\bibfield  {journal} {\bibinfo
  {journal} {J. Chem. Phys.}\ }\textbf {\bibinfo {volume} {113}},\ \bibinfo
  {pages} {6070--6081} (\bibinfo {year} {2000})}\BibitemShut {NoStop}%
\bibitem [{\citenamefont {Matubayasi}\ and\ \citenamefont
  {Nakahara}(2002)}]{matubayasi2002Theory}%
  \BibitemOpen
  \bibfield  {author} {\bibinfo {author} {\bibfnamefont {N.}~\bibnamefont
  {Matubayasi}}\ and\ \bibinfo {author} {\bibfnamefont {M.}~\bibnamefont
  {Nakahara}},\ }\bibfield  {title} {\enquote {\bibinfo {title} {Theory of
  solutions in the energy representation. {{II}}. {{Functional}} for the
  chemical potential},}\ }\href {https://doi.org/10.1063/1.1495850} {\bibfield
  {journal} {\bibinfo  {journal} {J. Chem. Phys.}\ }\textbf {\bibinfo {volume}
  {117}},\ \bibinfo {pages} {3605--3616} (\bibinfo {year} {2002})}\BibitemShut
  {NoStop}%
\bibitem [{\citenamefont {Matubayasi}\ and\ \citenamefont
  {Nakahara}(2003)}]{matubayasi2003Theory}%
  \BibitemOpen
  \bibfield  {author} {\bibinfo {author} {\bibfnamefont {N.}~\bibnamefont
  {Matubayasi}}\ and\ \bibinfo {author} {\bibfnamefont {M.}~\bibnamefont
  {Nakahara}},\ }\bibfield  {title} {\enquote {\bibinfo {title} {Theory of
  solutions in the energy representation. {{III}}. {{Treatment}} of the
  molecular flexibility},}\ }\href {https://doi.org/10.1063/1.1613938}
  {\bibfield  {journal} {\bibinfo  {journal} {J. Chem. Phys.}\ }\textbf
  {\bibinfo {volume} {119}},\ \bibinfo {pages} {9686--9702} (\bibinfo {year}
  {2003})}\BibitemShut {NoStop}%
\bibitem [{\citenamefont {Sakuraba}\ and\ \citenamefont
  {Matubayasi}(2014)}]{sakuraba2014Ermod}%
  \BibitemOpen
  \bibfield  {author} {\bibinfo {author} {\bibfnamefont {S.}~\bibnamefont
  {Sakuraba}}\ and\ \bibinfo {author} {\bibfnamefont {N.}~\bibnamefont
  {Matubayasi}},\ }\bibfield  {title} {\enquote {\bibinfo {title} {Ermod:
  {{Fast}} and versatile computation software for solvation free energy with
  approximate theory of solutions},}\ }\href
  {https://doi.org/10.1002/jcc.23651} {\bibfield  {journal} {\bibinfo
  {journal} {J. Comput. Chem.}\ }\textbf {\bibinfo {volume} {35}},\ \bibinfo
  {pages} {1592--1608} (\bibinfo {year} {2014})}\BibitemShut {NoStop}%
\bibitem [{\citenamefont
  {Matubayasi}(2019)}]{matubayasi2019EnergyRepresentation}%
  \BibitemOpen
  \bibfield  {author} {\bibinfo {author} {\bibfnamefont {N.}~\bibnamefont
  {Matubayasi}},\ }\bibfield  {title} {\enquote {\bibinfo {title}
  {Energy-{{Representation Theory}} of {{Solutions}}: {{Its Formulation}} and
  {{Application}} to {{Soft}}, {{Molecular Aggregates}}},}\ }\href
  {https://doi.org/10.1246/bcsj.20190246} {\bibfield  {journal} {\bibinfo
  {journal} {Bull. Chem. Soc. Jpn.}\ }\textbf {\bibinfo {volume} {92}},\
  \bibinfo {pages} {1910--1927} (\bibinfo {year} {2019})}\BibitemShut {NoStop}%
\bibitem [{\citenamefont {Karino}\ and\ \citenamefont
  {Matubayasi}(2013)}]{karino2013Interactioncomponent}%
  \BibitemOpen
  \bibfield  {author} {\bibinfo {author} {\bibfnamefont {Y.}~\bibnamefont
  {Karino}}\ and\ \bibinfo {author} {\bibfnamefont {N.}~\bibnamefont
  {Matubayasi}},\ }\bibfield  {title} {\enquote {\bibinfo {title}
  {Interaction-component analysis of the urea effect on amino acid analogs},}\
  }\href {https://doi.org/10.1039/c3cp43346c} {\bibfield  {journal} {\bibinfo
  {journal} {Phys. Chem. Chem. Phys.}\ }\textbf {\bibinfo {volume} {15}},\
  \bibinfo {pages} {4377} (\bibinfo {year} {2013})}\BibitemShut {NoStop}%
\bibitem [{\citenamefont {Eppenga}\ and\ \citenamefont
  {Frenkel}(1984)}]{eppenga1984Monte}%
  \BibitemOpen
  \bibfield  {author} {\bibinfo {author} {\bibfnamefont {R.}~\bibnamefont
  {Eppenga}}\ and\ \bibinfo {author} {\bibfnamefont {D.}~\bibnamefont
  {Frenkel}},\ }\bibfield  {title} {\enquote {\bibinfo {title} {Monte {{Carlo}}
  study of the isotropic and nematic phases of infinitely thin hard
  platelets},}\ }\href {https://doi.org/10.1080/00268978400101951} {\bibfield
  {journal} {\bibinfo  {journal} {Mol. Phys.}\ }\textbf {\bibinfo {volume}
  {52}},\ \bibinfo {pages} {1303--1334} (\bibinfo {year} {1984})}\BibitemShut
  {NoStop}%
\end{thebibliography}
%aipnum4-2.bst 2019-01-14 (MD) hand-edited version of apsrev4-1.bst

\end{document}

\setcounter{equation}{0}
\setcounter{figure}{0}
\setcounter{table}{0}
\setcounter{page}{1}

\noindent{\bf\Large Supplementary Material}
\vspace{5mm}
\begin{center}
\textbf{\large  
Atomistic analysis of nematic phase transition in
 4-cyano-4$^{\prime}$-$n$-alkyl biphenyl liquid crystals: 
Sampling for the first-order phase transition and the free-energy 
decomposition}\\

\vspace{5mm}
 
{Shunsuke Ogita,$^1$ Yoshiki Ishii,$^2$ Go Watanabe,$^2$ Hitoshi Washizu,$^{3}$
Kang Kim,$^1$
and Nobuyuki Matubayasi$^1$}
\\

\vspace{5mm}

\noindent
\textit{
$^{1)}$Division of Chemical Engineering, Graduate School of Engineering Science, Osaka University, Osaka 560-8531, Japan}\\
\textit{
$^{2)}$Department of Data Science, School of Frontier Engineering, Kitasato University, Sagamihara, Kanagawa 252-0373, Japan}\\
\textit{
$^{3)}$Graduate School of Information Science, University of Hyogo, Hyogo 650-0047, Japan}
\end{center}

%%%%%%%%% Fig. S1 %%%%%%%%%%%%%%%%%%%%%%%%%%%%%%%%%%%%%%%%%%%%%%%%%%%%%%%%%%%%%%%%%%%%%%%%%
\begin{figure}[H]
\centering
\includegraphics[width=0.8\textwidth]{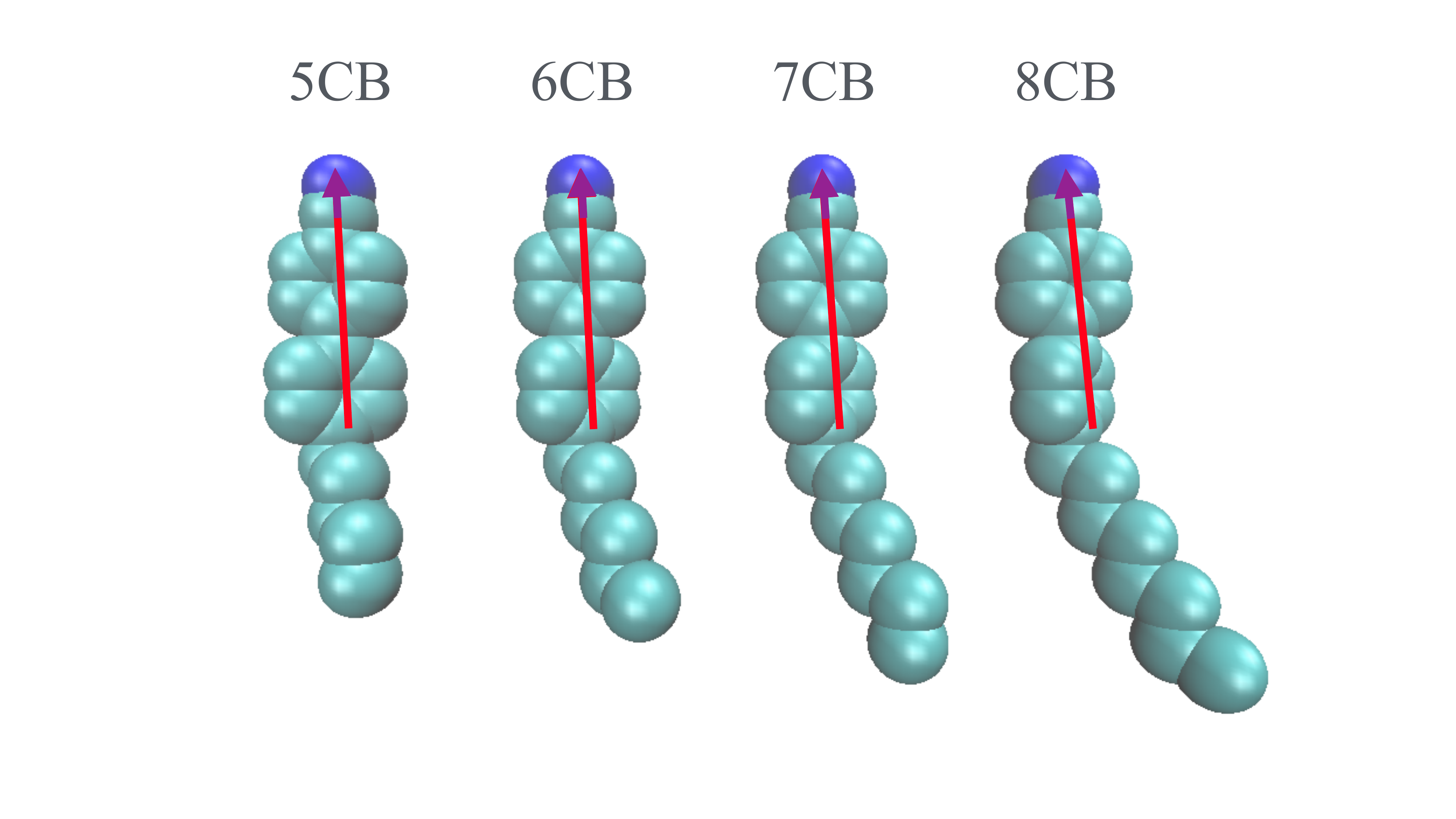}
\caption{Molecular structures of $n$CB ($n=5$, 6, 7, and 8) are shown,
 with 
purple and red vectors representing the unit vectors as the molecular axes
 for calculating the orientational order parameter tensor $\bm{Q}$.
}
\end{figure}
%%%%%%%%%%%%%%%%%%%%%%%%%%%%%%%%%%%%%%%%%%%%%%%%%%%%%%%%%%%%%%%%%%%%%%%%%%%%%%%%%%%%%%%%%%

%%%%%%%%% Table S1 %%%%%%%%%%%%%%%%%%%%%%%%%%%%%%%%%%%%%%%%%%%%%%%%%%%%%%%%%%%%%%%%%%%%%%%%%
\begin{table}[H]
\centering
\caption{Parameters, $T_0$, $T_M$, $\hat{H}_1$, $\hat{H}_M$,
 and $\gamma$, 
 for gREM calculations for $n$CB ($n=5$, 6, 7, and 8).}
  \begin{tabular}{lcccc}
  \hline
  \hline
   & 5CB & 6CB & 7CB & 8CB \\ \hline
  $T_1$ (K) & 300 & 290 & 310 & 310 \\ 
  $T_M$ (K) & 320 & 310 & 330 & 330 \\ 
  $\tilde{H}_0$ (kcal/mol) & 275901.8 & 253970.7 & 258457.6 & 278985.2 \\
  $\tilde{H}_M$ (kcal/mol) & 284902.2 & 262995.8 & 269201.7 & 289908 \\
  $\gamma$ ((K $\cdot $mol)/kcal) & -0.00222 & -0.00222 & -0.00186 & -0.00183 \\ 
  \hline
  \hline
  \end{tabular}
\end{table}
%%%%%%%%%%%%%%%%%%%%%%%%%%%%%%%%%%%%%%%%%%%%%%%%%%%%%%%%%%%%%%%%%%%%%%%%%%%%%%%%%%%%%%%%%%

\newpage
%%%%%%%%% Fig. S2 %%%%%%%%%%%%%%%%%%%%%%%%%%%%%%%%%%%%%%%%%%%%%%%%%%%%%%%%%%%%%%%%%%%%%%%%%
\begin{figure}[t]
\centering
\includegraphics[width=0.4\textwidth]{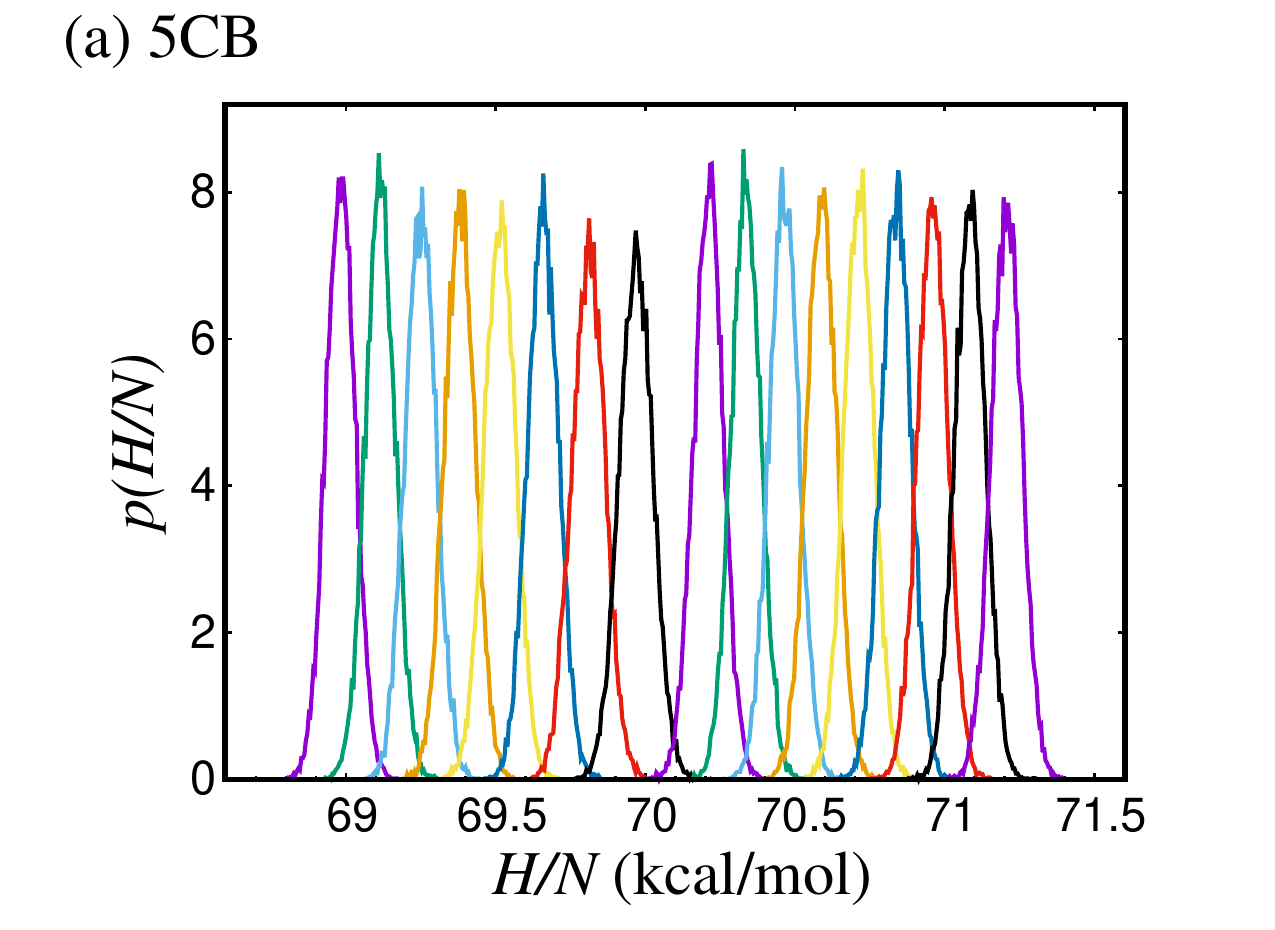}
\includegraphics[width=0.4\textwidth]{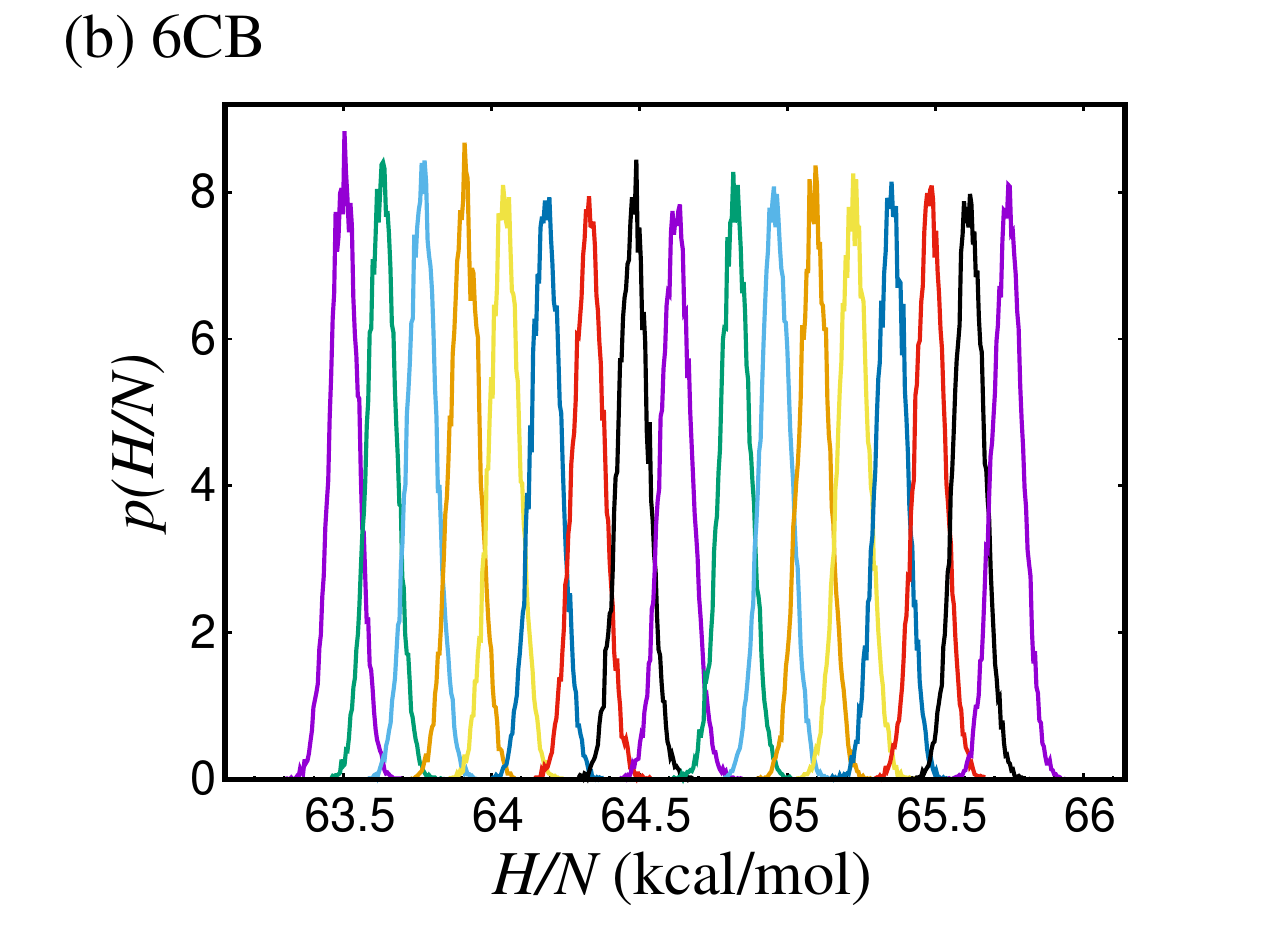}
\includegraphics[width=0.4\textwidth]{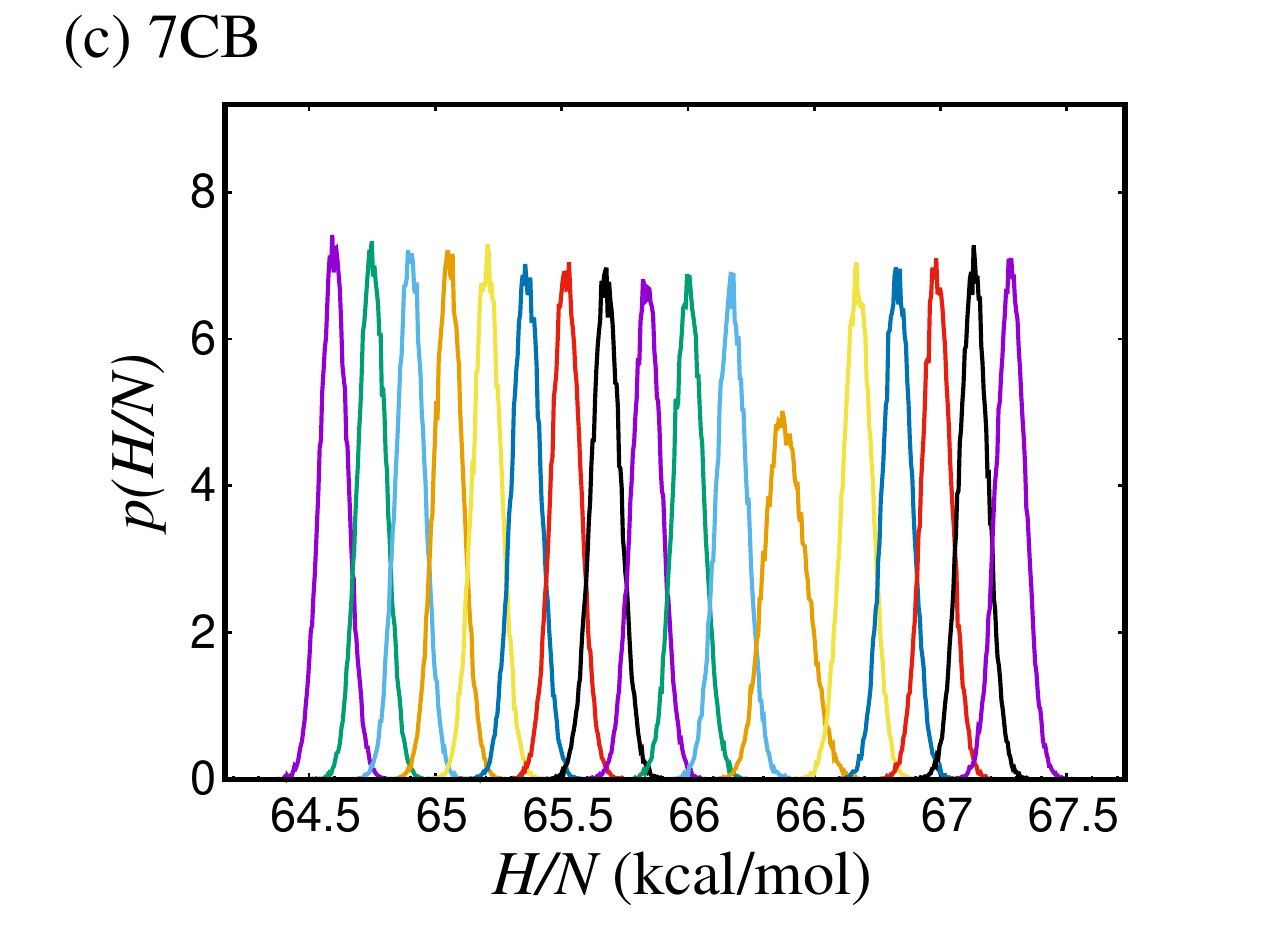}
\includegraphics[width=0.4\textwidth]{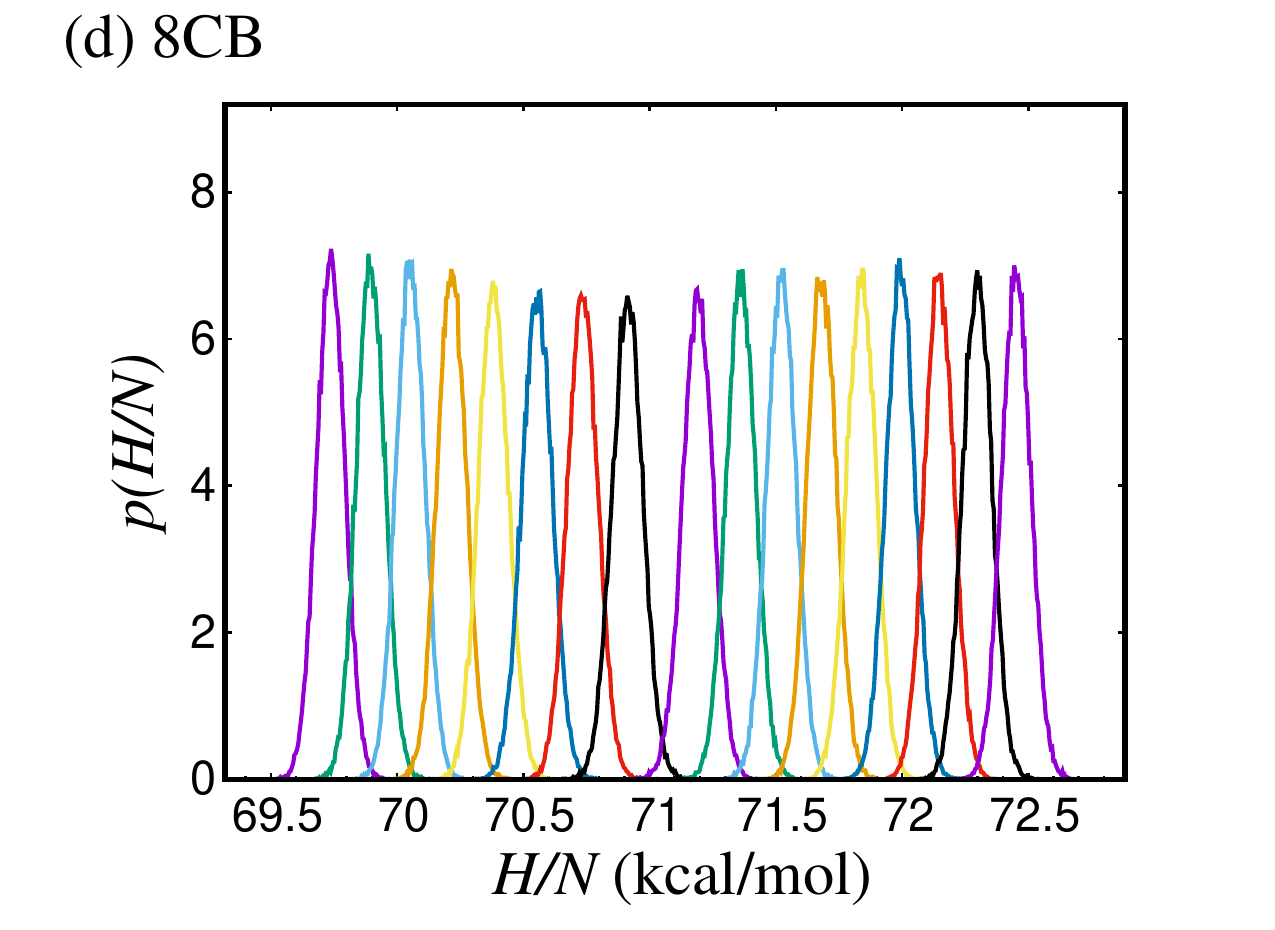}
\caption{Probability distributions of enthalpy $H$ of each replica in
 gREM simulations for
 5CB (a), 6CB (b), 7CB (c), and 8CB (d).
From left to right, the replica index $\alpha$ ranges from 1 to $M=17$. 
The enthalpy is normalized by the number of molecules $N=4000$.
}
\end{figure}
%%%%%%%%%%%%%%%%%%%%%%%%%%%%%%%%%%%%%%%%%%%%%%%%%%%%%%%%%%%%%%%%%%%%%%%%%%%%%%%%%%%%%%%%%%

\newpage
%%%%%%%%% Fig. S3 %%%%%%%%%%%%%%%%%%%%%%%%%%%%%%%%%%%%%%%%%%%%%%%%%%%%%%%%%%%%%%%%%%%%%%%%%
\begin{figure}[t]
\centering
\includegraphics[width=0.4\textwidth]{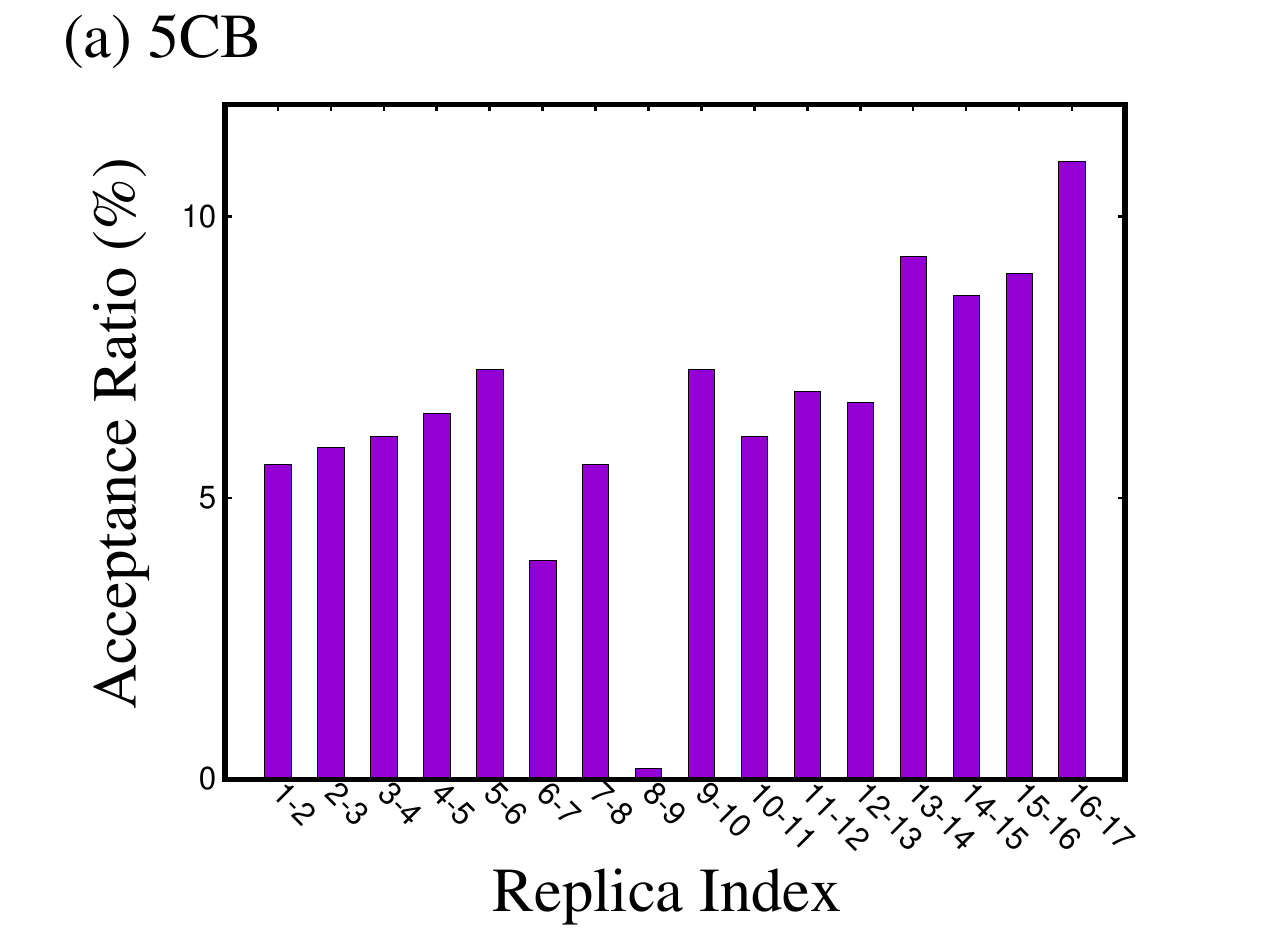}
\includegraphics[width=0.4\textwidth]{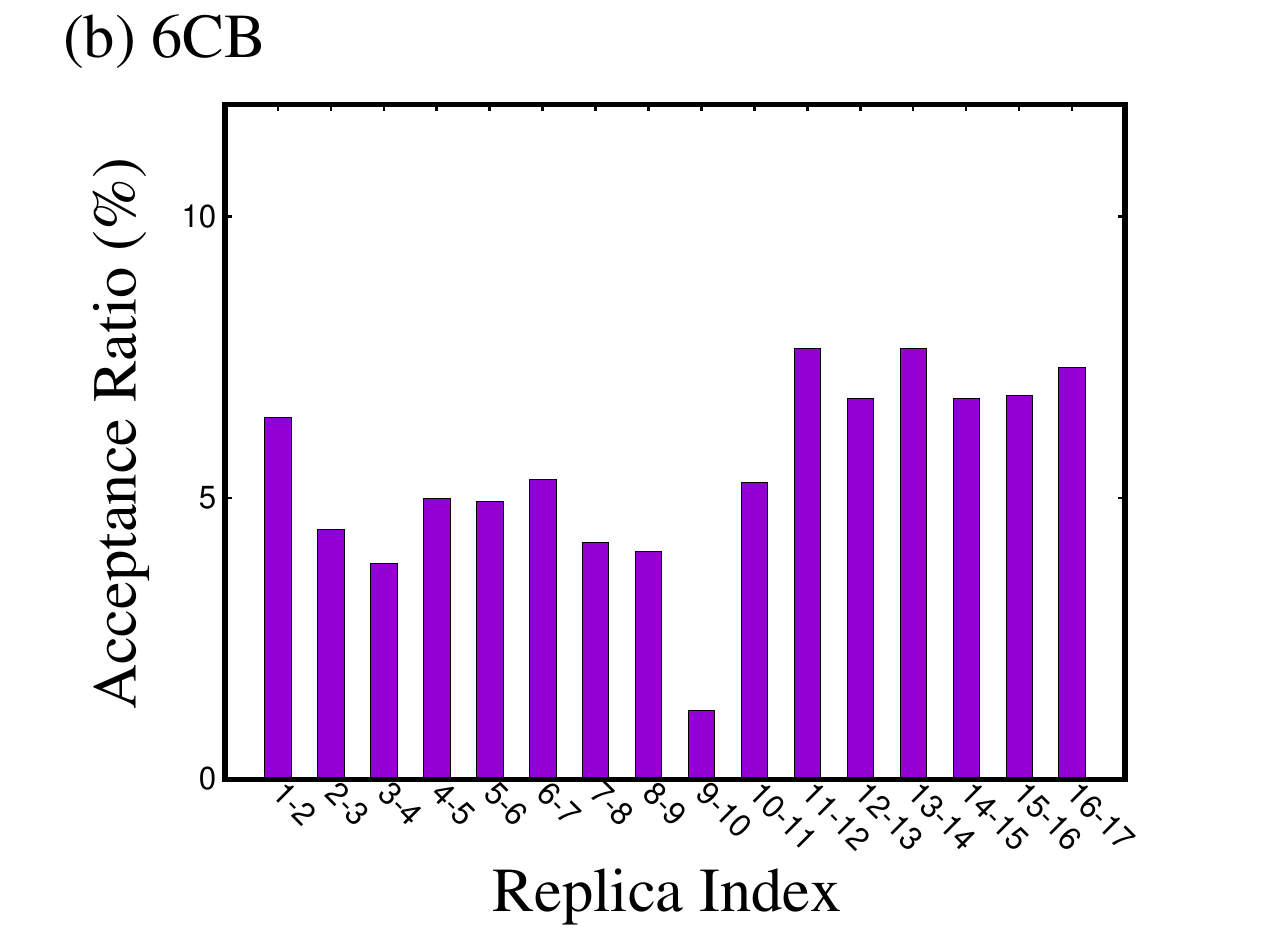}
\includegraphics[width=0.4\textwidth]{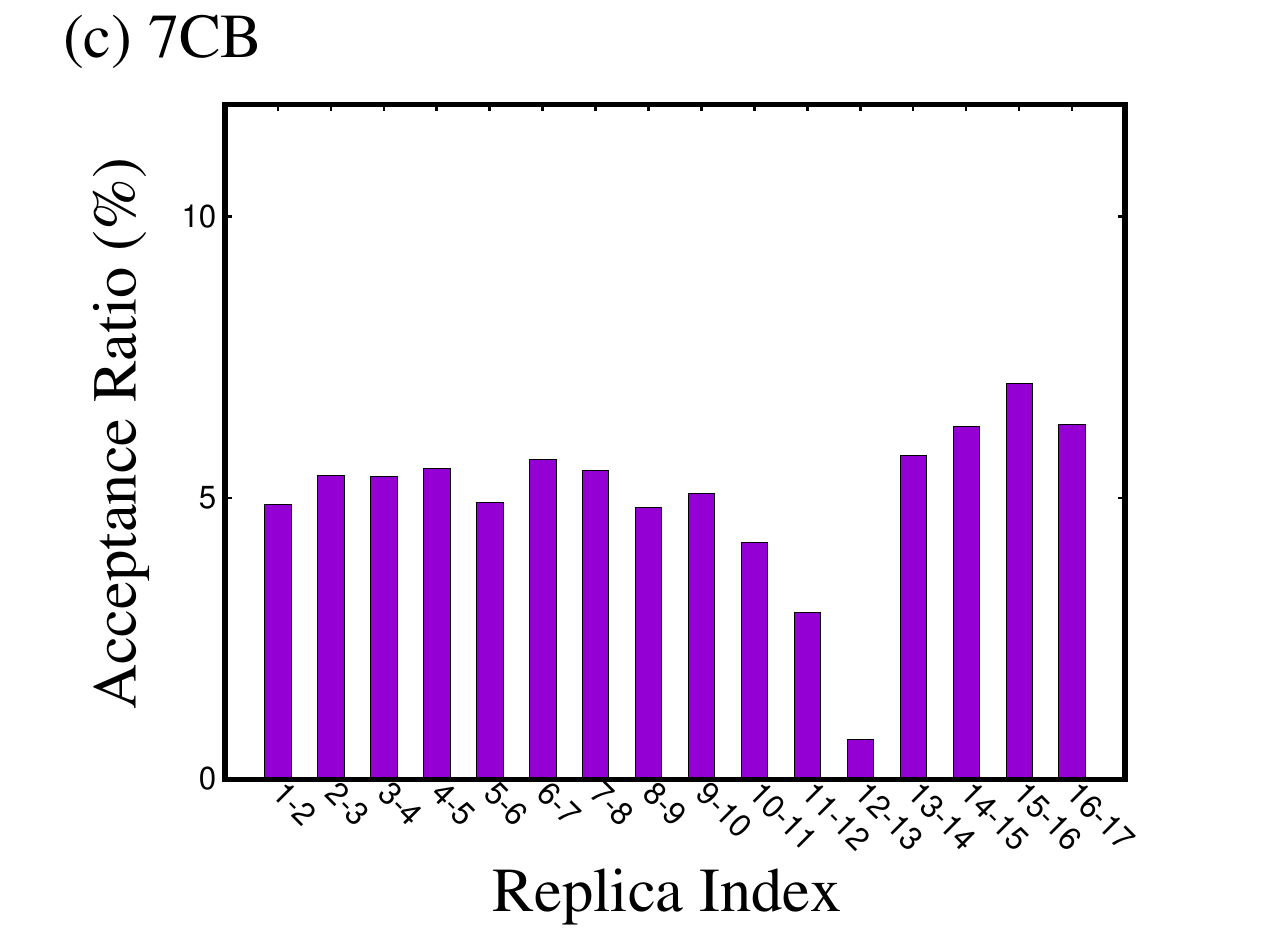}
\includegraphics[width=0.4\textwidth]{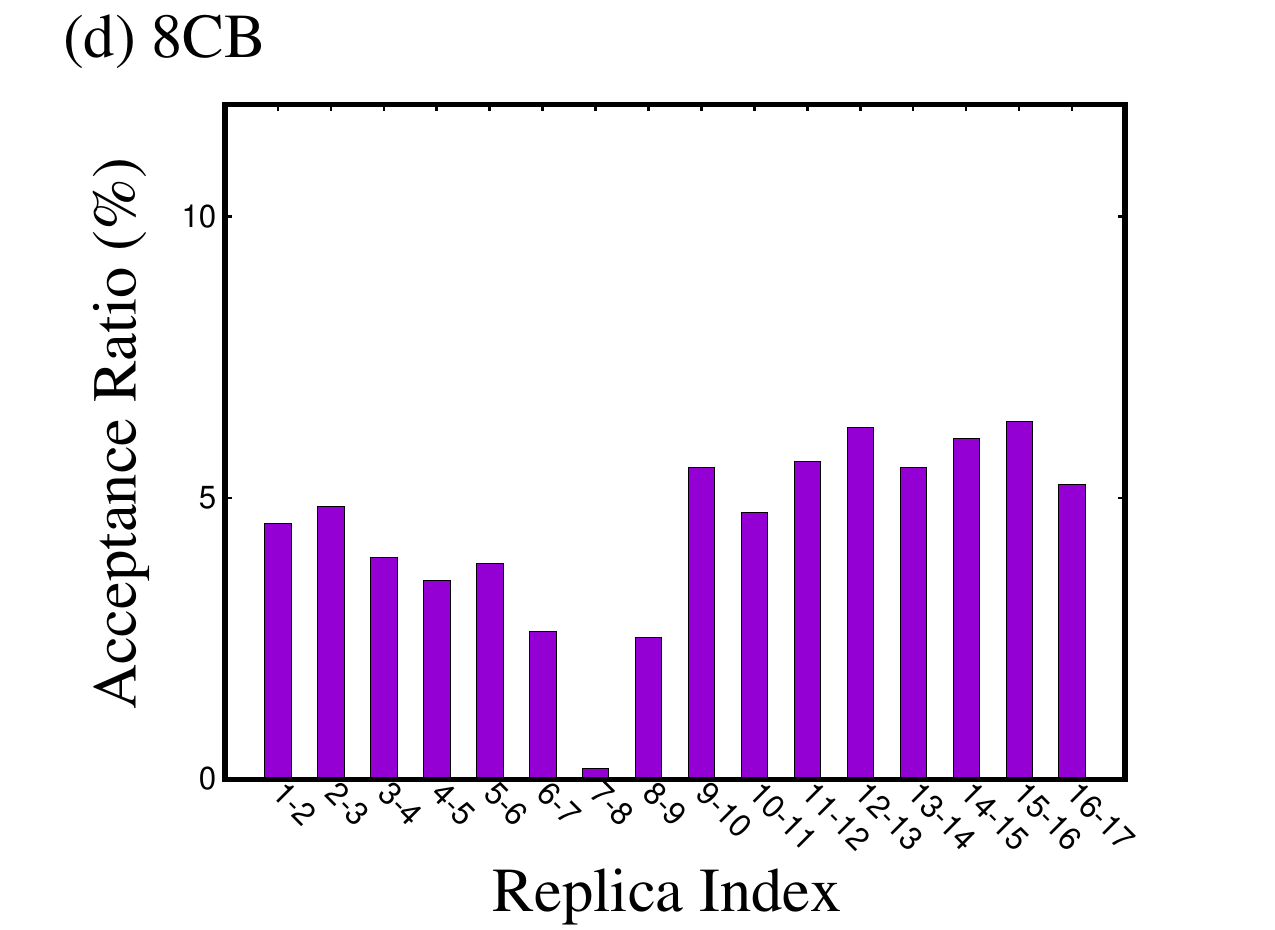}
\caption{Acceptance ratios of replica exchanges in gREM simulations for
 5CB (a), 6CB (b), 7CB (c), and 8CB (d).
}
\end{figure}
%%%%%%%%%%%%%%%%%%%%%%%%%%%%%%%%%%%%%%%%%%%%%%%%%%%%%%%%%%%%%%%%%%%%%%%%%%%%%%%%%%%%%%%%%%

\newpage
%%%%%%%% Fig. S4 %%%%%%%%%%%%%%%%%%%%%%%%%%%%%%%%%%%%%%%%%%%%%%%%%%%%%%%%%%%%%%%%%%%%%%%%%
\begin{figure}[t]
\centering
\includegraphics[width=0.4\textwidth]{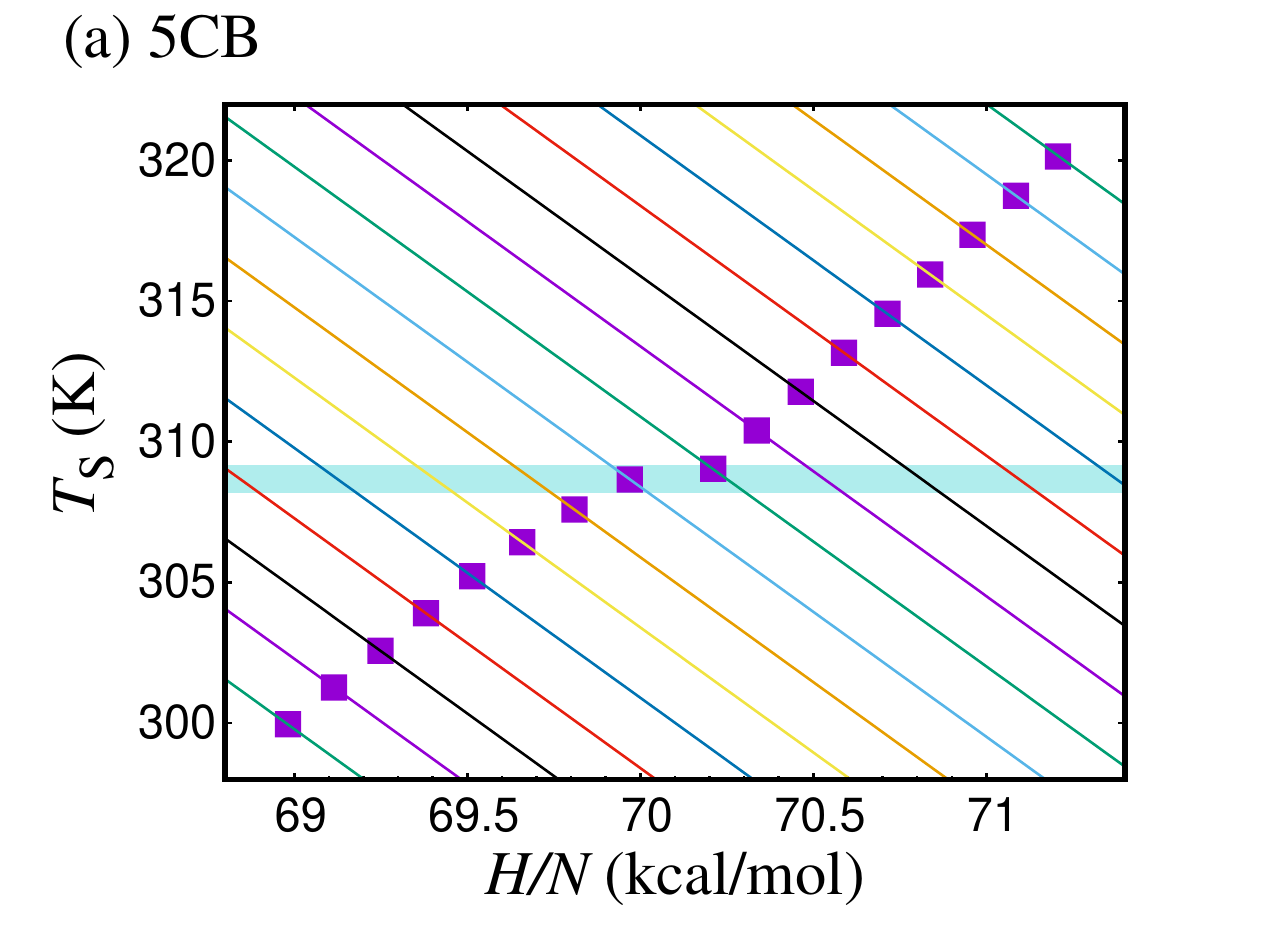}
\includegraphics[width=0.4\textwidth]{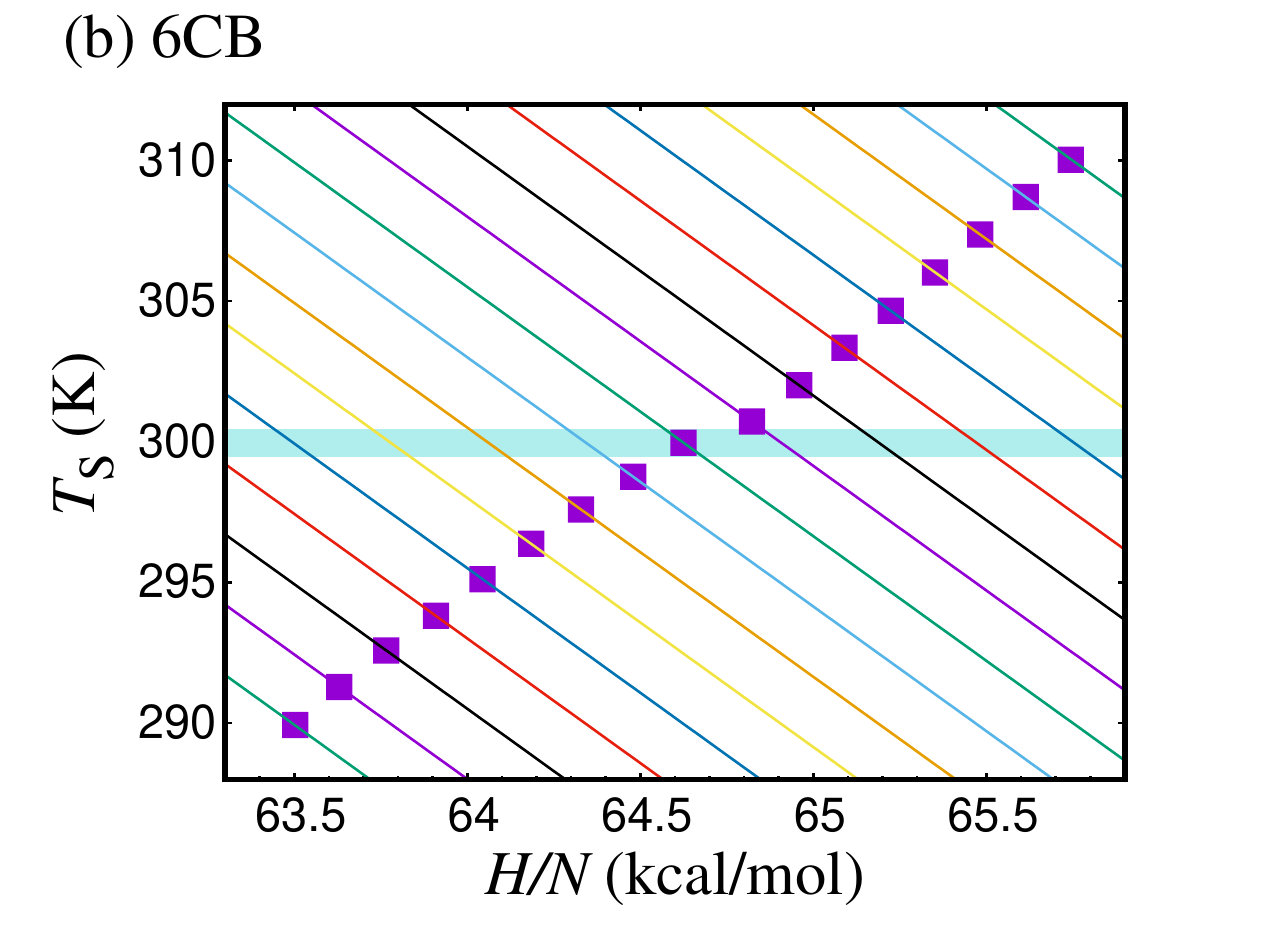}
\includegraphics[width=0.4\textwidth]{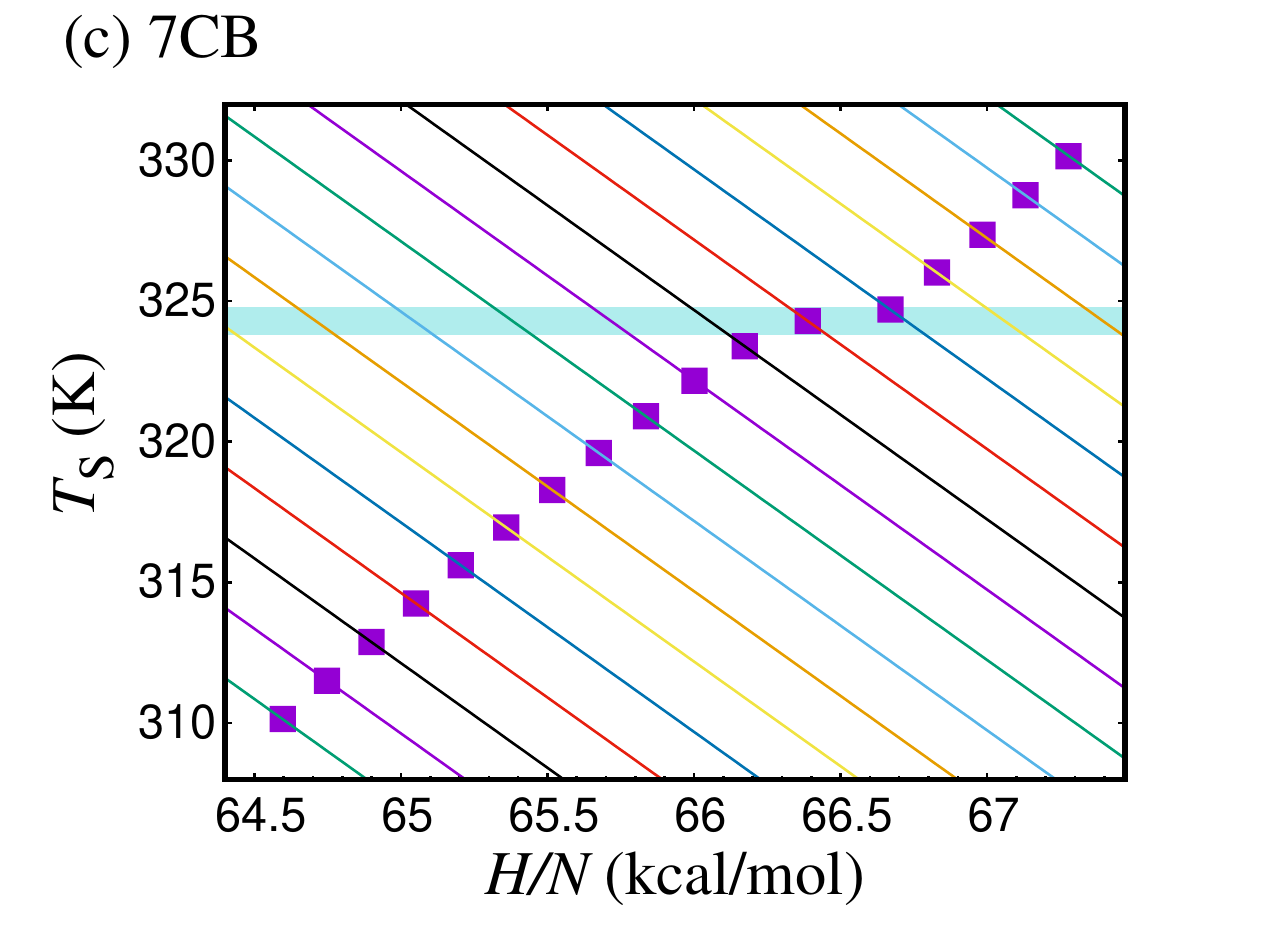}
\includegraphics[width=0.4\textwidth]{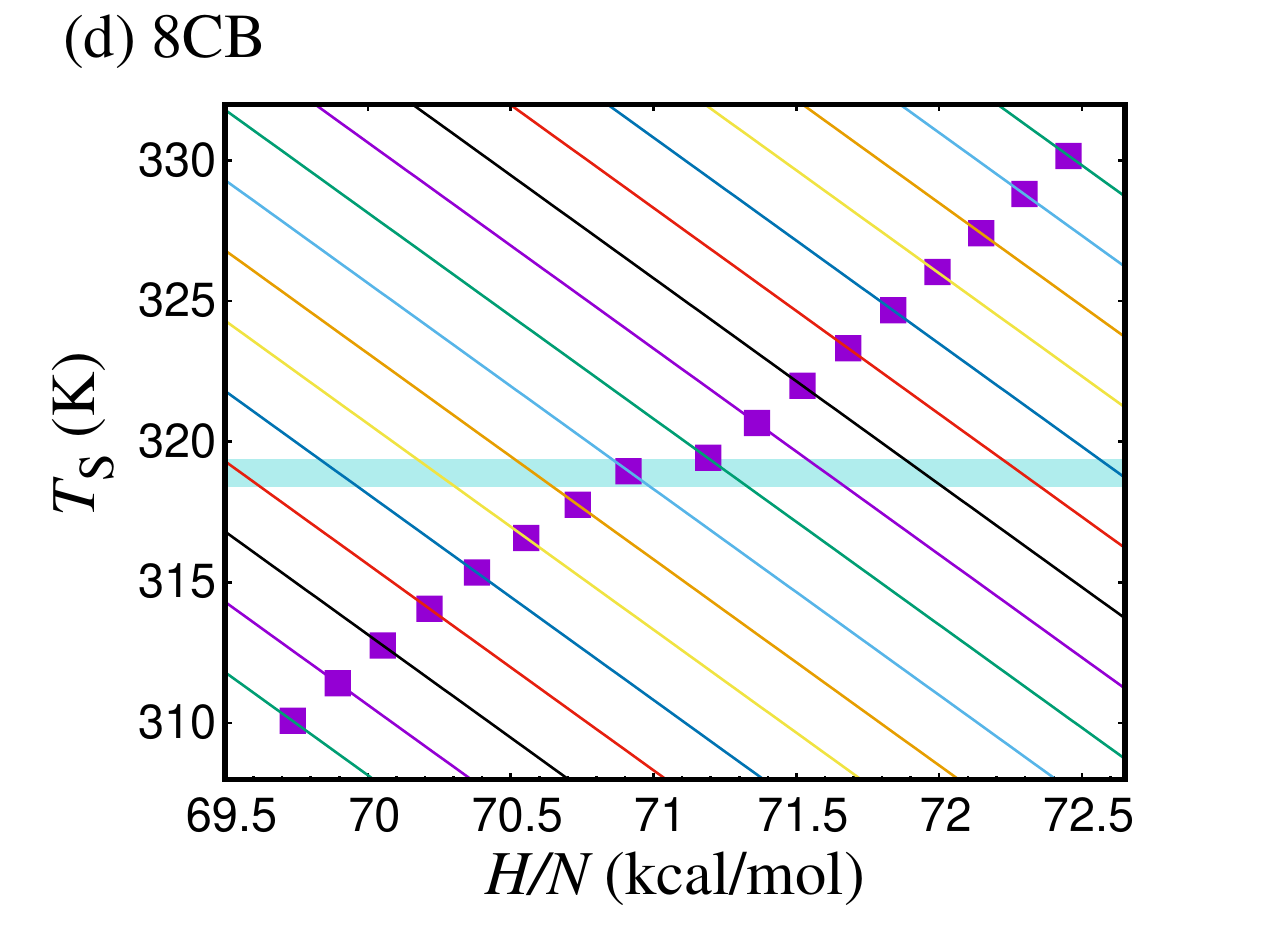}
\caption{Statistical temperature $T_\mathrm{S}$ as a function of 
enthalpy $H$ at each replica in
 gREM simulations for 5CB (a), 6CB (b), 7CB (c), and 8CB (d).
From left to right, the replica index $\alpha$ ranges from 1 to $M=17$. 
The enthalpy is normalized by the number of molecules $N=4000$.
The horizontal color bar indicates the NI transition temperature $T_\mathrm{NI}$.
The straight line represents the linear effective 
temperature at each replica, given by $T_\alpha (H)=\lambda_\alpha+\gamma (H-H_0)$ (see
 Eq.~(5) in the main text).
}
\end{figure}
%%%%%%%%%%%%%%%%%%%%%%%%%%%%%%%%%%%%%%%%%%%%%%%%%%%%%%%%%%%%%%%%%%%%%%%%%%%%%%%%%%%%%%%%%%

\newpage
%%%%%%%% Fig. S5 %%%%%%%%%%%%%%%%%%%%%%%%%%%%%%%%%%%%%%%%%%%%%%%%%%%%%%%%%%%%%%%%%%%%%%%%%
\begin{figure}[t]
\centering
\includegraphics[width=0.4\textwidth]{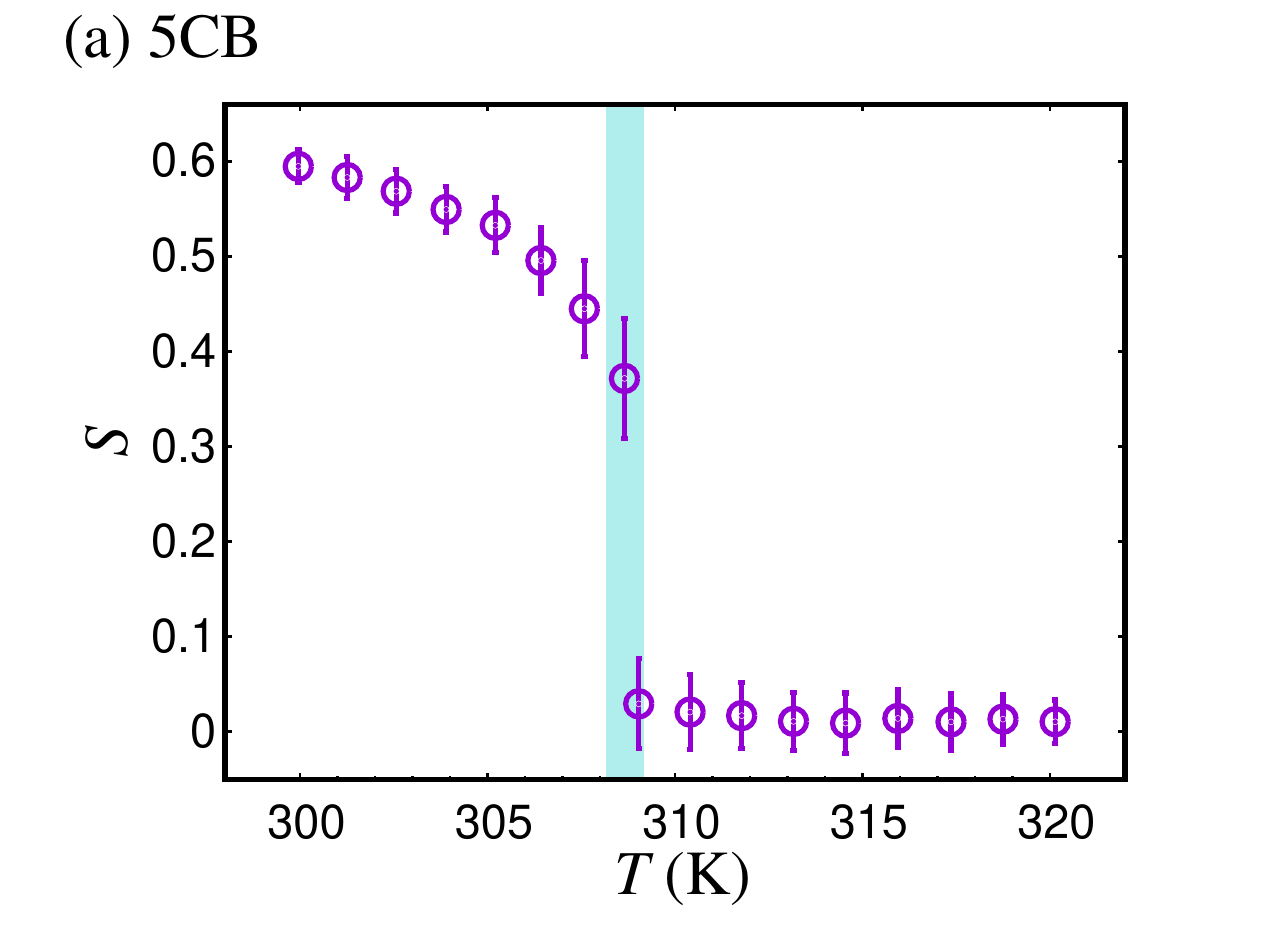}
\includegraphics[width=0.4\textwidth]{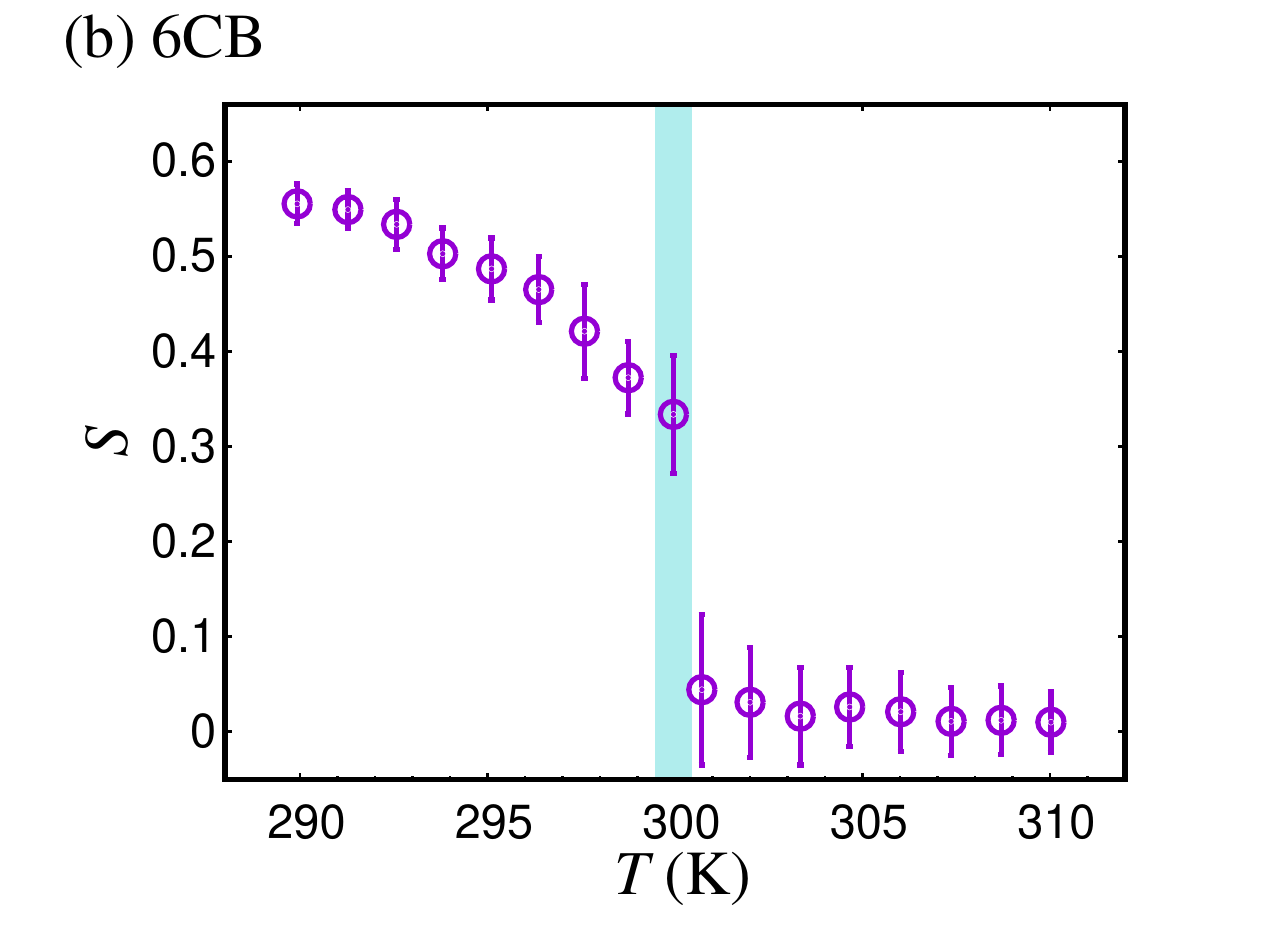}
\includegraphics[width=0.4\textwidth]{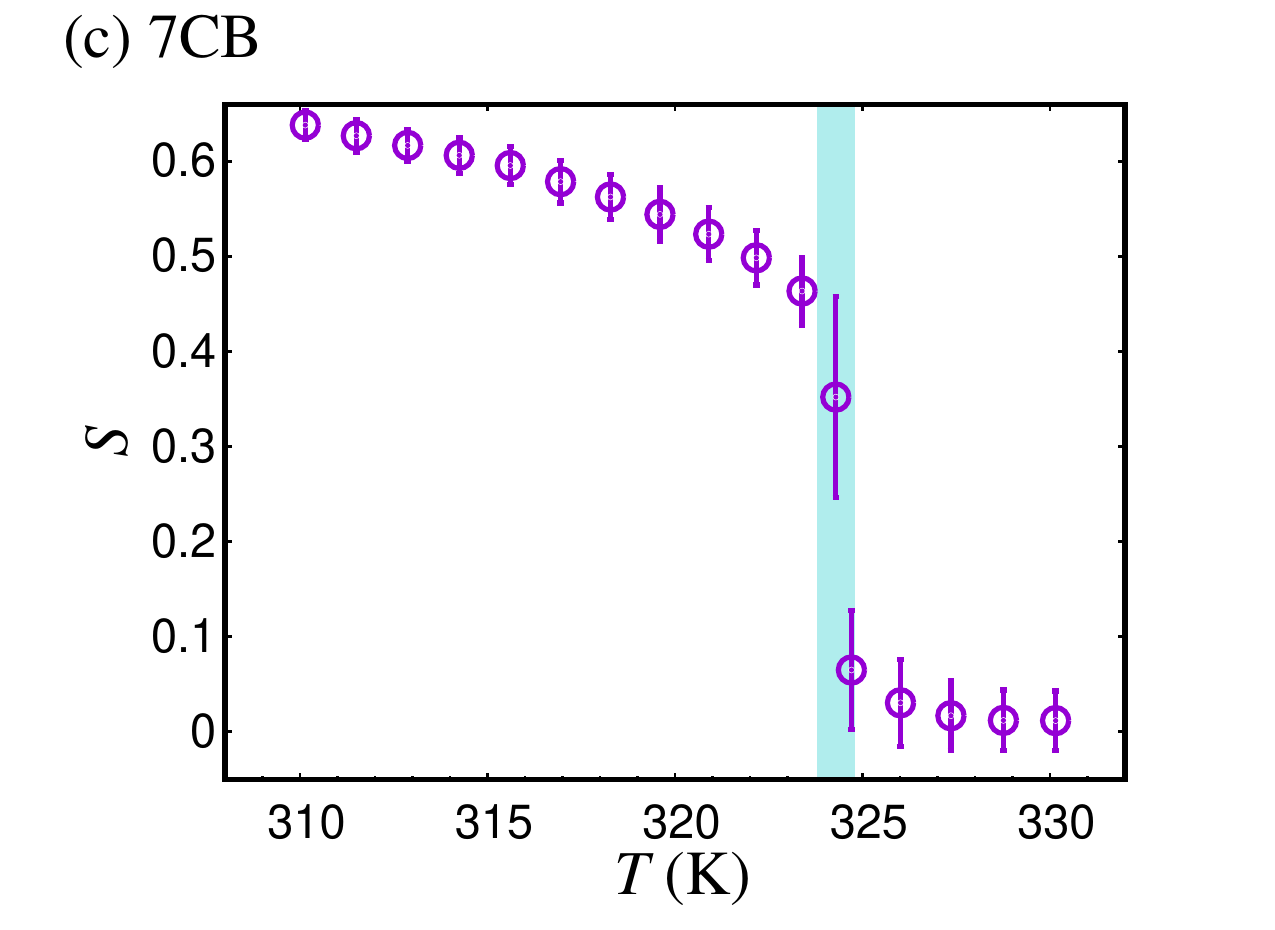}
\includegraphics[width=0.4\textwidth]{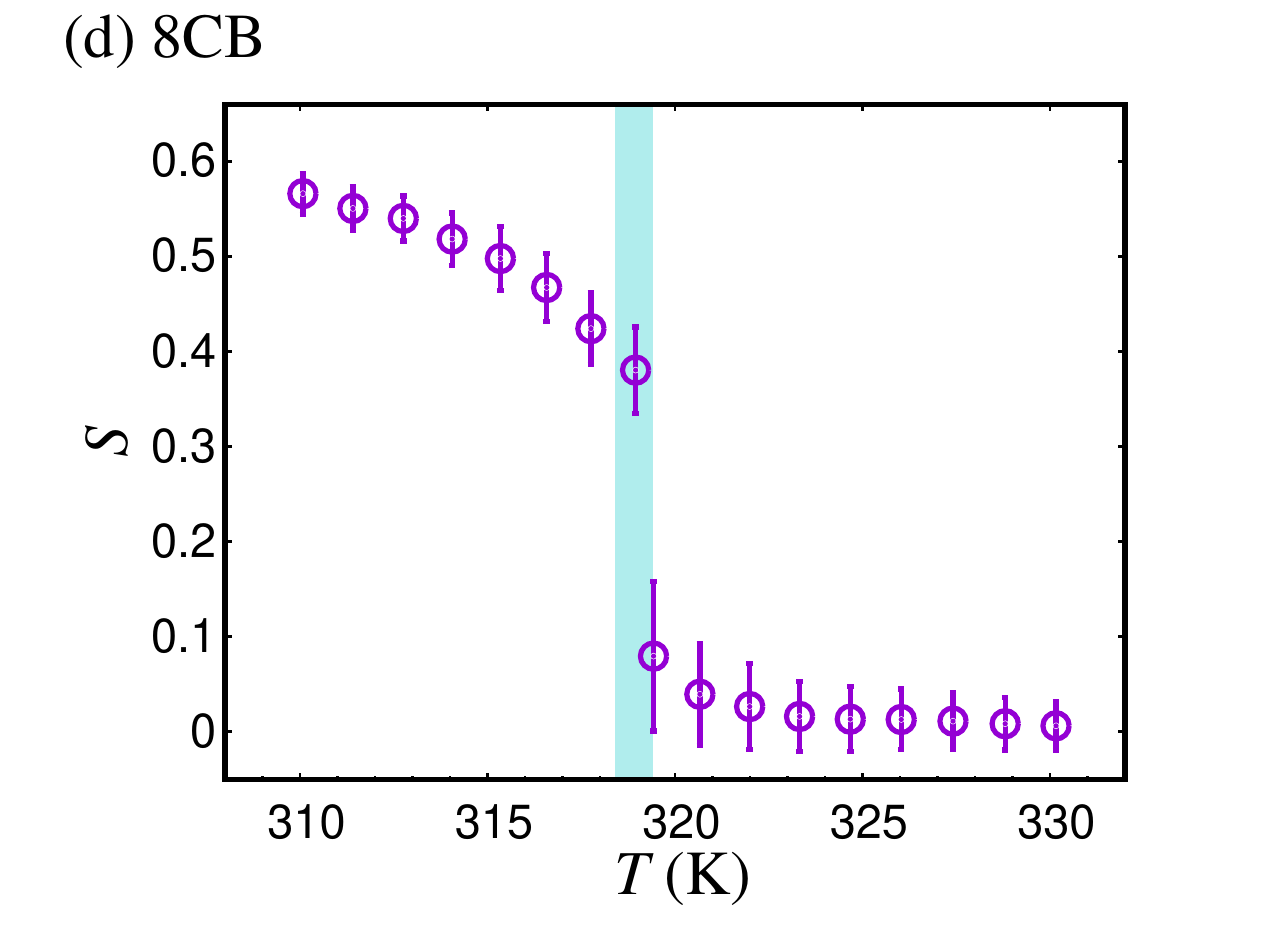}
\caption{Temperature dependence of orientational order parameter $S$ for
 5CB (a), 6CB (b), 7CB (c), and 8CB (d).
The molecular axis is defined using the line
connecting N and C of the phenyl ring, represented by the
 red vector in Fig.~S1 of the supplementary material.
The error bar at each temperature corresponds to the standard deviation.
The vertical color bar indicates the NI transition temperature $T_\mathrm{NI}$.}
\end{figure}
%%%%%%%%%%%%%%%%%%%%%%%%%%%%%%%%%%%%%%%%%%%%%%%%%%%%%%%%%%%%%%%%%%%%%%%%%%%%%%%%%%%%%%%%%%

\newpage
%%%%%%%% Fig. S6 %%%%%%%%%%%%%%%%%%%%%%%%%%%%%%%%%%%%%%%%%%%%%%%%%%%%%%%%%%%%%%%%%%%%%%%%%
\begin{figure}[t]
\centering
\includegraphics[width=0.4\textwidth]{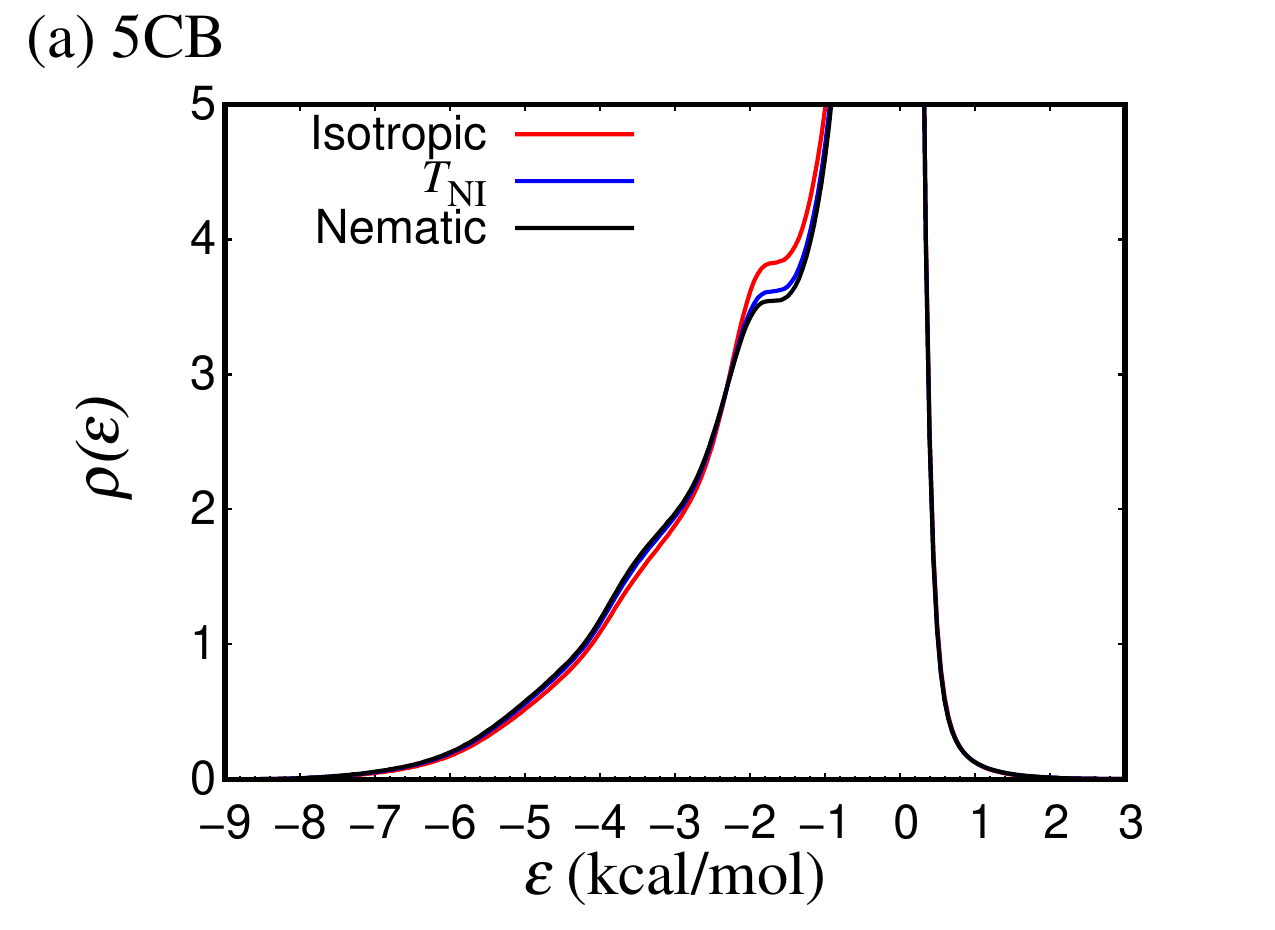}
\includegraphics[width=0.4\textwidth]{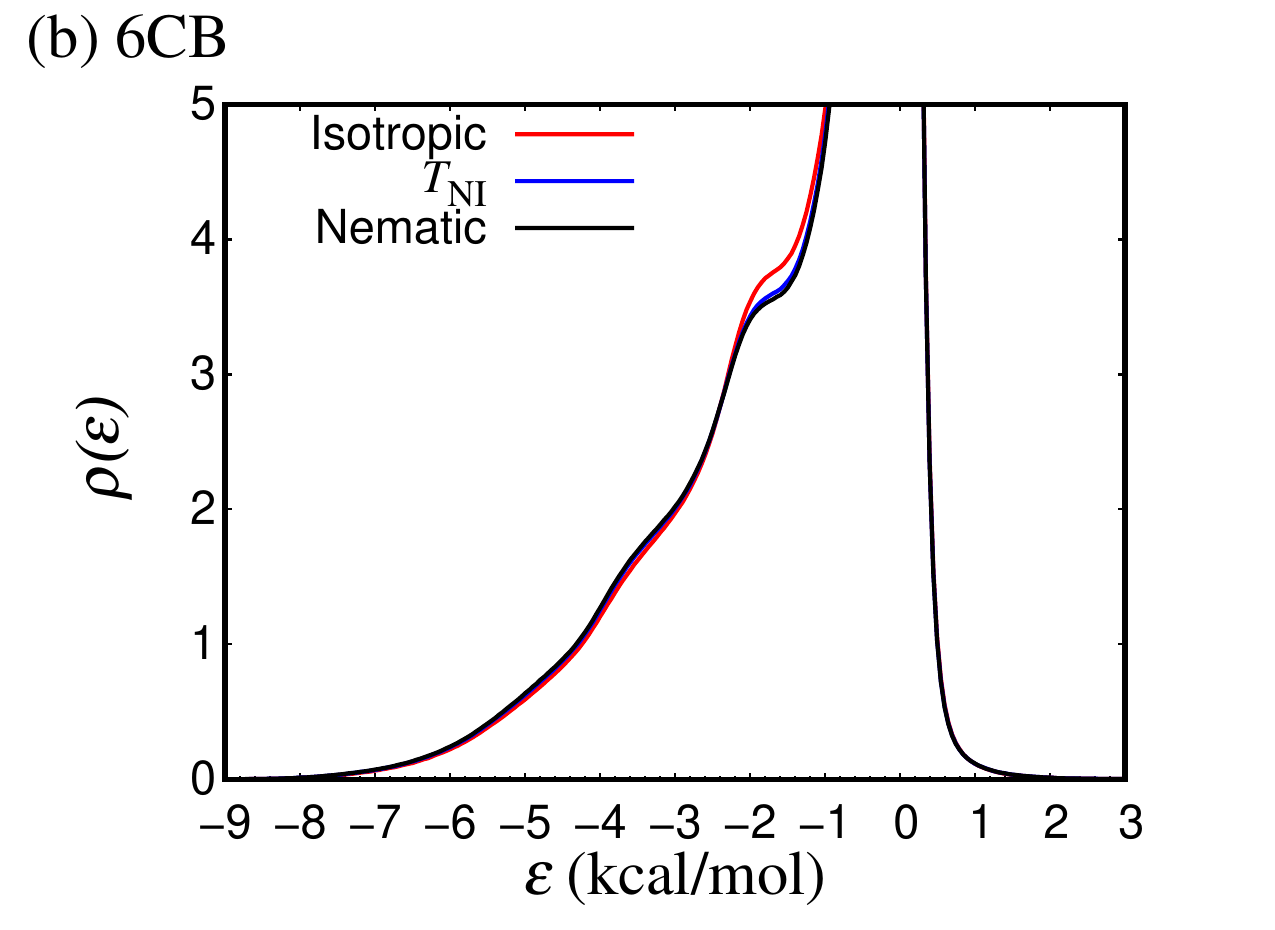}
\includegraphics[width=0.4\textwidth]{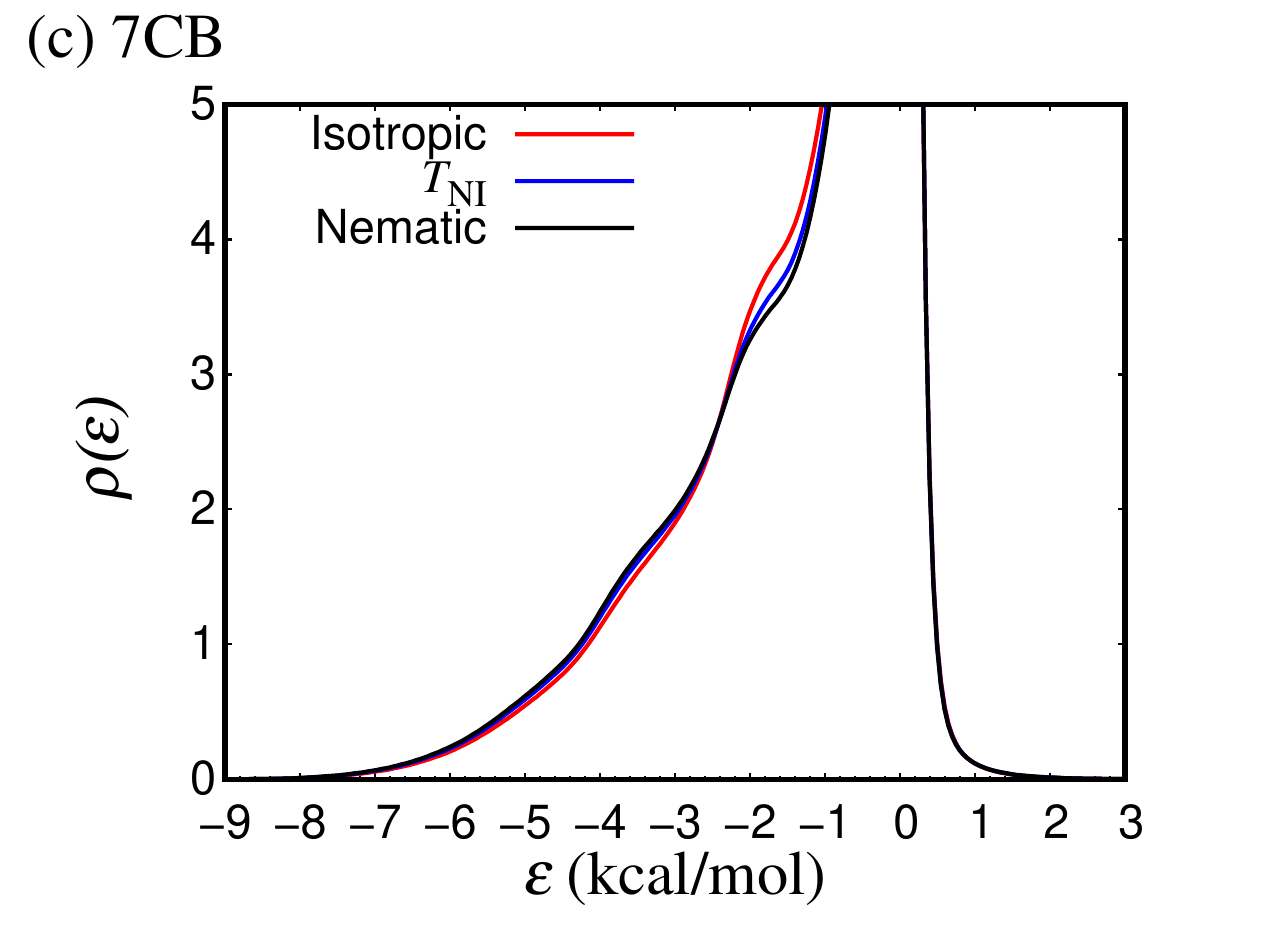}
\includegraphics[width=0.4\textwidth]{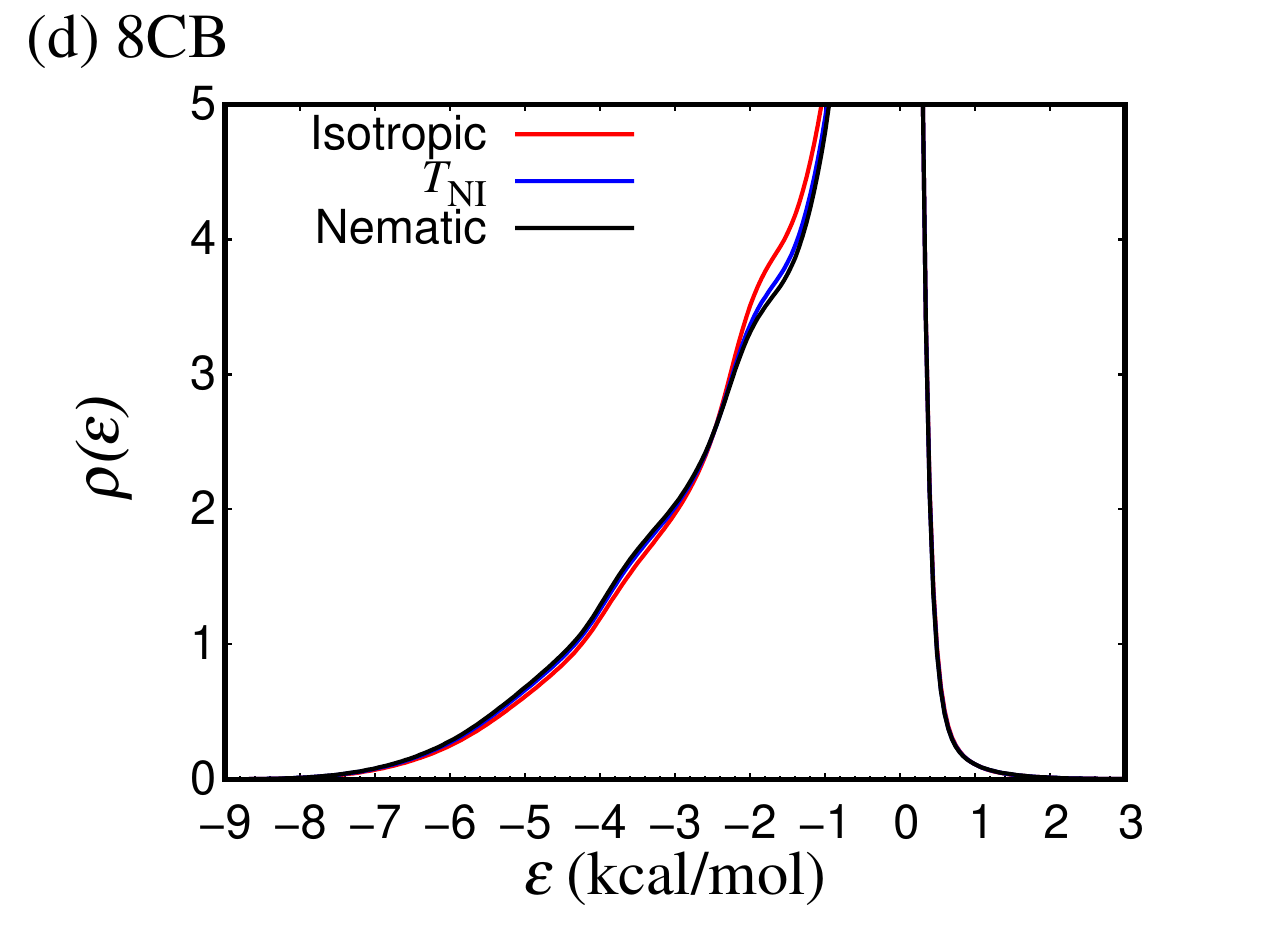}
\caption{Density profiles of intermolecular interaction energy,
 $\rho(\epsilon)$, obtained from MD simulations for
 5CB (a), 6CB (b), 7CB (c), and 8CB (d).
The results are plotted for replicas at $T_\mathrm{NI}$, as well as for
 replicas above and below $T_\mathrm{NI}$ (isotropic and nematic phases).
}
\end{figure}
%%%%%%%%%%%%%%%%%%%%%%%%%%%%%%%%%%%%%%%%%%%%%%%%%%%%%%%%%%%%%%%%%%%%%%%%%%%%%%%%%%%%%%%%%%

\newpage
%%%%%%%% Fig. S7 %%%%%%%%%%%%%%%%%%%%%%%%%%%%%%%%%%%%%%%%%%%%%%%%%%%%%%%%%%%%%%%%%%%%%%%%%
\begin{figure}[t]
\centering
\includegraphics[width=0.4\textwidth]{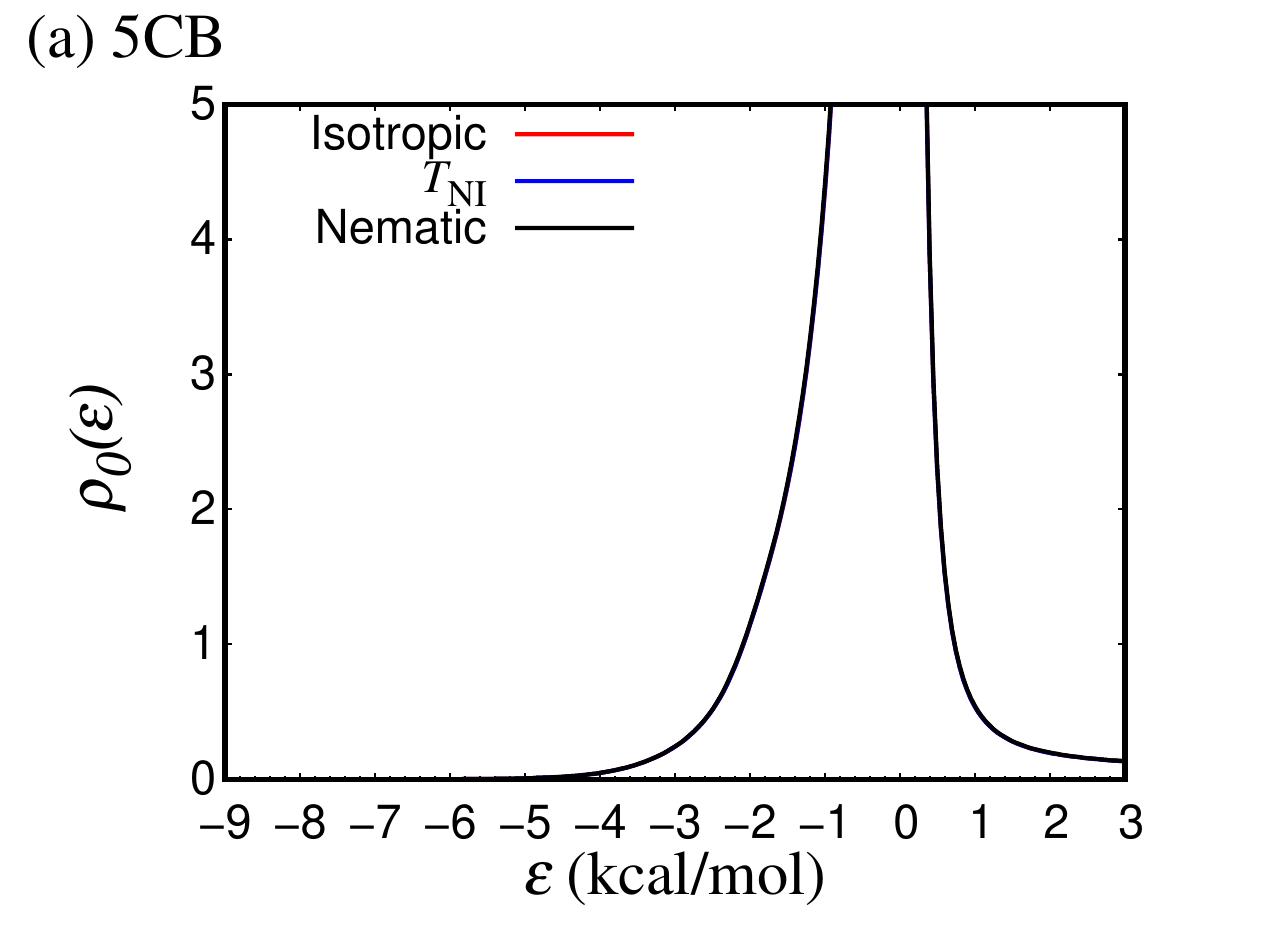}
\includegraphics[width=0.4\textwidth]{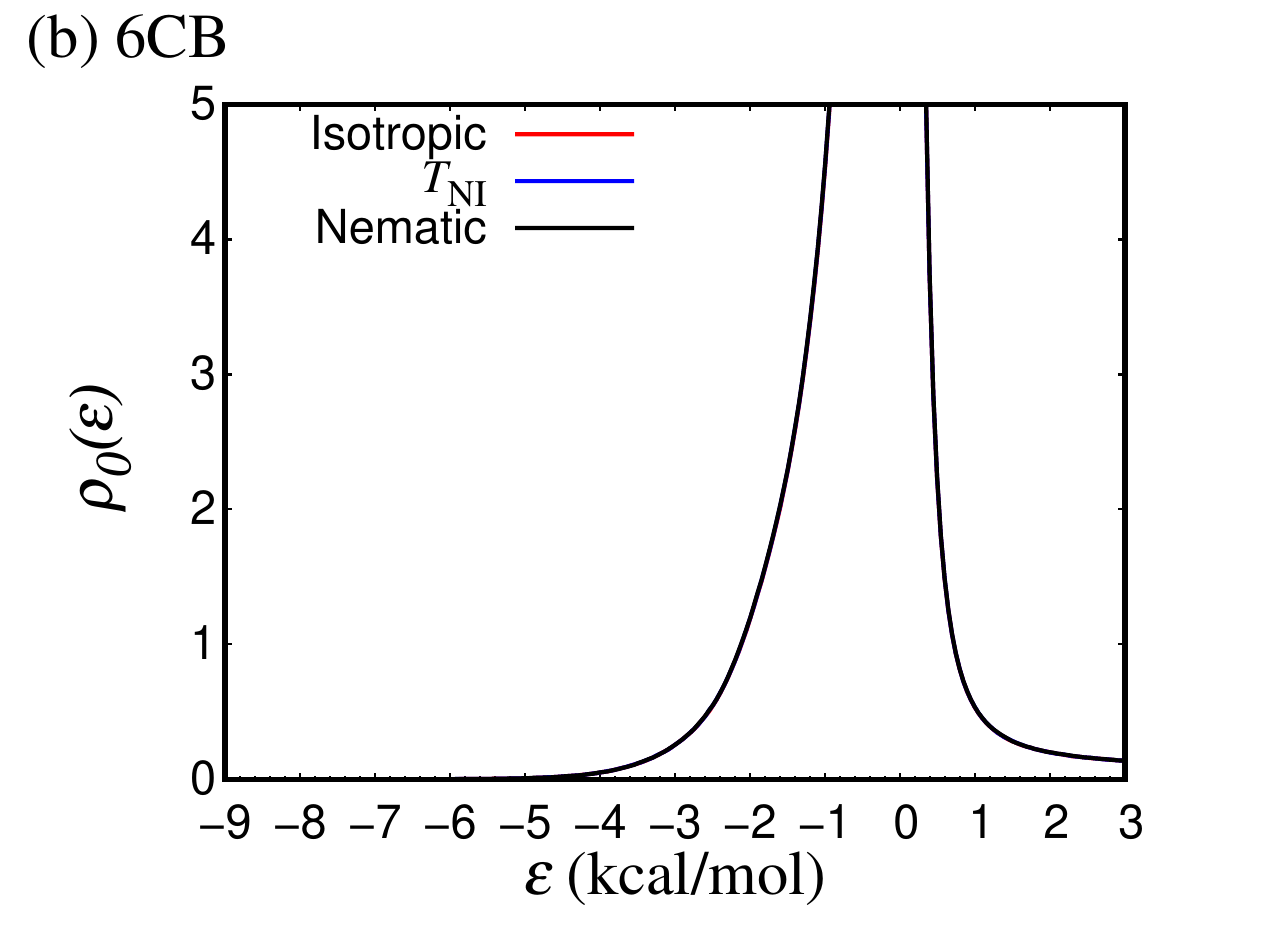}
\includegraphics[width=0.4\textwidth]{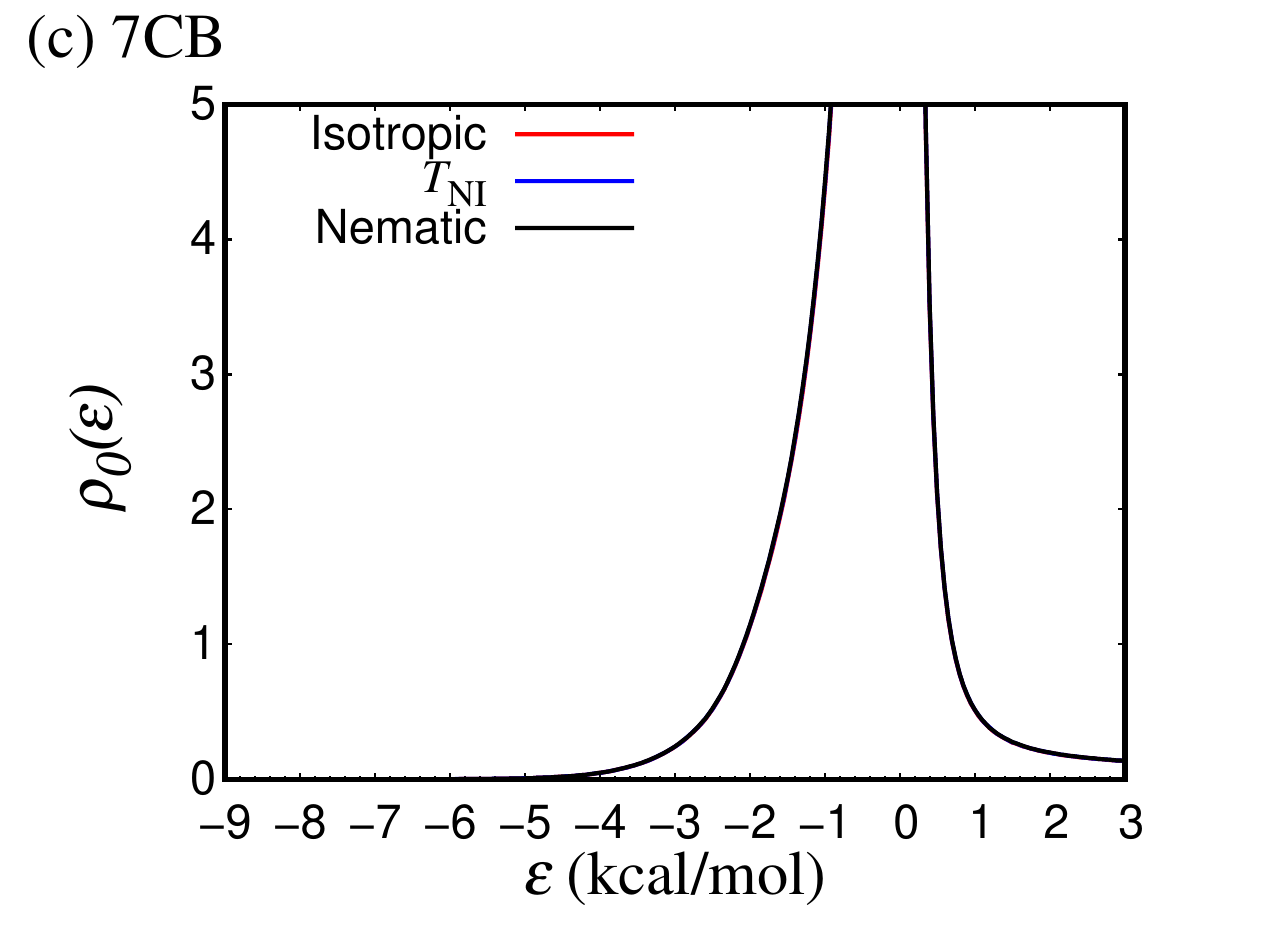}
\includegraphics[width=0.4\textwidth]{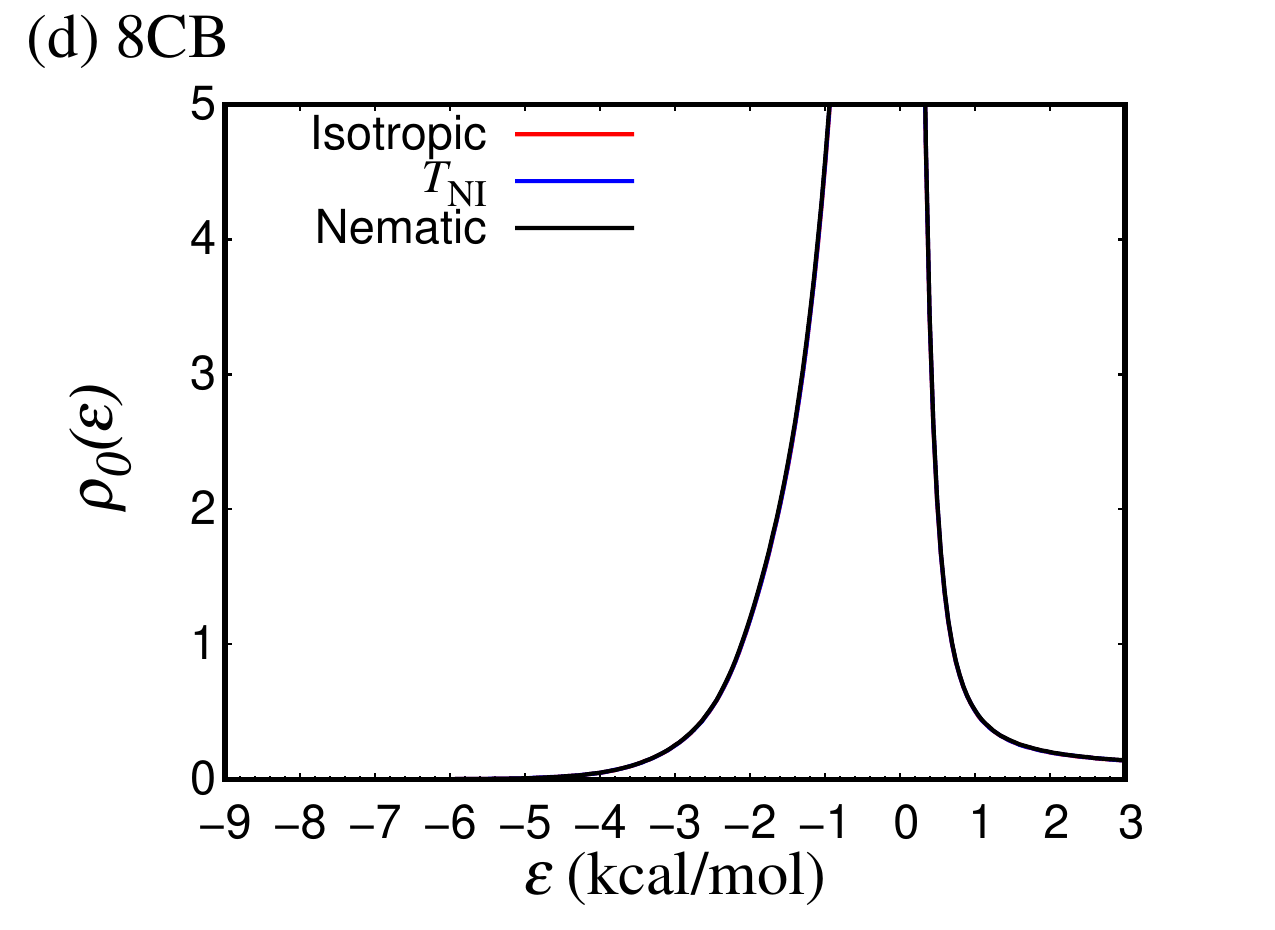}
\caption{Density profiles of intermolecular interaction energy,
 $\rho_0(\epsilon)$, obtained from MD simulations for
 5CB (a), 6CB (b), 7CB (c), and 8CB (d).
The results are plotted for replicas at $T_\mathrm{NI}$, as well as for
 replicas above and below $T_\mathrm{NI}$ (isotropic and nematic phases).
}
\end{figure}
%%%%%%%%%%%%%%%%%%%%%%%%%%%%%%%%%%%%%%%%%%%%%%%%%%%%%%%%%%%%%%%%%%%%%%%%%%%%%%%%%%%%%%%%%%

\newpage
%%%%%%%% Fig. S8 %%%%%%%%%%%%%%%%%%%%%%%%%%%%%%%%%%%%%%%%%%%%%%%%%%%%%%%%%%%%%%%%%%%%%%%%%
\begin{figure}[t]
\centering
\includegraphics[width=0.4\textwidth]{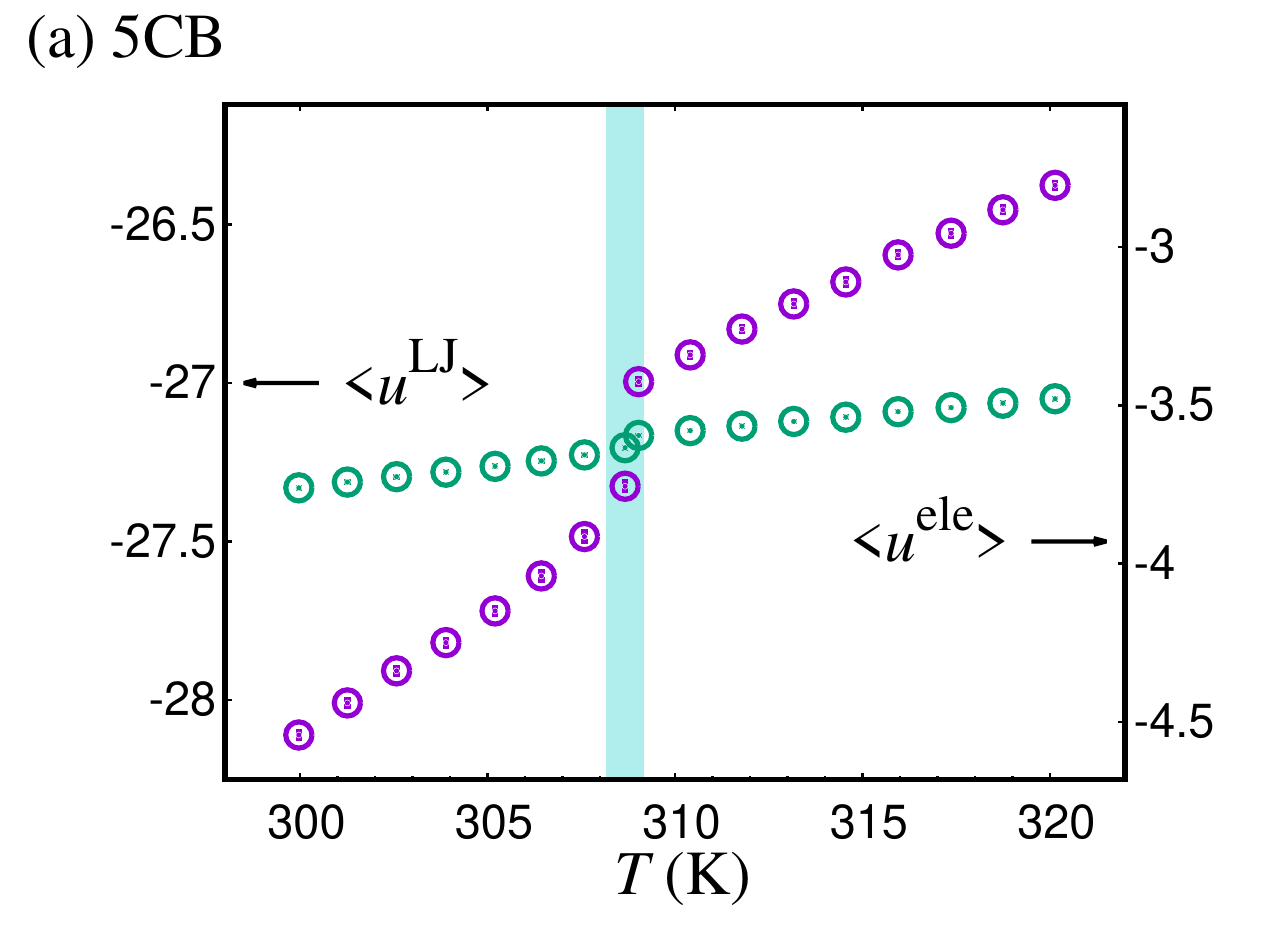}
\includegraphics[width=0.4\textwidth]{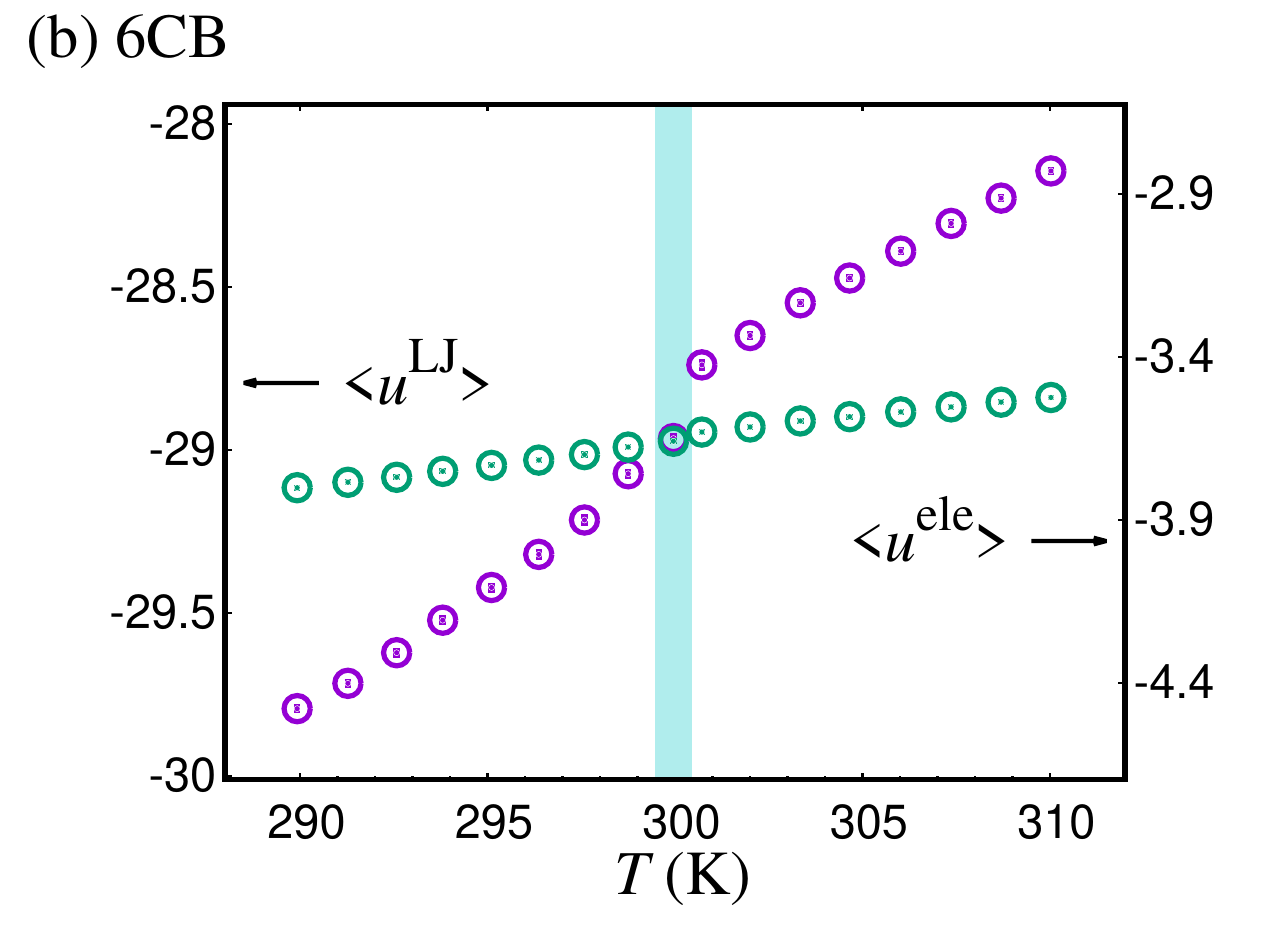}
\includegraphics[width=0.4\textwidth]{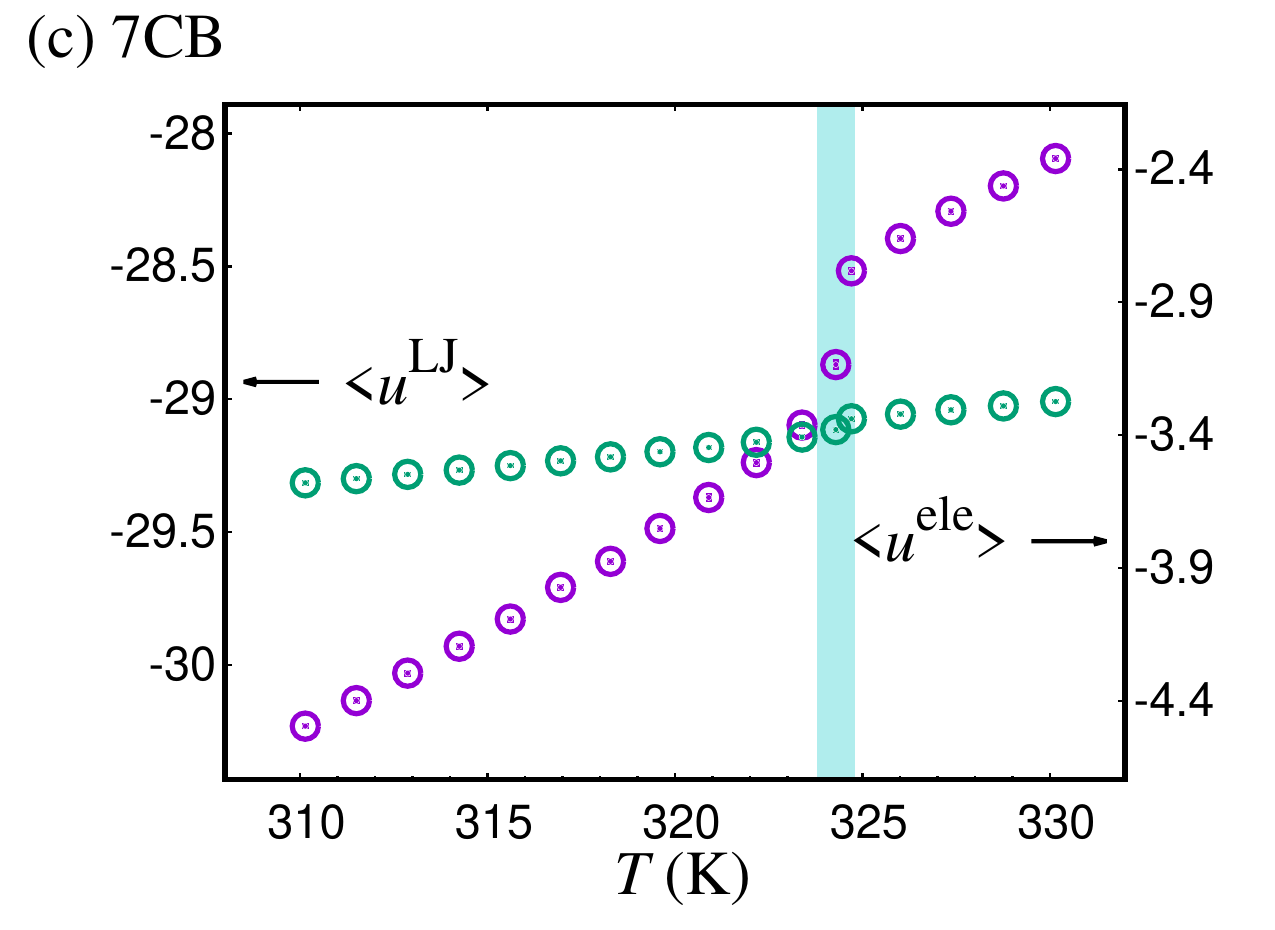}
\includegraphics[width=0.4\textwidth]{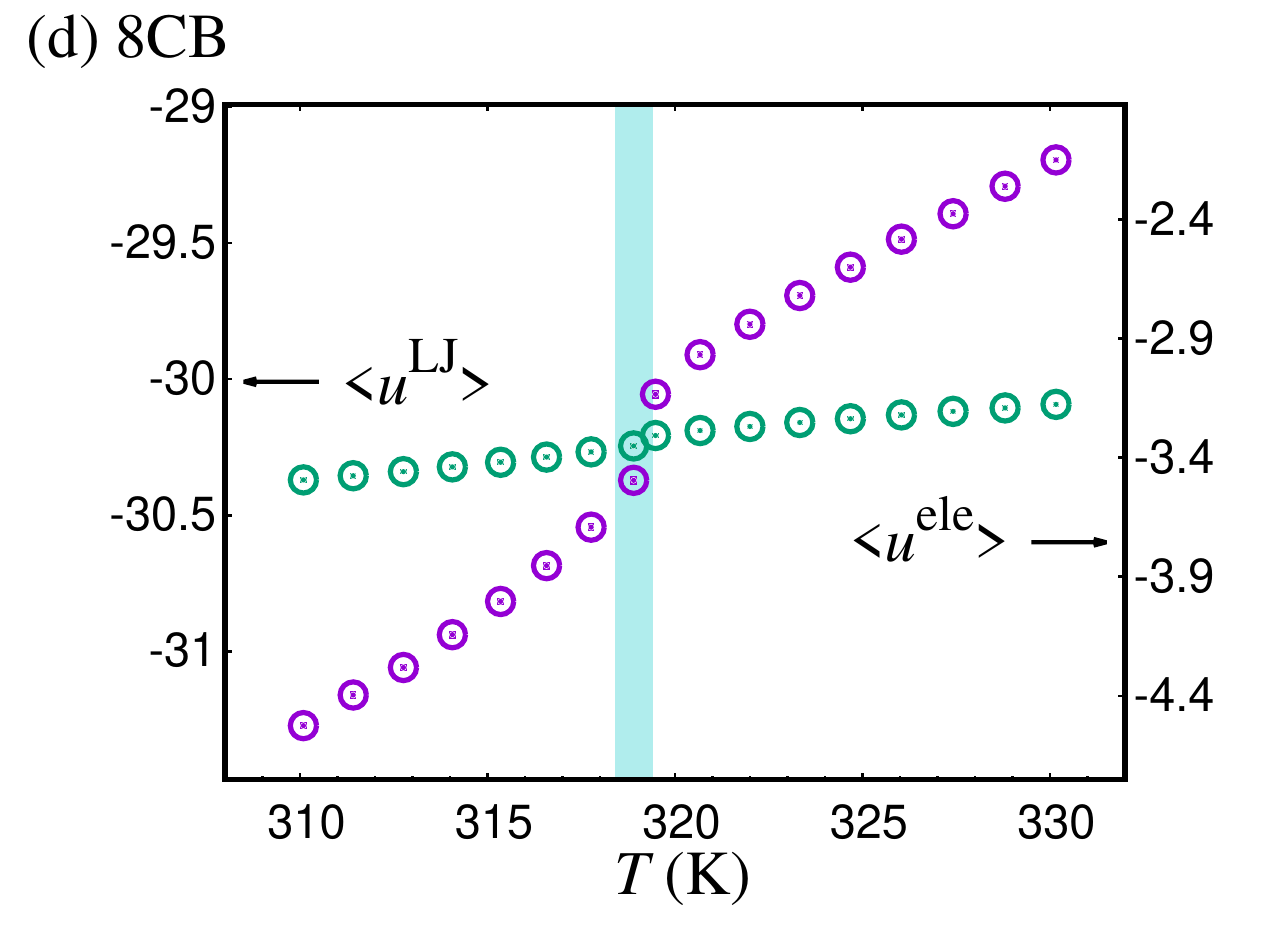}
\caption{Temperature dependence of the decomposition of the
 intermolecular interaction energy $\langle u\rangle$ into LJ and
 electrostatic
 contributions, $\langle u^\mathrm{LJ}\rangle$
and $\langle u^\mathrm{ele}\rangle$, 
for 5CB (a), 6CB (b), 7CB (c), and 8CB (d).
Note that $\langle u^\mathrm{LJ}\rangle$ (purple)
and $\langle u^\mathrm{ele}\rangle$ (green) are
 plotted on 
the left and right $y$-axes, respectively, in units of kcal/mol.
The error bars at each temperature represent the standard deviation.
The vertical color bar indicates the NI transition temperature $T_\mathrm{NI}$.
}
\end{figure}
%%%%%%%%%%%%%%%%%%%%%%%%%%%%%%%%%%%%%%%%%%%%%%%%%%%%%%%%%%%%%%%%%%%%%%%%%%%%%%%%%%%%%%%%%%

%%%%%%%%% Fig. S9 %%%%%%%%%%%%%%%%%%%%%%%%%%%%%%%%%%%%%%%%%%%%%%%%%%%%%%%%%%%%%%%%%%%%%%%%%
\begin{figure}[t]
\centering
\includegraphics[width=0.4\textwidth]{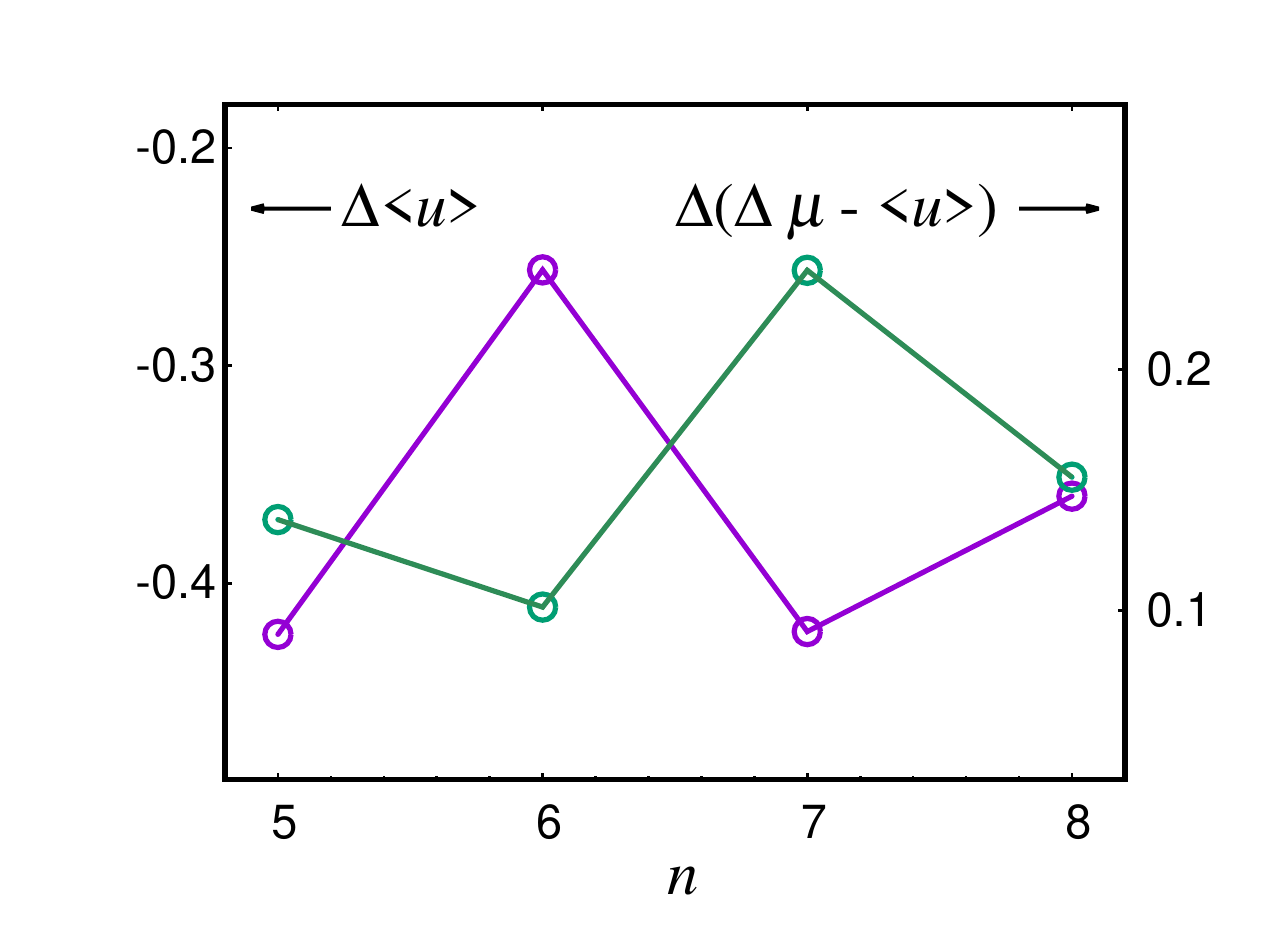}
\caption{Dependence of the energy gap $\Delta \langle u \rangle$ and
 entropy gap $\Delta (\Delta
\mu- \langle u\rangle)$ across the NI transition on the number of carbons in
 the alkyl chain, $n$.
Note that $\Delta \langle u \rangle$ (purple) and $\Delta (\Delta \mu-
 \langle u\rangle)$ (green) are plotted on the left and right $y$-axes,
 respectively, in units of kcal/mol.}
\end{figure}
%%%%%%%%%%%%%%%%%%%%%%%%%%%%%%%%%%%%%%%%%%%%%%%%%%%%%%%%%%%%%%%%%%%%%%%%%%%%%%%%%%%%%%%%%%